%% file: main.tex
\documentclass{article}
\usepackage{iclr2027_conference,times}

\input{math_commands.tex}

\usepackage{hyperref}
\usepackage{url}
\usepackage{booktabs}
\usepackage{enumitem}
\usepackage{subcaption}
\usepackage{graphicx}
\usepackage{lipsum}
\usepackage{wrapfig}

\title{
    Event-guided Neural Video Compression
}

\author{
    Jiyun Kong\textsuperscript{\rm 1*},
    Jungwoo Kim\textsuperscript{\rm 2*}, 
    Enes Eray Demirtas\textsuperscript{\rm 3},
    Touradj Ebrahimi\textsuperscript{\rm 3$\dagger$}, 
    Jong-Seok Lee\textsuperscript{\rm 2$\dagger$} \\
    \textsuperscript{\rm 1}Department of Artificial Intelligence, Yonsei University, Republic of Korea \\
    \textsuperscript{\rm 2}School of Integrated Technology, Yonsei University, Republic of Korea \\
    \textsuperscript{\rm 3}Multimedia Signal Processing Group, EPFL, Switzerland \\
    \texttt{\{jiyun.kong, kjungwoo, jong-seok.lee\}@yonsei.ac.kr} \\
    \texttt{\{enes.demirtas, touradj.ebrahimi\}@epfl.ch}
}

\iclrfinalcopy

\begin{document}

\maketitle

\begingroup
\renewcommand\thefootnote{}
\footnotetext{\textsuperscript{*}Equal contribution.}
\footnotetext{\textsuperscript{$\dagger$}Corresponding authors.}
\endgroup

\input{sec/0_abstract}
\input{sec/1_intro}
\input{sec/2_rework}
\input{sec/3_dataset}
\input{sec/4_method}

\input{sec/5_experiments}

\input{sec/6_discussion}
\input{sec/7_conclusion}

\newpage
\clearpage

\bibliography{iclr2027_conference}
\bibliographystyle{iclr2027_conference}

\newpage
\clearpage

\appendix
\input{sec/a_scenario}
\input{sec/b_training_details}
\input{sec/c_case_analysis}
\input{sec/d_real_event}
\input{sec/e_gate_analysis}
\input{sec/f_sensitivity_to_event_perturbations}
\input{sec/g_bit_allocation.tex}
\input{sec/h_additional_datasets_quality_metrics}
\input{sec/i_additional_examples}

\end{document}

%% file: math_commands.tex
\usepackage{amsmath,amsfonts,bm}

\def\eqref#1{equation~\ref{#1}}

\def\1{\bm{1}}

\DeclareMathAlphabet{\mathsfit}{\encodingdefault}{\sfdefault}{m}{sl}
\SetMathAlphabet{\mathsfit}{bold}{\encodingdefault}{\sfdefault}{bx}{n}



%% file: sec/0_abstract.tex
\begin{abstract}
Neural video codecs derive motion and temporal contexts mainly from RGB frames, leaving room for cross-modal guidance from complementary temporal observations.
Event streams can provide such observations by recording brightness changes between frames.
In this work, we propose an \textbf{Event-guided Neural Video Codec (ENVC)} that uses events shared by the encoder and decoder to improve RGB compression efficiency.
For motion coding, ENVC forms an event-guided motion prior and codes the remaining motion residual.
For frame coding, an event-conditioned predictor supplies multi-scale features for gated temporal context refinement.
To support training and evaluation on standard video datasets, we synthesize paired RGB--event data and assess its predictive utility through comparisons with real events.
Across six benchmarks, ENVC achieves average BD-rate savings of \textbf{39.13\%} using PSNR-RGB and \textbf{67.63\%} using LPIPS relative to DCMVC.
Further analyses show that our gains persist on large-motion sequences and that ENVC effectively learns to integrate event information.
These results demonstrate the potential of events as a complementary modality for reducing the RGB coding rate.
Our model and code are available at \url{https://github.com/kjungwoo03/ENVC}.
\end{abstract}

%% file: sec/1_intro.tex
\section{Introduction}
\label{sec:intro}

Video accounts for a large share of network traffic, making efficient video compression a central problem in visual communication.
Neural video codecs (NVCs) optimize their coding modules jointly under an end-to-end rate--distortion (RD) objective~\citep{lu2019dvc}.
Conditional coding methods, which code the current frame using temporal contexts built from previously decoded information, have demonstrated RD gains over conventional codecs~\citep{li2021dcvc,li2023dcvcdc}.
The quality of these contexts therefore strongly affects the bitrate needed for a given reconstruction quality.
Accordingly, recent studies have refined context modeling via multi-scale propagation~\citep{sheng2022temporal}, reference-guided modulation~\citep{tang2025dcmvc}, and non-local multi-reference mining~\citep{jiang2025ecvc}.

\begin{figure*} 
    \centering
    \begin{subfigure}[t]{0.32\textwidth}
        \centering
        \includegraphics[width=\linewidth]{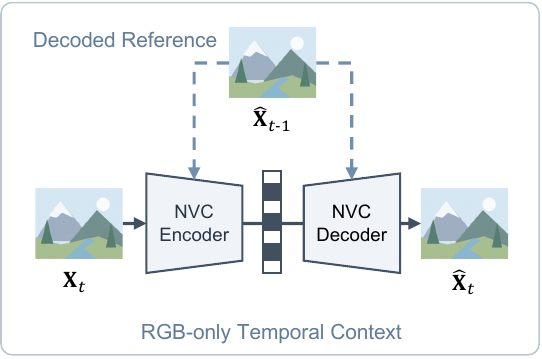}
        \caption{RGB-only Video Coding}
        \label{fig:scheme-nvc}
    \end{subfigure}\hfill
    \begin{subfigure}[t]{0.32\textwidth}
        \centering
        \includegraphics[width=\linewidth]{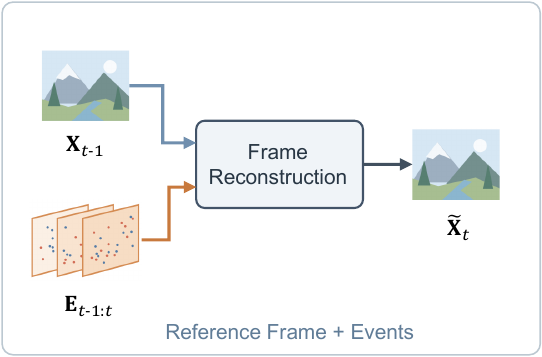}
        \caption{Event-based Reconstruction}
        \label{fig:scheme-event}
    \end{subfigure}\hfill
    \begin{subfigure}[t]{0.32\textwidth}
        \centering
        \includegraphics[width=\linewidth]{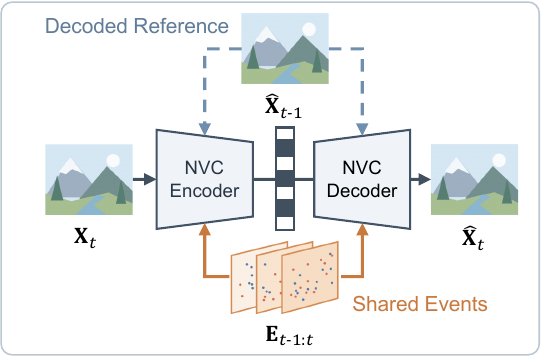}
        \caption{Event-guided Video Coding}
        \label{fig:scheme-ours}
    \end{subfigure}
    \vspace{-0.5em}
    \caption{
        \textbf{Comparison of Video Coding and Reconstruction Settings.}
        (a) RGB-only video coding reconstructs the current frame from a decoded reference and an RGB coding bitstream.
        (b) Event-based reconstruction uses a reference frame and inter-frame events without an additional current-frame RGB bitstream.
        (c) Event-guided video coding uses events shared by the encoder and decoder to construct temporal contexts and code the current frame under a rate--distortion objective.
    }
    \label{fig:scheme}
    \vspace{-1.5em}
\end{figure*}

Despite these advances, temporal contexts in the RGB-only framework (Fig.~\ref{fig:scheme-nvc}) depend only on previously decoded frames and propagated features.
Diverse motion patterns can make prediction from these references less reliable, while reconstruction errors can accumulate over long prediction chains.
In both cases, more RGB information may be needed to maintain reconstruction quality.
We therefore ask whether complementary observations of inter-frame changes can improve temporal contexts and reduce the required RGB bitrate.

To investigate this possibility, we focus on event streams, which asynchronously record brightness changes between video frames.
An event is triggered whenever the magnitude of the log-intensity change at a pixel exceeds a contrast threshold, with its polarity indicating the sign of the change.
This mechanism yields microsecond-level temporal resolution and a high dynamic range~\citep{gallego2022event}.
These properties have motivated a range of recent event-based methods for motion analysis~\citep{gehrig2021raft,wang2024evimo} and video reconstruction and interpolation~\citep{rebecq2019events,weng2021etnet,sun2023refid}.

A reference frame and inter-frame events can be used to reconstruct the current frame directly, as illustrated in Fig.~\ref{fig:scheme-event}.
However, without additional current-frame RGB information, prediction errors can limit reconstruction fidelity~\citep{zhu2024vfpsie,kong2025dessert}.
We instead consider RGB coding with events shared by the encoder and decoder (Fig.~\ref{fig:scheme-ours}).
Both sides use these events to construct temporal contexts, while a bitstream carries additional current-frame RGB information to improve reconstruction quality.
Similar shared-side-information settings have been explored in prior work on cross-modal compression~\citep{gnutti2024lidar, jiang2024sparse, song2026lowlatency}, and we discuss practical scenarios in App.~\ref{app:scenario}.

In this work, we investigate how shared event observations can reduce the RGB rate required for reconstruction.
We propose an \textbf{Event-guided Neural Video Codec (ENVC)} that incorporates these observations into motion coding and temporal context construction.
ENVC forms an event-guided motion prior at both the encoder and decoder and codes the remaining motion residual.
For frame coding, an event-conditioned predictor generates a frame prediction and multi-scale features.
Learned gates inject these features into temporal contexts, which are further refined using the frame prediction.

To train and evaluate ENVC on standard NVC datasets, we generate paired events from their RGB videos through frame interpolation and event simulation (Sec.~\ref{sec:dataset}).
We assess the utility of the synthetic events by comparing event statistics and next-frame prediction quality with those obtained using real events.
Across six standard video coding benchmarks paired with synthetic events, ENVC achieves average BD-rate savings of 39.13\% using PSNR-RGB and 67.63\% using LPIPS~\citep{zhang2018unreasonable} relative to DCMVC~\citep{tang2025dcmvc} under the shared-event setting.
Our evaluation further analyzes event utilization, motion-dependent performance, sensitivity to event perturbations, and RD differences between real and synthetic event inputs.

Our contributions can be summarized as follows:
\begin{itemize}[leftmargin=10pt, topsep=0.8pt, itemsep=1.8pt, parsep=0pt]
    \item We investigate shared events as a complementary modality for RGB video compression and propose ENVC, which combines event-guided residual motion coding with gated temporal context refinement.
    \item We construct paired RGB--event datasets for standard NVC training and evaluation, and assess the synthetic events through comparisons with real events in terms of event statistics and next-frame prediction quality.
    \item We demonstrate substantial RGB coding gains across six standard benchmarks using PSNR-RGB and LPIPS under the shared-event setting. 
    These gains persist on large-motion sequences, supporting the use of events as temporal guidance for RGB video coding.
\end{itemize}

%% file: sec/2_rework.tex
\section{Related Work}
\label{sec:rework}

\paragraph{Neural Video Compression.}
Beyond conventional hybrid codecs~\citep{wiegand2003overview,sullivan2012overview,bross2021overview}, NVCs learn motion and residual-based coding~\citep{lu2019dvc,agustsson2020scale,hu2021fvc,rippel2021elfvc} or code frames conditioned on temporal contexts~\citep{ladune2021conditional,ho2022canf,li2021dcvc}.
Context modeling has advanced through multi-scale mining~\citep{sheng2022temporal}, spatio-temporal entropy modeling~\citep{li2022hybrid,li2023dcvcdc}, reference modulation~\citep{tang2025dcmvc}, and multi-reference aggregation~\citep{jiang2025ecvc}.
Other works address quality control and temporal stability~\citep{li2024neural}, speed~\citep{jia2025towards,li2026ultra}, buffer requirements~\citep{chen2025hytip}, and model capacity~\citep{tang2026nvc1b}.
ENVC uses shared event observations to complement the RGB-derived temporal information used by these methods.

\vspace{-0.8em}
\paragraph{Event-based Frame Reconstruction.}
Event-only methods reconstruct intensity frames from event streams~\citep{rebecq2019events,weng2021etnet,zhu2024temporal}, although events do not directly measure absolute intensity~\citep{gallego2022event}.
Frame-guided methods combine RGB images and events for interpolation~\citep{tulyakov2021time,tulyakov2022time} or prediction from a single reference~\citep{zhu2024vfpsie,kong2025dessert,wang2025continuitycam}.
Single-reference prediction is compatible with sequential decoding, as a previously reconstructed frame can serve as the reference without requiring future RGB frames.
ENVC uses event-based prediction to guide compression and codes additional current-frame RGB information under an RD objective.

\vspace{-0.8em}
\paragraph{Cross-modal Guidance for Compression.}
Joint visible--infrared and RGB--depth codecs exploit inter-modality redundancy~\citep{lu2022learning,zheng2024end}.
Auxiliary modalities also guide RGB compression as geometric~\citep{gnutti2024lidar,jiang2024sparse} or semantic side information~\citep{jiang2023multi,lee2024neural,hu2026movie}, including semantic residual coding~\citep{resulic2025} and text-conditioned latent refinement for machine vision~\citep{lee2025slim}.
Depth-assisted video compression uses geometric priors for temporal context and entropy modeling~\citep{sheng2026depth}.
Our work first investigates shared events as temporal side information, using their observations of inter-frame brightness changes to guide motion coding and temporal context construction.

%% file: sec/3_dataset.tex
\begin{figure}
    \centering
    \includegraphics[width=\linewidth]{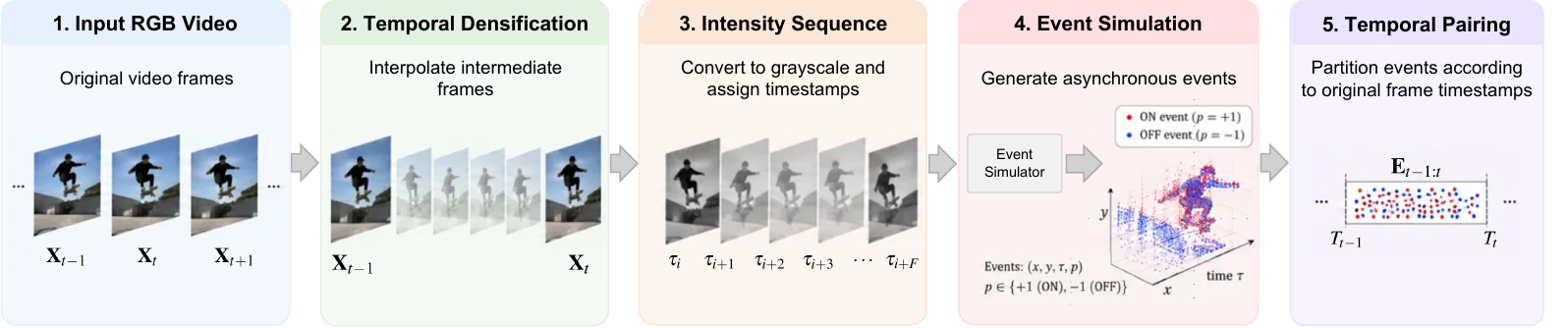}
    \vspace{-1.0em}
    \caption{
        \textbf{Synthetic RGB--Event Dataset Construction.}
        Temporally densified frames are used for event synthesis, while the original RGB frames are retained for coding.
        The resulting event stream is partitioned at the original frame timestamps into groups $\mathbf{E}_{t-1:t}$ aligned with each coding interval.
    }
    \label{fig:pipeline}
    \vspace{-1.0em}
\end{figure}

\section{Synthetic RGB--Event Dataset Construction} 
\label{sec:dataset} 

Training and evaluating RGB compression with shared events requires paired RGB--event data.
Real-event datasets provide such pairs~\citep{tulyakov2021time,tulyakov2022time,kim2023event}, while paired events are unavailable for Vimeo-90K~\citep{xue2019video}, a widely used dataset for NVC training, and standard video coding benchmarks~\citep{flynn2013common,wang2016mcl,mercat2020uvg,cheon2018mcml4k}.
We therefore synthesize events from these RGB datasets using the pipeline in Fig.~\ref{fig:pipeline}, enabling RD comparisons with prior work on the same video content.

\vspace{-0.5em}
\subsection{Event Synthesis Pipeline}
\label{sec:dataset_pipeline}

\paragraph{Temporal Densification.}
We use Super-SloMo~\citep{jiang2018super} to insert $F-1$ intermediate frames between each adjacent pair in the original sequence $\{\mathbf{X}_t\}_{t=1}^{N}$.
Given the source frame rate $f_\mathrm{src}$, we set $F = \lceil f_\mathrm{target}/f_\mathrm{src} \rceil$ with $f_\mathrm{target}=200$~fps, yielding a nominal frame rate of at least $f_\mathrm{target}$.
This gives a densified sequence $\{\mathbf{Y}_i\}_{i=1}^{F(N-1)+1}$, which we convert to $8$-bit grayscale and assign timestamps $\tau_i = i \cdot \lfloor 10^{6} / (F f_\mathrm{src}) \rceil$ in microseconds, yielding an intensity sequence $I(\tau)$ at each pixel.
The densified grayscale frames are used only for event synthesis.

\vspace{-0.8em}
\paragraph{Event Simulation.}
We generate events from $I(\tau)$ with DVS-Voltmeter~\citep{lin2022dvs}, a stochastic simulator that models per-pixel voltage changes and event triggering.
We use the DVS346\footnote{\url{https://github.com/Lynn0306/DVS-Voltmeter}.} preset of the simulator without additional calibration.

\vspace{-0.8em}
\paragraph{Temporal Pairing.}
Each event is stored as a tuple $(x, y, \tau, p)$ with pixel location $(x, y)$, timestamp $\tau$, and polarity $p \in \{+1, -1\}$.
Let $T_t = \tau_{1+F(t-1)}$ denote the timestamp assigned to the original frame $\mathbf{X}_t$.
We partition the event stream into
\begin{equation}
    \mathbf{E}_{t-1:t}
    =
    \{(x, y, \tau, p) \mid T_{t-1} < \tau \le T_t\},
    \qquad t=2,\ldots,N.
\end{equation}
Each coding interval is thus associated with $(\mathbf{X}_{t-1}, \mathbf{E}_{t-1:t}, \mathbf{X}_t)$.

\begin{figure*}[t]
    \centering
    \begin{subfigure}[t]{\linewidth}
        \centering
        \includegraphics[width=0.185\linewidth]
            {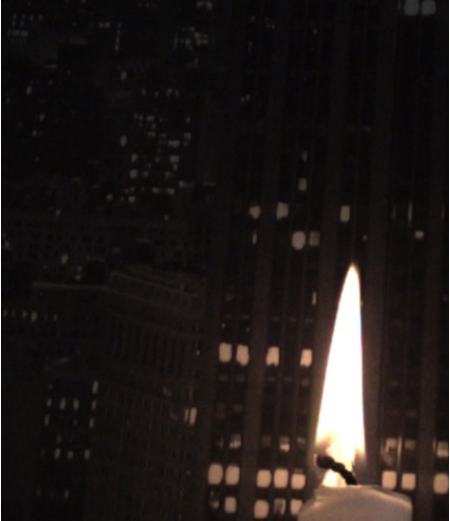}
        \includegraphics[width=0.185\linewidth]
            {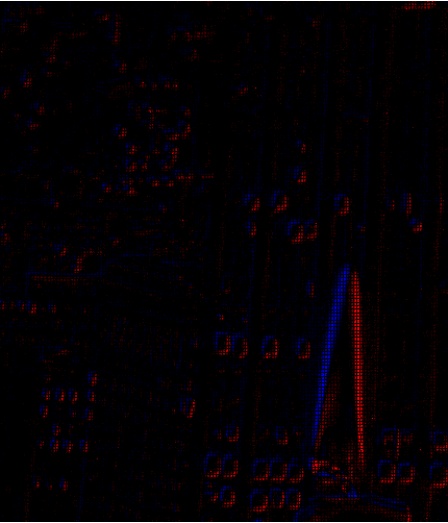}
        \includegraphics[width=0.185\linewidth]
            {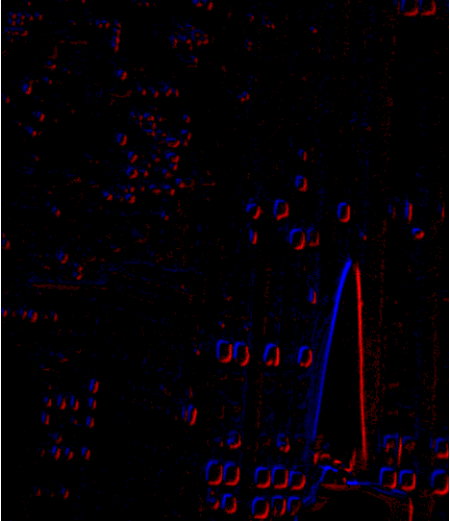}
        \includegraphics[width=0.185\linewidth]
            {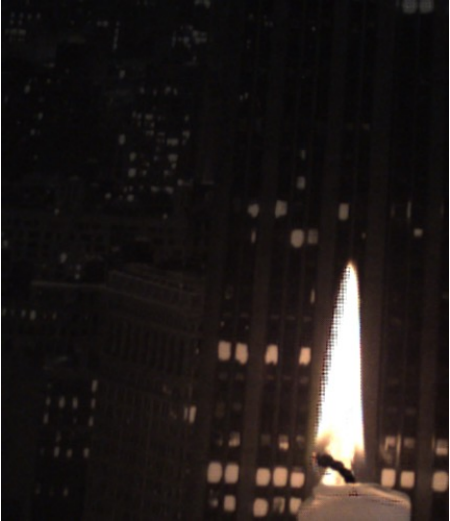}
        \includegraphics[width=0.185\linewidth]
            {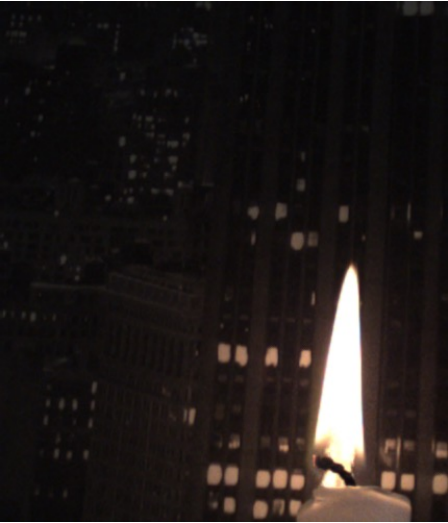}
    \end{subfigure}
    \vspace{0.2em}
    \begin{subfigure}[t]{\linewidth}
        \centering
        \includegraphics[width=0.185\linewidth]
            {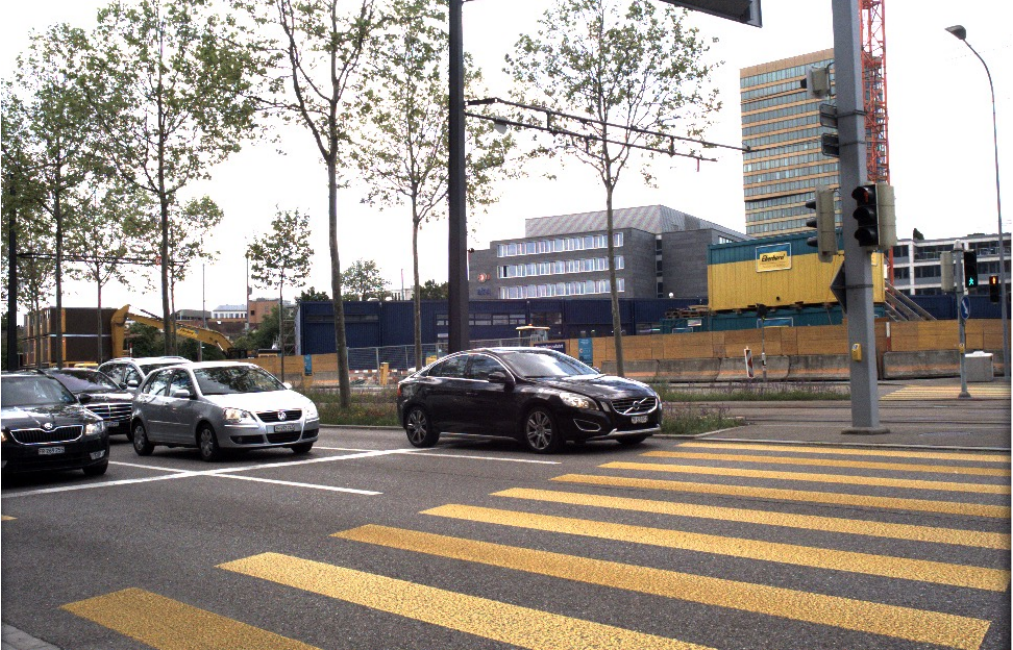}   
        \includegraphics[width=0.185\linewidth]
            {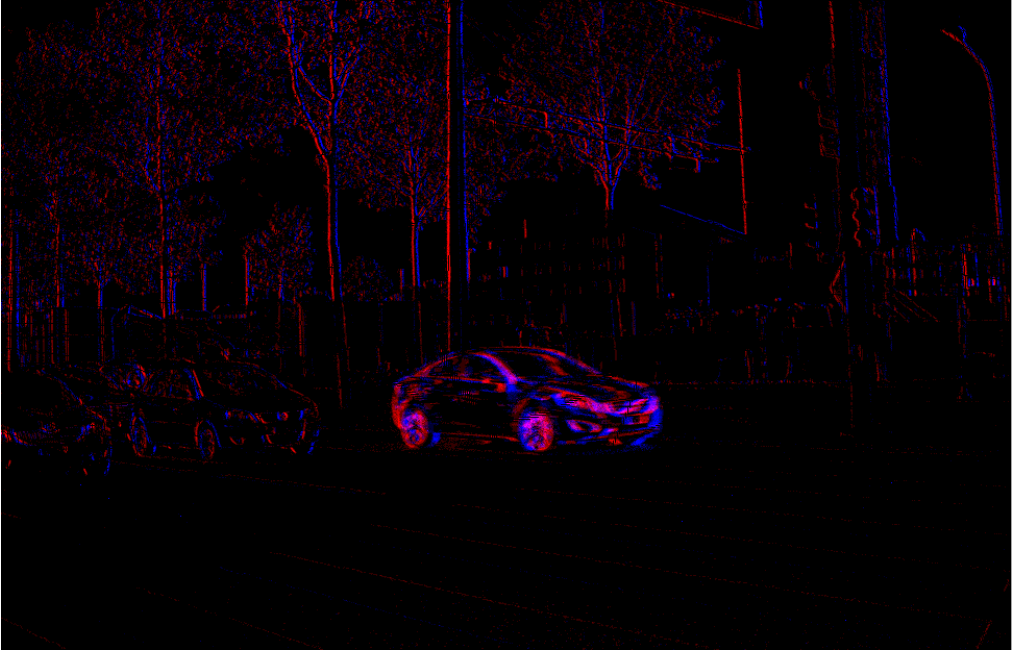}
        \includegraphics[width=0.185\linewidth]
            {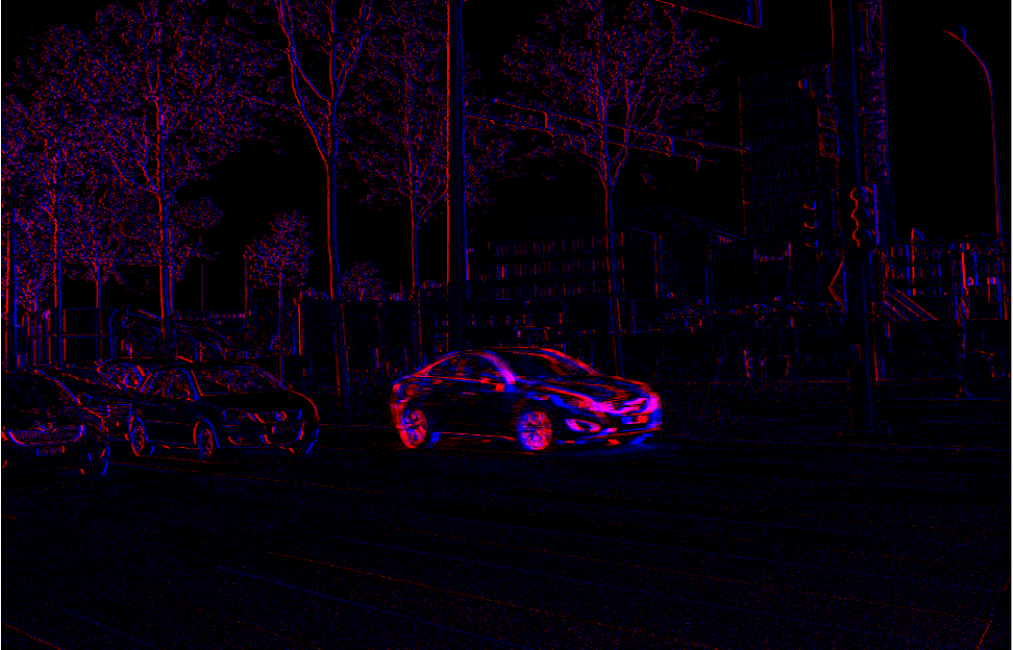}
        \includegraphics[width=0.185\linewidth]
            {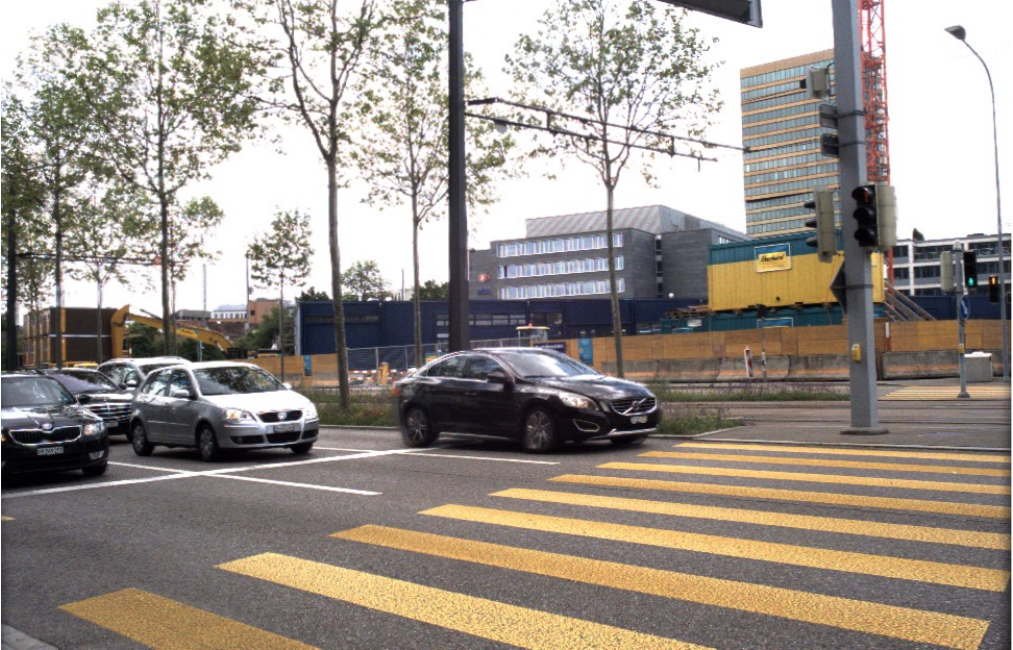}
        \includegraphics[width=0.185\linewidth]
            {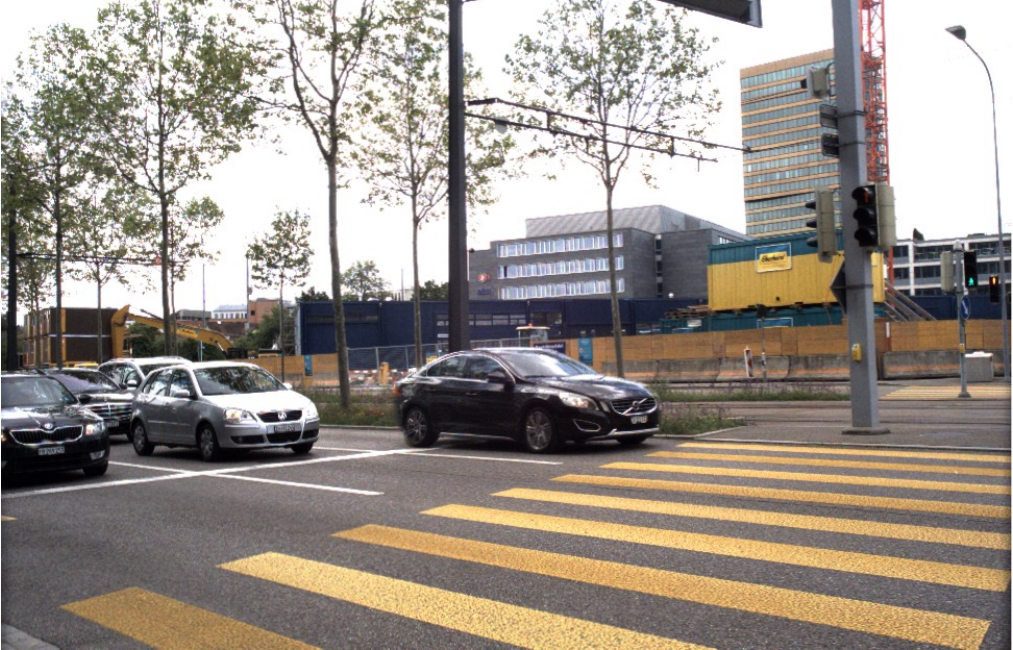}
    \end{subfigure}
    \vspace{-1.0em}
    \makebox[0.185\linewidth]{\small Target RGB}
    \makebox[0.185\linewidth]{\small Real Events}
    \makebox[0.185\linewidth]{\small Synthetic Events}
    \makebox[0.185\linewidth]{\small Pred. (Real)}
    \makebox[0.185\linewidth]{\small Pred. (Synthetic)}
    \vspace{0.5em}
    \caption{
        \textbf{Qualitative Comparison.}
        The top and bottom rows show HS-ERGB and BS-ERGB examples, respectively.
        VFPSIE predicts the target frame $\mathbf{X}_t$ from the same
        reference frame $\mathbf{X}_{t-1}$ using either real or
        synthetic events from the corresponding interval.
    }
    \label{fig:event_quality}
    \vspace{-1.0em}
\end{figure*}

\vspace{-0.5em}
\subsection{Synthetic Event Quality Assessment}
\label{sec:dataset_quality}

We assess whether synthetic events provide useful temporal information for frame prediction.
Using the pipeline in Sec.~\ref{sec:dataset_pipeline}, we synthesize events from the RGB frames of 15 HS-ERGB test sequences~\citep{tulyakov2021time} and 17 BS-ERGB validation sequences~\citep{tulyakov2022time}, and compare them with real events over the same intervals.
For each sequence, we evaluate the 31 adjacent intervals within its first 32 frames.

\begin{wraptable}{r}{0.55\textwidth}
    \vspace{-0.8em}
    \centering
    \small
    \caption{
        \textbf{Real and Synthetic Events.}
        PSNR-RGB is measured for the reference frame (Ref.) and
        VFPSIE predictions using real or synthetic events (Real/Syn.).
    }
    \label{tab:event_quality}
    \vspace{-0.5em}
    \setlength{\tabcolsep}{3pt}
    \renewcommand{\arraystretch}{1.1}
    \resizebox{\linewidth}{!}{
        \begin{tabular}{lrrrrr}
            \toprule
            Dataset
            & \multicolumn{2}{c}{Event statistics}
            & \multicolumn{3}{c}{PSNR-RGB (dB)} \\
            \cmidrule(lr){2-3}
            \cmidrule(lr){4-6}
            & Count Ratio
            & Spatial Corr.
            & Ref.
            & Real
            & Syn. \\
            \midrule
            HS-ERGB
            & 1.543 & 0.507
            & 29.27 & 31.78 & 34.13 \\
            BS-ERGB
            & 1.587 & 0.606
            & 22.05 & 23.91 & 24.91 \\
            \bottomrule
        \end{tabular}
    }
    \vspace{-0.8em}
\end{wraptable}

We compare event counts and their spatial distributions (see Tab.~\ref{tab:event_quality}).
For each interval, spatial count maps record event counts per $4\times4$ cell for HS-ERGB and per pixel for BS-ERGB.
We report the pooled synthetic-to-real event count ratio and median per-interval correlation between the corresponding spatial count maps.

Using $\mathbf{X}_{t-1}$ directly as a prediction of $\mathbf{X}_t$ gives a reference-only baseline (Ref.).
We compare this baseline with predictions from the same pretrained VFPSIE~\citep{zhu2024vfpsie}, conditioned on $\mathbf{X}_{t-1}$ and either real or synthetic events, reporting PSNR-RGB against $\mathbf{X}_t$ averaged across intervals.
The PSNR-RGB gain over Ref. reflects the predictive benefit of inter-frame events beyond the information available in the previous RGB frame.

Synthetic events are 1.54 and 1.59 times as numerous as real events on HS-ERGB and BS-ERGB, respectively, while exhibiting spatial agreement with real events.
They enable VFPSIE to achieve prediction quality comparable to that obtained with real events (Tab.~\ref{tab:event_quality}).
The examples in Fig.~\ref{fig:event_quality} further show that synthetic events closely resemble real events in their dominant spatial structures, capturing similar object contours and regions of concentrated event activity.
Together, these results support the utility of synthetic events as temporal guidance for frame prediction.
Additional qualitative examples are provided in App.~\ref{app:examples}.

%% file: sec/4_method.tex
\vspace{-0.5em}
\section{Proposed Method}
\label{sec:methods}

\begin{figure*}[t]
    \centering
    \includegraphics[width=\textwidth]{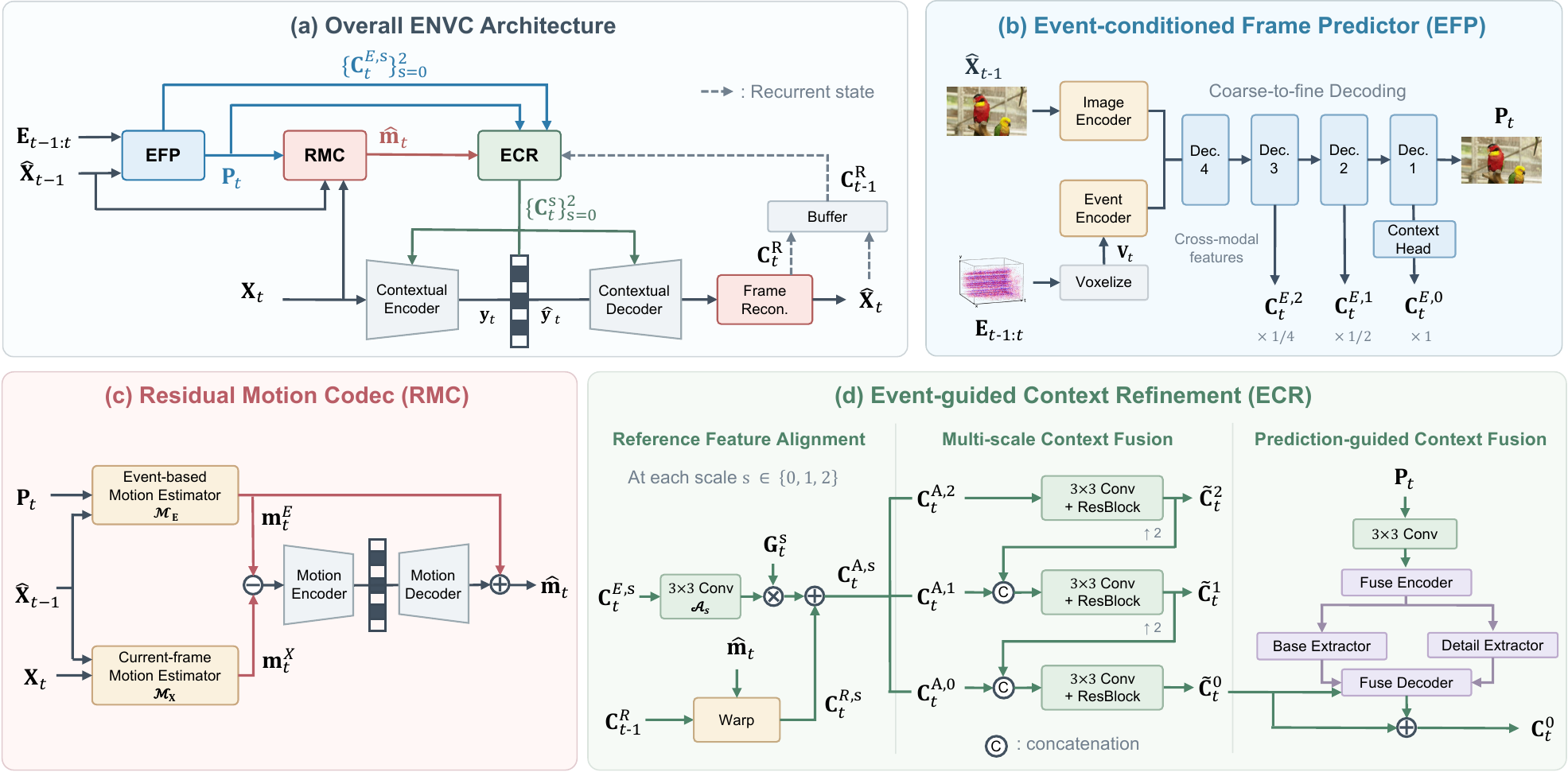}
    \vspace{-1.5em}
    \caption{
        \textbf{Overview Pipeline of ENVC.}
        The overall framework (a) combines event-conditioned frame prediction (b), residual motion coding (c), and event-guided context refinement (d).
    }
    \label{fig:elvc-framework}
    \vspace{-1.0em}
\end{figure*}

In this section, we introduce \textbf{Event-guided Neural Video Codec (ENVC)}, and describe how ENVC leverages shared events to predict motion and refine temporal contexts for conditional frame coding.

\vspace{-0.5em}
\subsection{Framework Overview}
\label{sec:method_overview}

Fig.~\hyperref[fig:elvc-framework]{\ref*{fig:elvc-framework}a} illustrates the overview pipeline of ENVC, inspired by the conditional coding framework of DCVC-DC~\citep{li2023dcvcdc} and DCMVC~\citep{tang2025dcmvc}.
Both encoder and decoder use the reconstructed reference $\hat{\mathbf{X}}_{t-1}$ and shared events $\mathbf{E}_{t-1:t}$.
The \textbf{E}vent-conditioned \textbf{F}rame \textbf{P}redictor (\textbf{EFP}) produces a frame prediction $\mathbf{P}_t$ and multi-scale event-conditioned contexts.
We estimate motion from $\mathbf{P}_t$ relative to $\hat{\mathbf{X}}_{t-1}$, while the \textbf{R}esidual \textbf{M}otion \textbf{C}odec (\textbf{RMC}) codes the difference between the current-frame and event-based motion estimates.
Adding the decoded residual to the event-based estimate yields the reconstructed motion (Sec.~\ref{sec:method_event_motion}).

\textbf{E}vent-guided \textbf{C}ontext \textbf{R}efinement (\textbf{ECR}) aligns reference features using the reconstructed motion, then refines them through gated event injection and multi-scale fusion.
\textbf{P}rediction-guided \textbf{C}ontext \textbf{F}usion (\textbf{PCF}) further incorporates features extracted from $\mathbf{P}_t$ into the $\times 1$ context (Sec.~\ref{sec:method_context_refinement}).
The resulting temporal contexts condition frame encoding, entropy modeling, and reconstruction (Sec.~\ref{sec:method_conditional_coding}).

\vspace{-0.5em}
\subsection{Event-predictive Motion Coding}
\label{sec:method_event_motion}

\paragraph{Event-conditioned Frame Prediction (EFP).}
EFP is built on the VFPSIE backbone~\citep{zhu2024vfpsie}, with outputs for both frame prediction and multi-scale context construction.
We represent $\mathbf{E}_{t-1:t}$ as a signed event voxel grid $\mathbf{V}_t \in \mathbb{R}^{10 \times H \times W}$, distributing each event between neighboring temporal bins with linear weights.
Given $\hat{\mathbf{X}}_{t-1}$ and $\mathbf{V}_t$, EFP produces
\begin{equation}
    \left(\mathbf{P}_t,\{\mathbf{C}_t^{E,s}\}_{s=0}^{2}\right)
    =
    \mathcal{P}_{\theta}
    \left(\hat{\mathbf{X}}_{t-1},\mathbf{V}_t\right),
    \label{eq:ecfp}
\end{equation}
where $\mathcal{P}_{\theta}$ denotes EFP and $\mathbf{C}_t^{E,s}$ denotes its event-conditioned context at scale $s$.

As shown in Fig.~\hyperref[fig:elvc-framework]{\ref*{fig:elvc-framework}b}, separate encoders extract image and event features.
A coarse-to-fine decoder combines these features through cross-modal attention to estimate flow and synthesize an RGB frame.
The prediction $\mathbf{P}_t$ is obtained by blending the synthesized frame with the forward-warped reference.

We extract $\mathbf{C}_t^{E,1}$ and $\mathbf{C}_t^{E,2}$ from the decoder outputs at $\times 1/2$ and $\times 1/4$ resolution, respectively.
A separate context head generates $\mathbf{C}_t^{E,0}$ at $\times 1$ resolution from intermediate features of the final decoding stage.

\vspace{-0.8em}
\paragraph{Residual Motion Coding (RMC).}
For reference alignment, we estimate backward motion rather than reuse the internal forward flow from EFP.
The \textbf{E}vent-based \textbf{M}otion \textbf{E}stimator (\textbf{EME}) $\mathcal{M}_E$ and \textbf{C}urrent-frame \textbf{M}otion \textbf{E}stimator (\textbf{CME}) $\mathcal{M}_X$ use separate SPyNet-based networks~\citep{ranjan2017optical}:
\begin{equation}
    \begin{aligned}
        \mathbf{m}_t^E
        = \mathcal{M}_E
        \left(\mathbf{P}_t,\hat{\mathbf{X}}_{t-1}\right), \qquad
        \mathbf{m}_t^X
        = \mathcal{M}_X
        \left(\mathbf{X}_t,\hat{\mathbf{X}}_{t-1}\right).
    \end{aligned}
    \label{eq:motion_estimates}
\end{equation}
The event-based estimate is computed at both the encoder and decoder, whereas the current-frame estimate is computed only at the encoder.

RMC codes their difference:
\begin{equation}
    \begin{aligned}
        \Delta\mathbf{m}_t
        = \mathbf{m}_t^X-\mathbf{m}_t^E, \qquad
        \hat{\mathbf{m}}_t
        = \mathbf{m}_t^E+\Delta\hat{\mathbf{m}}_t,
    \end{aligned}
    \label{eq:residual_motion}
\end{equation}
where $\Delta\hat{\mathbf{m}}_t$ is reconstructed from the quantized motion latent (Fig.~\hyperref[fig:elvc-framework]{\ref*{fig:elvc-framework}c}).
The motion entropy model uses a hyperprior and the previously reconstructed motion latent when available~\citep{li2023dcvcdc}.

\begin{table}[t]
    \centering
    \small
    \caption{
        \textbf{BD-Rate (\%) against DCMVC.}
        Each comparison uses PSNR-RGB (left) and LPIPS (right).
        ``--'' indicates no overlapping quality range.
        Averages are computed over datasets with valid BD-rate values.
    }
    \vspace{-0.5em}
    \label{tab:bd_rate}
    \setlength{\tabcolsep}{5pt}
    \renewcommand{\arraystretch}{1.1}
    \resizebox{\linewidth}{!}{%
        \begin{tabular}{lcccccc|c}
            \toprule
            Method
            & UVG
            & MCL-JCV
            & HEVC-B
            & HEVC-C
            & HEVC-D
            & HEVC-E
            & Average \\
            \midrule
            DCMVC
            & 0.00 / 0.00 & 0.00 / 0.00 & 0.00 / 0.00
            & 0.00 / 0.00 & 0.00 / 0.00 & 0.00 / 0.00
            & 0.00 / 0.00 \\
            \midrule
            DCVC
            & $+176.33$ / $+188.87$
            & $+133.70$ / $+139.12$
            & $+141.41$ / $+176.24$
            & $+222.23$ / $+168.15$
            & $+213.26$ / $+166.59$
            & -- / $+228.29$
            & $+177.39$ / $+177.88$ \\
            DCVC-TCM
            & $+59.17$ / $+29.03$
            & $+59.66$ / $+38.33$
            & $+49.37$ / $+38.54$
            & $+91.85$ / $+52.10$
            & $+84.92$ / $+55.88$
            & $+101.45$ / $+49.84$
            & $+74.40$ / $+43.95$ \\
            DCVC-DC
            & $+0.01$ / $-11.50$
            & $+1.05$ / $-9.55$
            & $-1.85$ / $-8.89$
            & $+3.03$ / $-5.70$
            & $-8.83$ / $-4.50$
            & $-3.64$ / $-14.50$
            & $-1.71$ / $-9.10$ \\
            DCVC-FM
            & $+6.16$ / $-11.46$
            & $+10.75$ / $-4.57$
            & $+8.14$ / $-8.73$
            & $+11.67$ / $-3.76$
            & $-10.62$ / $+2.28$
            & $-1.70$ / $-14.34$
            & $+4.07$ / $-6.76$ \\
            DCVC-RT
            & $+20.74$ / $-7.72$
            & $+28.64$ / $+4.58$
            & $+31.32$ / $+2.92$
            & $+66.42$ / $+13.64$
            & $+44.64$ / $+15.91$
            & $+20.16$ / $-5.49$
            & $+35.32$ / $+3.97$ \\
            \midrule
            \textbf{ENVC (Ours)}
            & $\mathbf{-29.73}$ / $\mathbf{-82.21}$
            & $\mathbf{-52.28}$ / $\mathbf{-84.64}$
            & $\mathbf{-51.96}$ / $\mathbf{-77.12}$
            & $\mathbf{-49.74}$ / $\mathbf{-60.65}$
            & $\mathbf{-43.30}$ / $\mathbf{-49.54}$
            & $\mathbf{-7.76}$ / $\mathbf{-51.60}$
            & $\mathbf{-39.13}$ / $\mathbf{-67.63}$ \\
            \bottomrule
        \end{tabular}%
    }
    \vspace{-1.5em}
\end{table}

\vspace{-0.5em}
\subsection{Event-guided Context Refinement}
\label{sec:method_context_refinement}
Fig.~\hyperref[fig:elvc-framework]{\ref*{fig:elvc-framework}d} shows how motion-aligned reference information is refined using EFP contexts and predictions.

\vspace{-0.8em}
\paragraph{Reference Feature Alignment.}
From the propagated decoder feature $\mathbf{C}_{t-1}^{R}$, we construct reference features at $\times 1$, $\times 1/2$, and $\times 1/4$ resolution.
When this feature is unavailable, we instead extract reference features from $\hat{\mathbf{X}}_{t-1}$.
For numerical stability, we also regenerate the reference features from $\hat{\mathbf{X}}_{t-1}$ when the maximum absolute value of the initial $\times 1$ context exceeds $5$.
We align the reference features by backward warping with $\hat{\mathbf{m}}_t$, adjusting the motion field and its displacement to each scale.
At $\times 1$, offset diversity~\citep{li2023dcvcdc} further refines the alignment with learned sampling offsets and masks.
We denote the aligned reference contexts by $\{\mathbf{C}_t^{R,s}\}_{s=0}^{2}$.

\vspace{-0.8em}
\paragraph{Gated Multi-scale Event Injection.}
We inject EFP contexts into the aligned reference contexts through gated residual connections:
\begin{equation}
    \mathbf{C}_t^{A,s}
    =
    \mathbf{C}_t^{R,s}
    +
    \mathbf{G}_t^s \odot
    \mathcal{A}_s\!\left(\mathbf{C}_t^{E,s}\right),
    \label{eq:event_injection}
\end{equation}
where $\mathcal{A}_s$ is a zero-initialized feature adapter and $\odot$ denotes element-wise multiplication.
Each gate $\mathbf{G}_t^s$ is a single-channel sigmoid map generated from $\mathbf{V}_t$ and $\hat{\mathbf{X}}_{t-1}$, broadcast across feature channels.
Multi-scale context fusion combines the augmented contexts along a top-down path from $\times 1/4$ through $\times 1/2$ to $\times 1$, producing $\{\widetilde{\mathbf{C}}_t^s\}_{s=0}^{2}$.

\vspace{-0.8em}
\paragraph{Prediction-guided Context Fusion (PCF).}
PCF further refines the $\times 1$ context ($\widetilde{\mathbf{C}}_t^0$) using the frame prediction $\mathbf{P}_t$.
Features extracted from $\mathbf{P}_t$ are fused with $\widetilde{\mathbf{C}}_t^0$ through the base/detail fusion module of DCMVC~\citep{tang2025dcmvc}, and the fused representation is added back to $\widetilde{\mathbf{C}}_t^0$.
The three contexts are finally clipped to $[-10,10]$, yielding $\{\mathbf{C}_t^s\}_{s=0}^{2}$ for conditional frame coding.

\vspace{-0.5em}
\subsection{Frame Coding and Optimization}
\label{sec:method_conditional_coding}

We use the conditional frame codec of DCVC-DC~\citep{li2023dcvcdc}, conditioned on $\{\mathbf{C}_t^s\}_{s=0}^{2}$.
The contextual encoder transforms $\mathbf{X}_t$ into a latent using contexts at the corresponding scales, and the contextual decoder reconstructs the frame from the quantized latent and the same contexts.
The $\times 1/4$ context also provides a temporal prior for entropy modeling.

The reconstructed frame and decoder feature form the reference state for the next frame.
Excluding shared events, the estimated RGB coding rate for each P-frame is
\begin{equation}
    R_t
    =
    R(\hat{\mathbf{y}}_t^m)
    +
    R(\hat{\mathbf{z}}_t^m)
    +
    R(\hat{\mathbf{y}}_t)
    +
    R(\hat{\mathbf{z}}_t),
\end{equation}
where $R(\cdot)$ is the entropy-model rate estimate, and $(\hat{\mathbf{y}}_t^m,\hat{\mathbf{z}}_t^m)$ and $(\hat{\mathbf{y}}_t,\hat{\mathbf{z}}_t)$ are the motion and frame latents with their hyperlatents, respectively.
The main rate--distortion term is
\begin{equation}
    \mathcal{L}_{\mathrm{RD},t}
    =
    R_t
    +
    \lambda D(\mathbf{X}_t,\hat{\mathbf{X}}_t),
\end{equation}
where $D$ measures reconstruction distortion and $\lambda$ controls the rate--distortion trade-off.
Stage-wise objectives, detailed training stages and optimization settings are provided in App.~\ref{app:training}.

%% file: sec/5_experiments.tex
\begin{figure*}[t]
    \centering
    \begin{subfigure}[t]{0.325\textwidth}
        \centering
        \includegraphics[width=\linewidth]{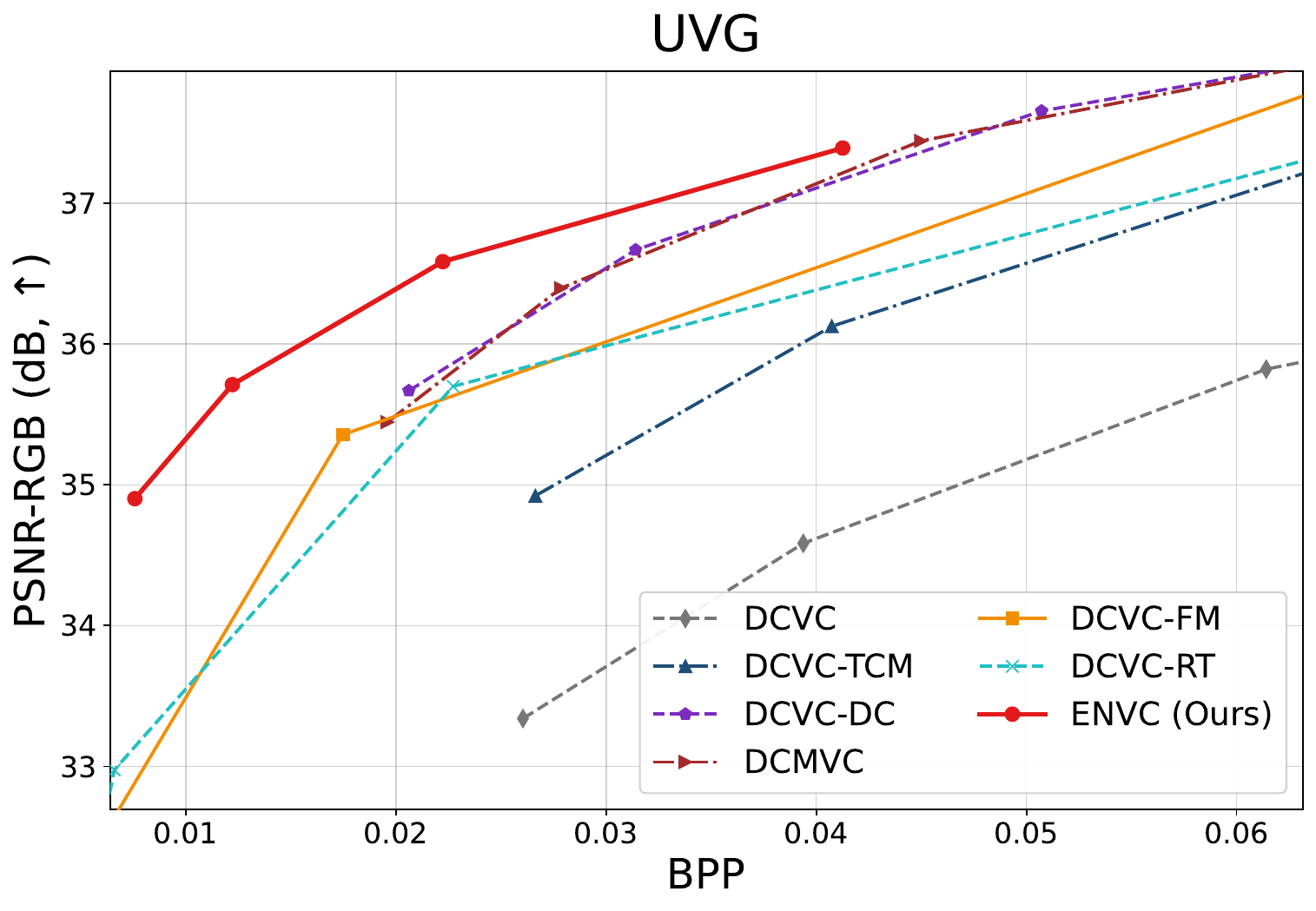}
        \vspace{-1.5em}
        \caption{UVG}
        \label{fig:rd_uvg_psnr}
    \end{subfigure}
    \begin{subfigure}[t]{0.325\textwidth}
        \centering
        \includegraphics[width=\linewidth]{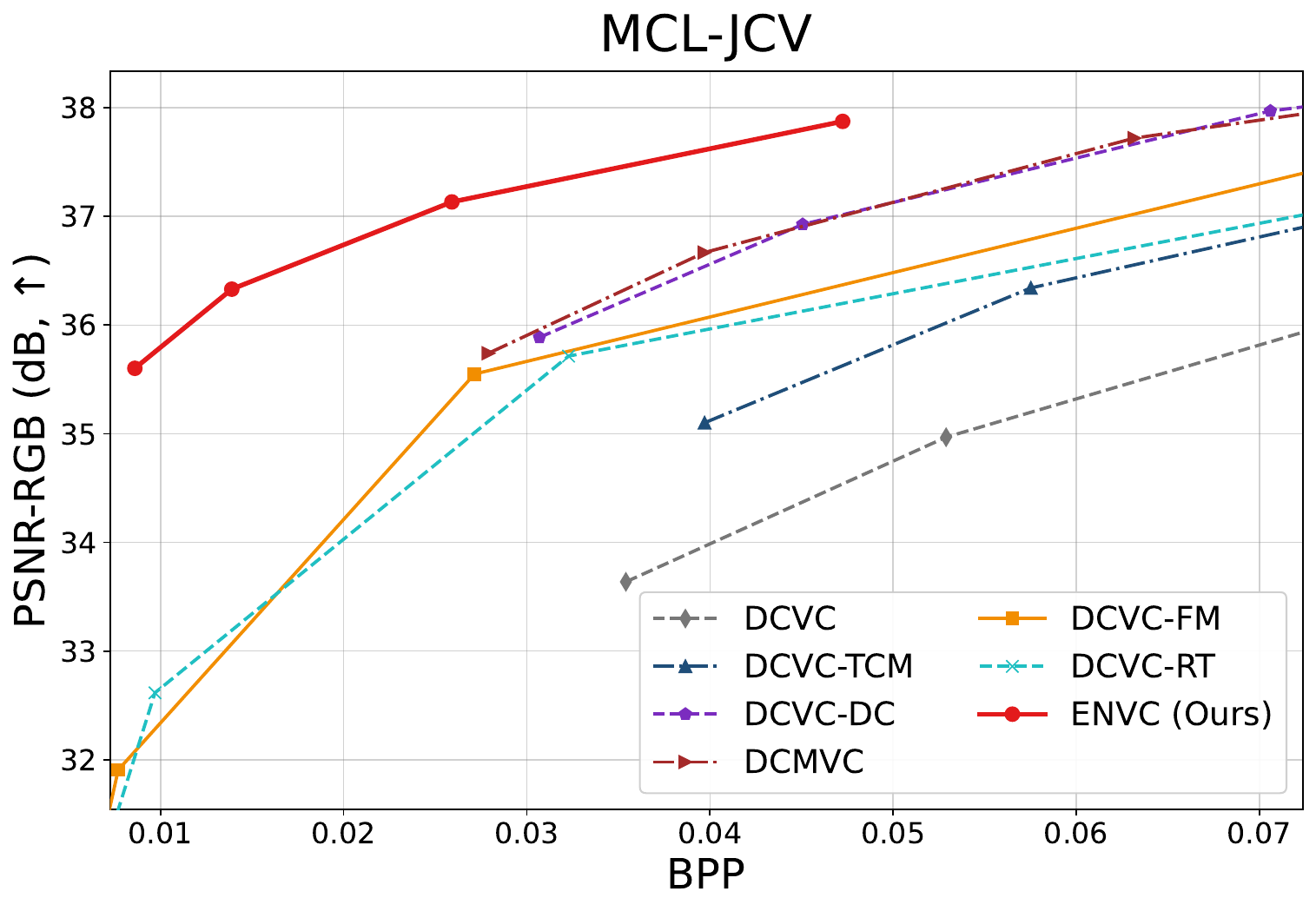}
        \vspace{-1.5em}
        \caption{MCL-JCV}
        \label{fig:rd_mcl_psnr}
    \end{subfigure}
    \begin{subfigure}[t]{0.325\textwidth}
        \centering
        \includegraphics[width=\linewidth]{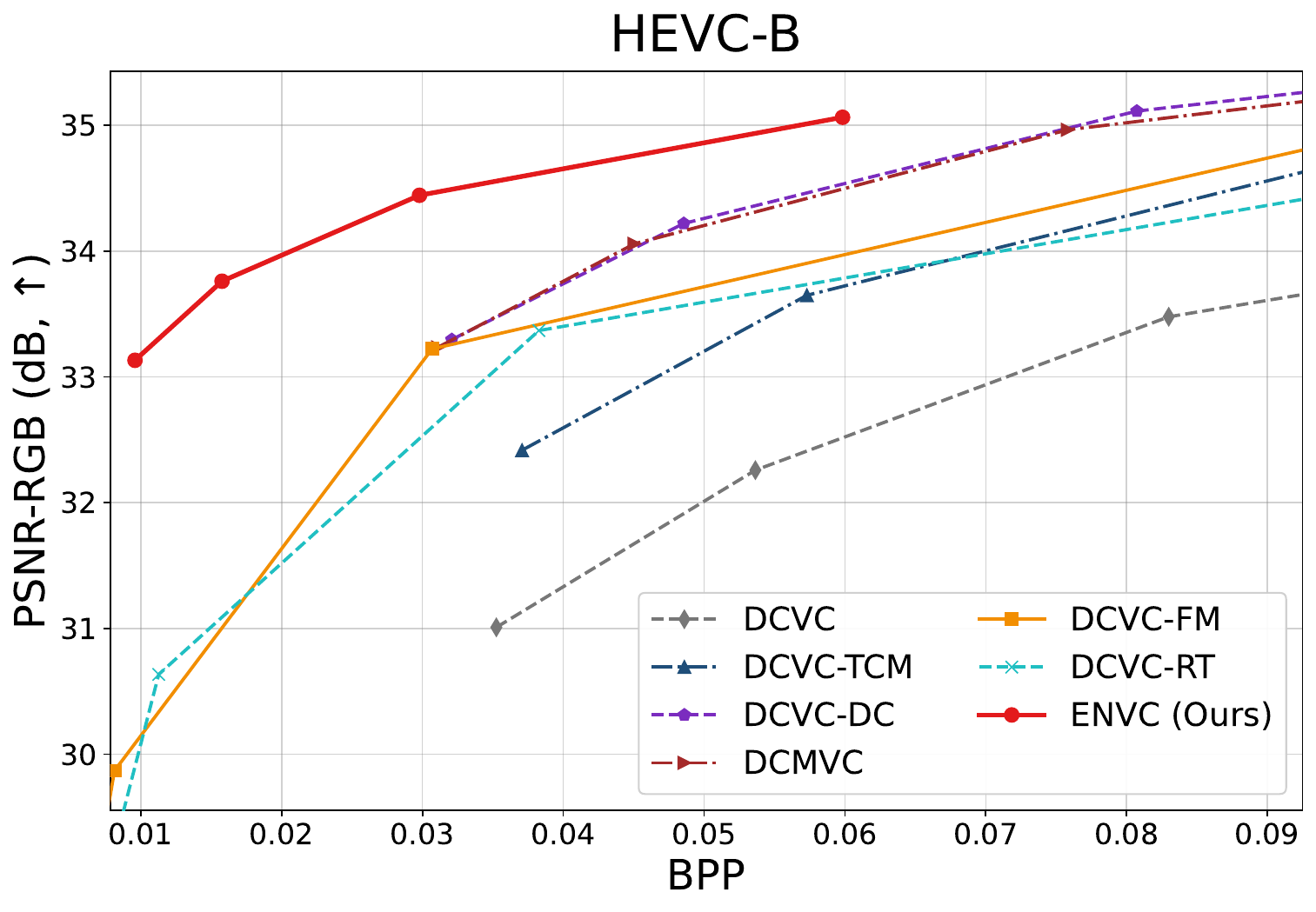}
        \vspace{-1.5em}
        \caption{HEVC-B}
        \label{fig:rd_hevcb_psnr}
    \end{subfigure}
    \vspace{0.25em}
    \begin{subfigure}[t]{0.325\textwidth}
        \centering
        \includegraphics[width=\linewidth]{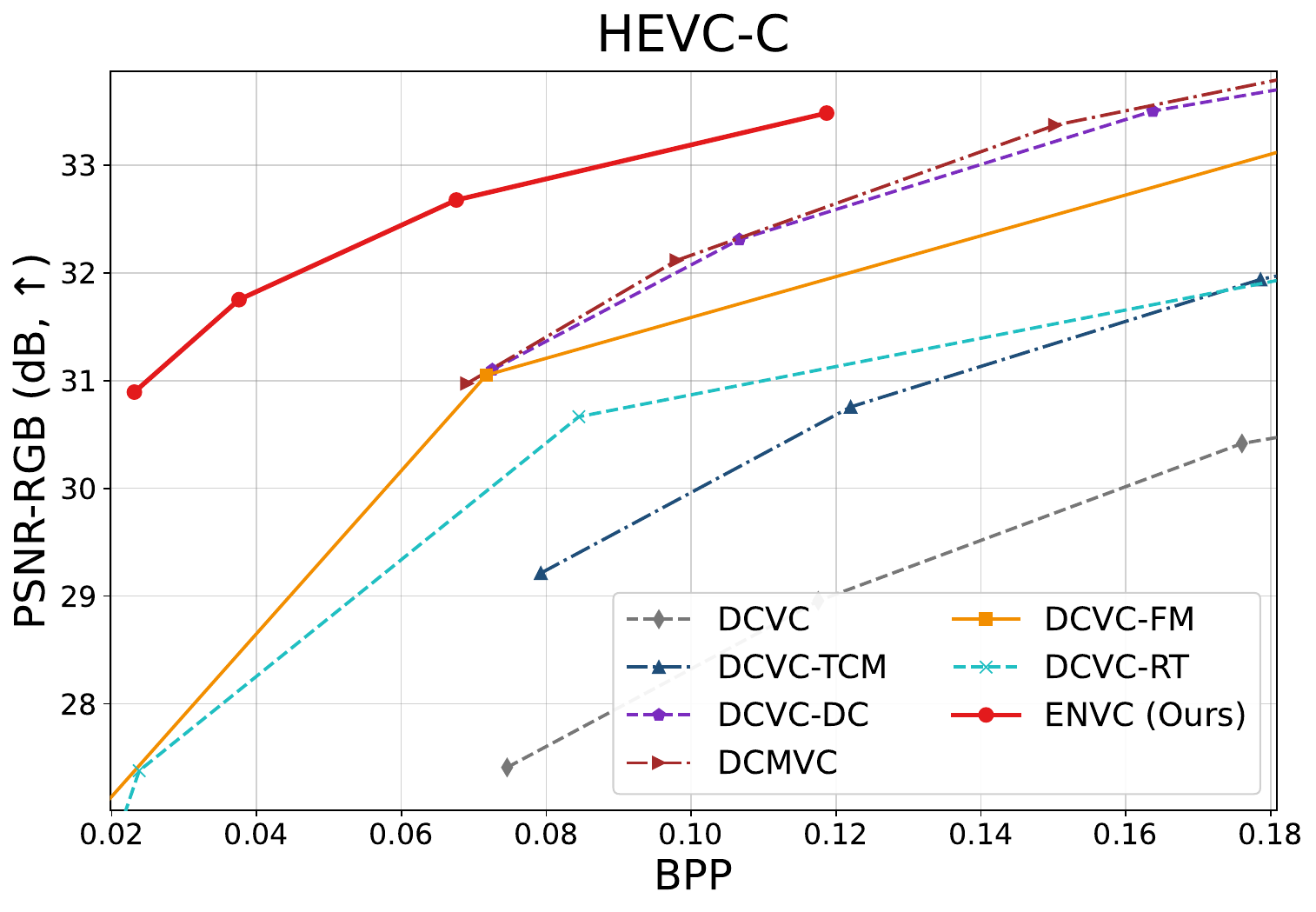}
        \vspace{-1.5em}
        \caption{HEVC-C}
        \label{fig:rd_hevcc_psnr}
    \end{subfigure}
    \begin{subfigure}[t]{0.325\textwidth}
        \centering
        \includegraphics[width=\linewidth]{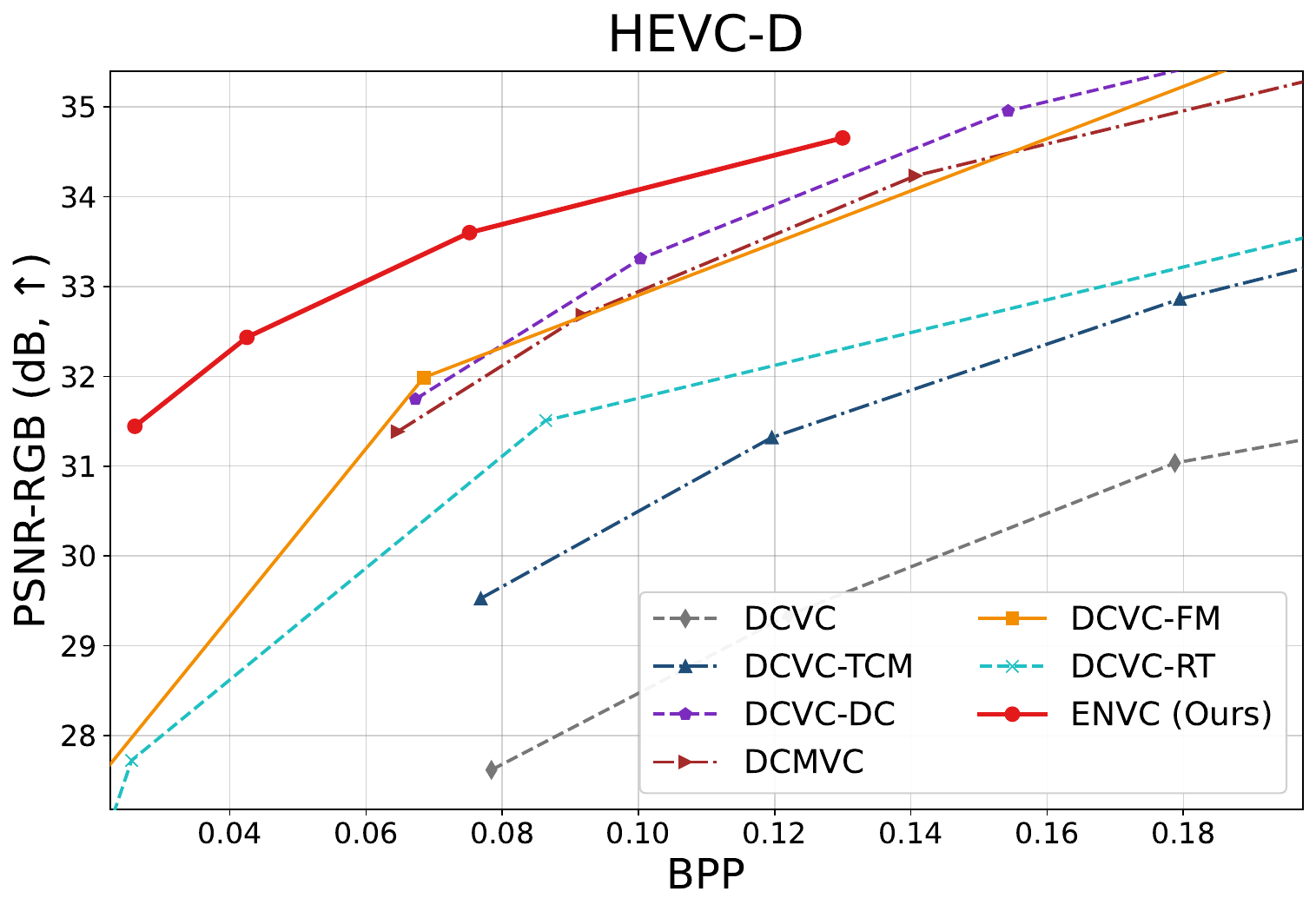}
        \vspace{-1.5em}
        \caption{HEVC-D}
        \label{fig:rd_hevcd_psnr}
    \end{subfigure}
    \begin{subfigure}[t]{0.325\textwidth}
        \centering
        \includegraphics[width=\linewidth]{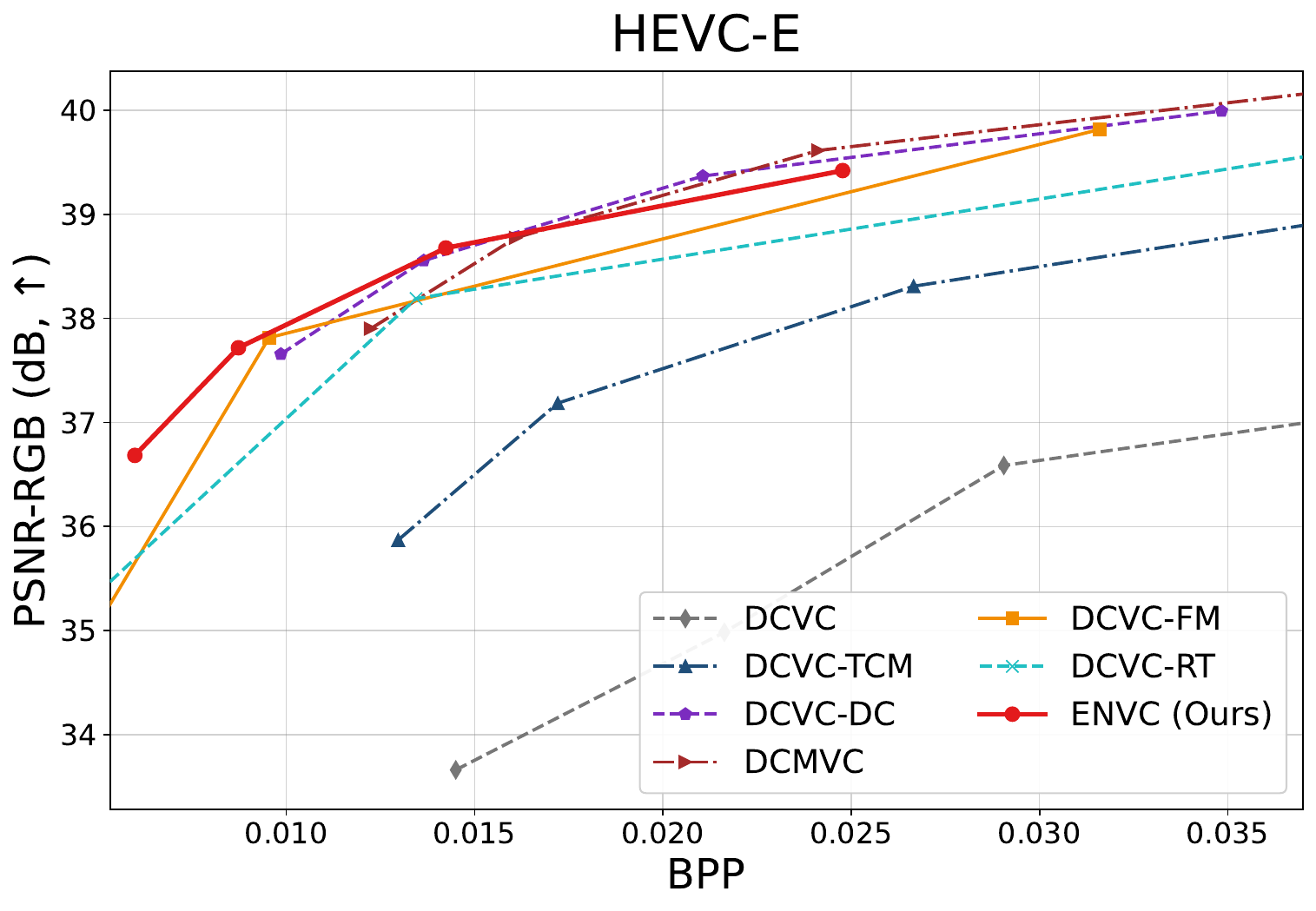}
        \vspace{-1.5em}
        \caption{HEVC-E}
        \label{fig:rd_hevce_psnr}
    \end{subfigure}
    \vspace{-0.5em}
    \caption{
        \textbf{Rate--Distortion Performance.}
        PSNR-RGB ($\uparrow$) versus bitrate on six benchmarks.
    }
    \label{fig:rd_curves_psnr}
    \vspace{-1.0em}
\end{figure*}

\section{Experiments}
\label{sec:experiments}

\subsection{Experimental Details}
\label{sec:exp_detail}

\paragraph{Training Details.}
We train ENVC on Vimeo-90K~\citep{xue2019video} septuplet sequences paired with synthetic events.
We adopt staged training followed by cascaded fine-tuning with up to 32 frames~\citep{li2024neural}, using longer clips extracted from the original Vimeo videos\footnote{\url{https://github.com/anchen1011/toflow/blob/master/data/original_vimeo_links.txt}.}.
The stage-wise objectives and training schedule are detailed in App.~\ref{app:training}.

\vspace{-0.8em}
\paragraph{Evaluation Details.}
We evaluate on UVG~\citep{mercat2020uvg}, MCL-JCV~\citep{wang2016mcl}, and HEVC Classes B, C, D, and E~\citep{flynn2013common}, using RGB frames paired with synthetic events.
Unless otherwise specified, all models are evaluated on the first 96 frames of each sequence with an intra period of 32.
ENVC uses the pretrained intra-frame codec of DCVC-DC~\citep{li2023dcvcdc} for I-frame coding.

We report PSNR-RGB and LPIPS~\citep{zhang2018unreasonable}, summarizing RD performance using BD-rate~\citep{bjontegaard2001calculation} relative to DCMVC~\citep{tang2025dcmvc}.
For all models, coding rates are estimated from entropy models and reported in bits per pixel (bpp), including both I- and P-frame coding.
Results using additional quality metrics and datasets are provided in App.~\ref{app:results}.

We compare ENVC with DCVC~\citep{li2021dcvc}, DCVC-TCM~\citep{sheng2022temporal}, DCVC-DC~\citep{li2023dcvcdc}, DCVC-FM~\citep{li2024neural}, DCMVC~\citep{tang2025dcmvc}, and DCVC-RT~\citep{jia2025towards}.

\vspace{-0.5em}
\subsection{Rate--Distortion Performance}
\label{sec:exp_rd}

\begin{figure*}[t]
    \centering
    \begin{subfigure}[t]{0.325\textwidth}
        \centering
        \includegraphics[width=\linewidth]{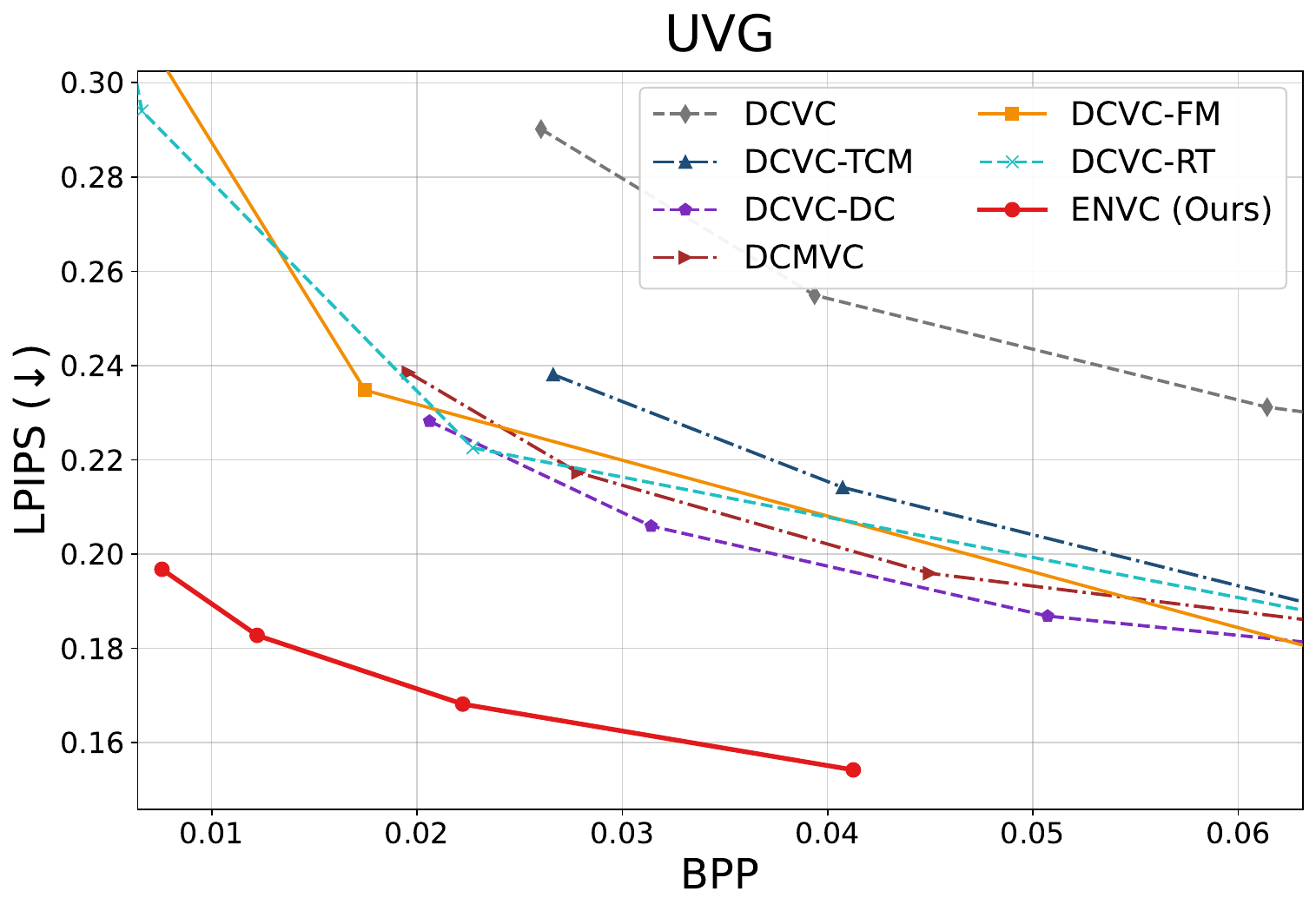}
        \vspace{-1.5em}
        \caption{UVG}
        \label{fig:rd_uvg_lpips}
    \end{subfigure}
    \begin{subfigure}[t]{0.325\textwidth}
        \centering
        \includegraphics[width=\linewidth]{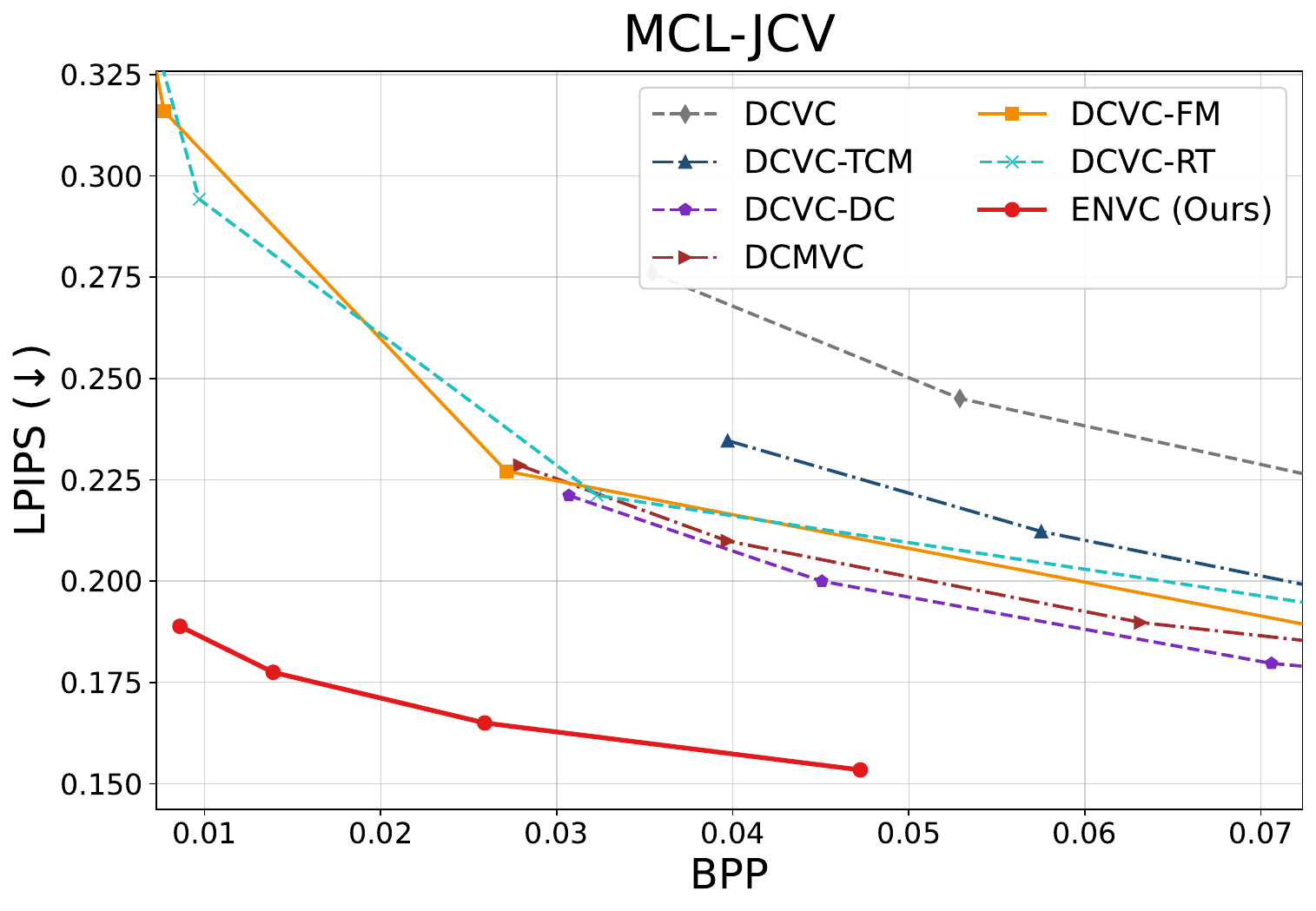}
        \vspace{-1.5em}
        \caption{MCL-JCV}
        \label{fig:rd_mcl_lpips}
    \end{subfigure}
    \begin{subfigure}[t]{0.325\textwidth}
        \centering
        \includegraphics[width=\linewidth]{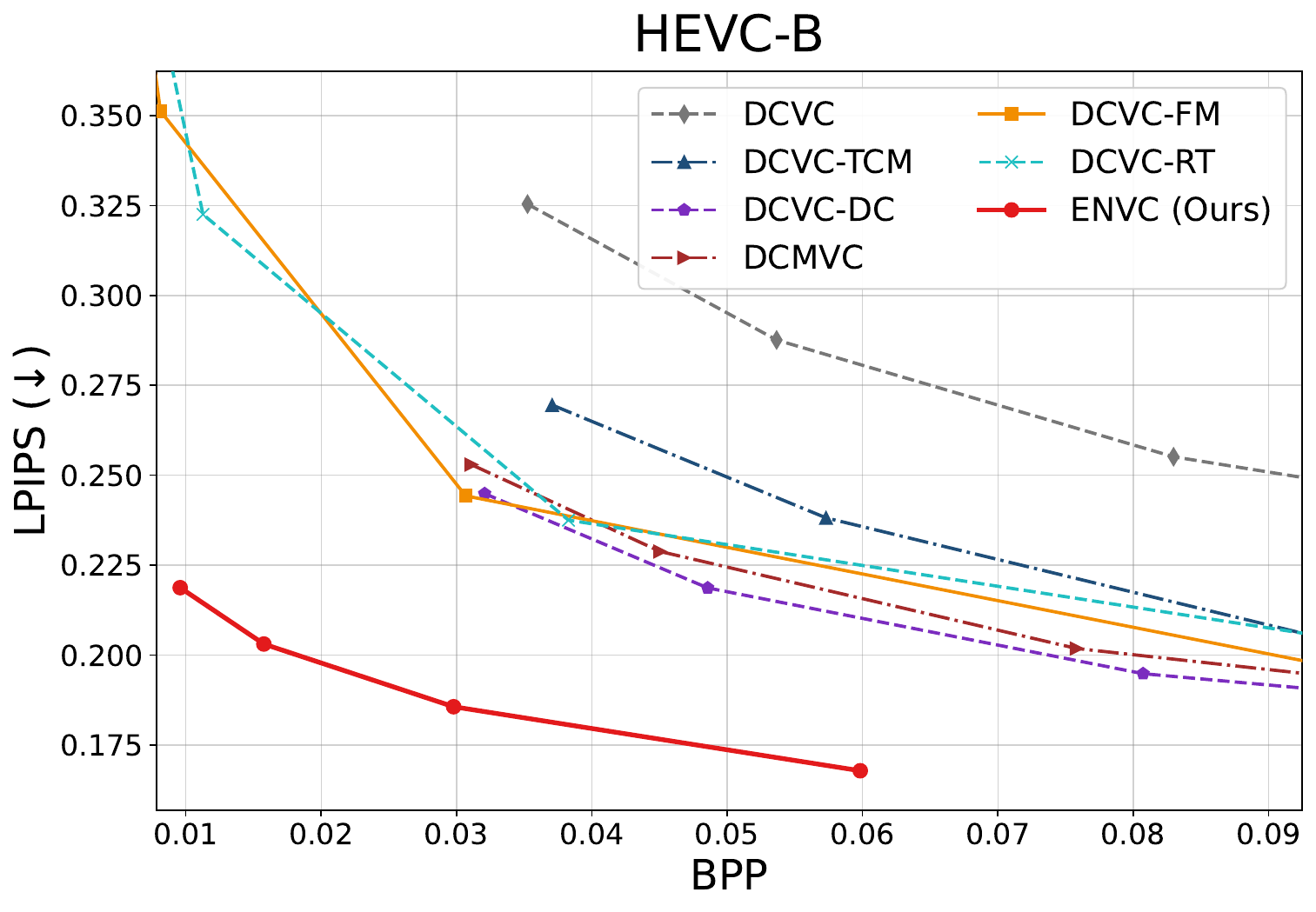}
        \vspace{-1.5em}
        \caption{HEVC-B}
        \label{fig:rd_hevcb_lpips}
    \end{subfigure}
    \vspace{0.25em}
    \begin{subfigure}[t]{0.325\textwidth}
        \centering
        \includegraphics[width=\linewidth]{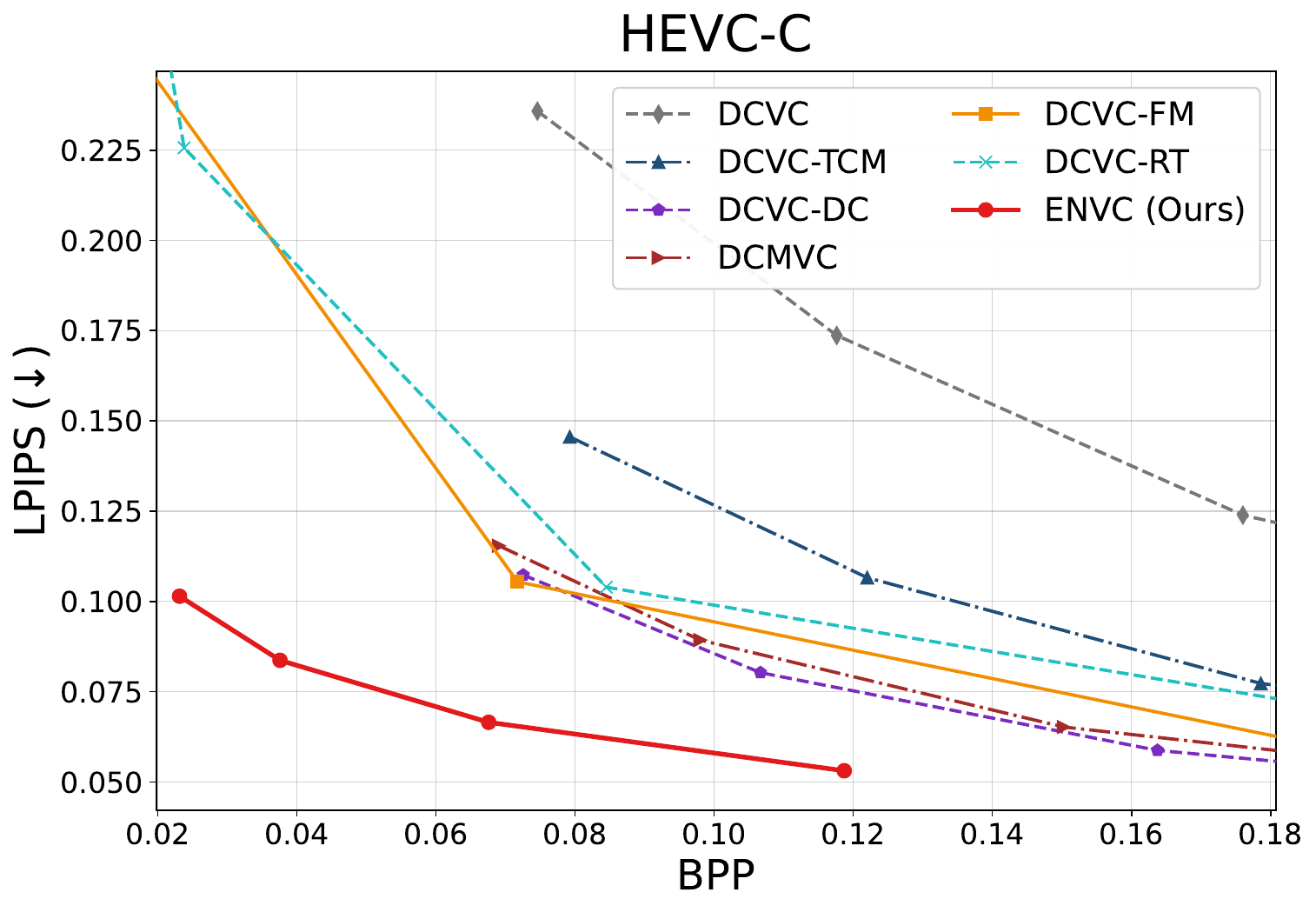}
        \vspace{-1.5em}
        \caption{HEVC-C}
        \label{fig:rd_hevcc_lpips}
    \end{subfigure}
    \begin{subfigure}[t]{0.325\textwidth}
        \centering
        \includegraphics[width=\linewidth]{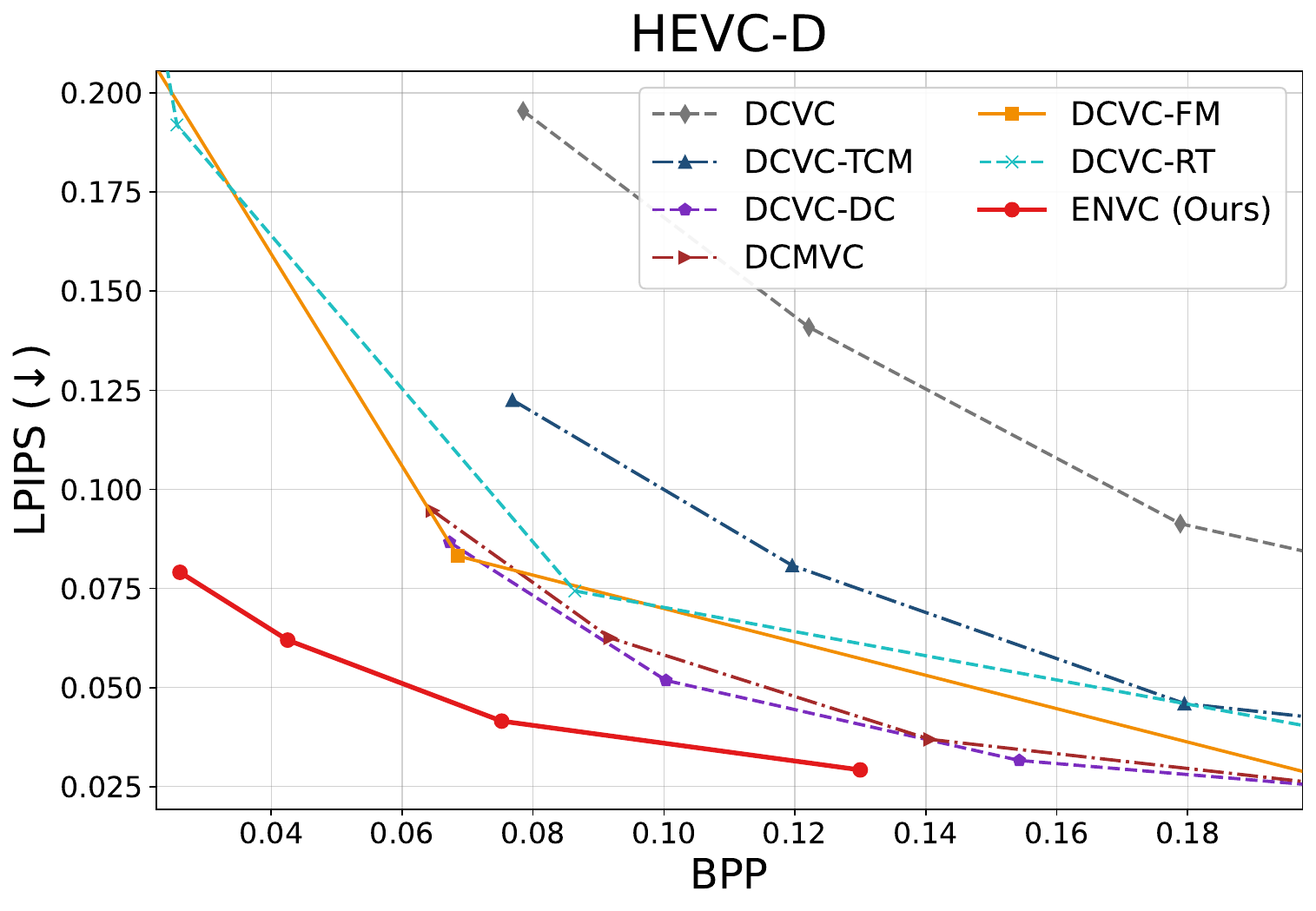}
        \vspace{-1.5em}
        \caption{HEVC-D}
        \label{fig:rd_hevcd_lpips}
    \end{subfigure}
    \begin{subfigure}[t]{0.325\textwidth}
        \centering
        \includegraphics[width=\linewidth]{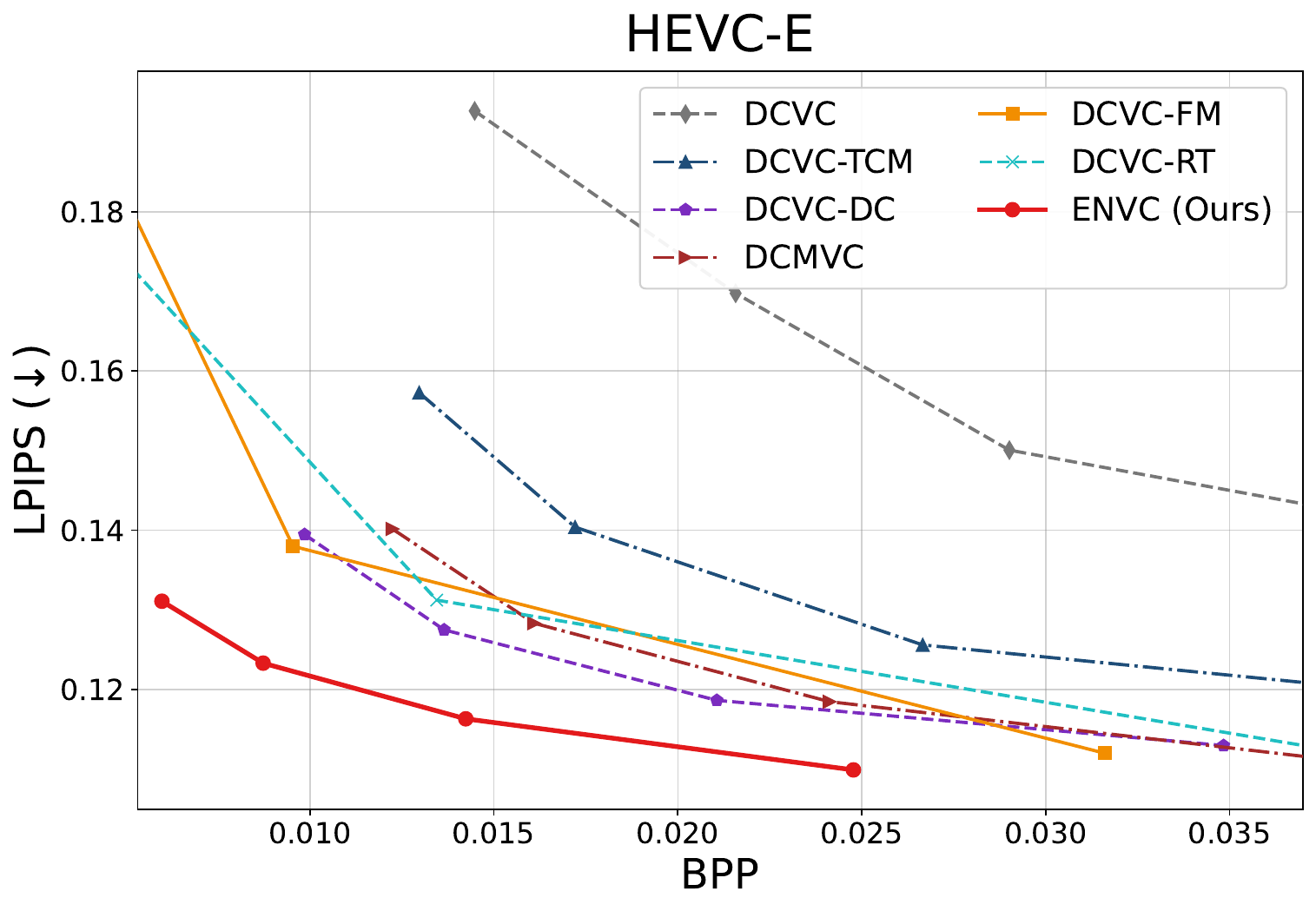}
        \vspace{-1.5em}
        \caption{HEVC-E}
        \label{fig:rd_hevce_lpips}
    \end{subfigure}
    \vspace{-0.5em}
    \caption{
        \textbf{Rate--Distortion Performance.}
        LPIPS ($\downarrow$) versus bitrate on six benchmarks.
    }
    \label{fig:rd_curves_lpips}
    \vspace{-1.0em}
\end{figure*}

Figs.~\ref{fig:rd_curves_psnr} and~\ref{fig:rd_curves_lpips}, together with Tab.~\ref{tab:bd_rate}, summarize the RD performance.
Across six benchmarks, ENVC achieves average BD-rate savings of 39.13\% using PSNR-RGB and 67.63\% using LPIPS relative to DCMVC.
These results show that shared events can reduce the RGB rate required for reconstruction under both pixel-wise and perceptual quality measures.
An analysis of P-frame bit allocation in App.~\ref{app:bit_allocation} further shows reduced motion coding rates relative to DCMVC.

On HEVC-E, PSNR-RGB savings are smaller (7.76\%), while LPIPS-based savings remain substantial (51.60\%).
The smaller PSNR-RGB gains are consistent with limited inter-frame motion and the high I-frame share of the total RGB rate (56.2\% at $Q_0$), which limits the impact of event-guided P-frame improvements.
Sequence-level gains and failure cases are further analyzed in App.~\ref{app:failure}.

Since ENVC is trained on synthetic events, we further examine how its RD performance changes when real events are used instead. Comparisons using real and synthetic event inputs on paired RGB–event datasets are provided in App.~\ref{app:real_event}.

\vspace{-0.5em}
\subsection{How Does ENVC Use Events?}
\label{sec:exp_gate}

To understand how ENVC incorporates events into its temporal context, we examine its multi-scale event gates ($\mathbf{G}_t^s$).
We use quality factor $Q_1$ and 12 sequences.

\begin{wrapfigure}{r}{0.59\textwidth}
    \centering
    \begin{minipage}[t]{0.49\linewidth}
        \vspace{0pt}
        \centering
        \includegraphics[width=\linewidth]{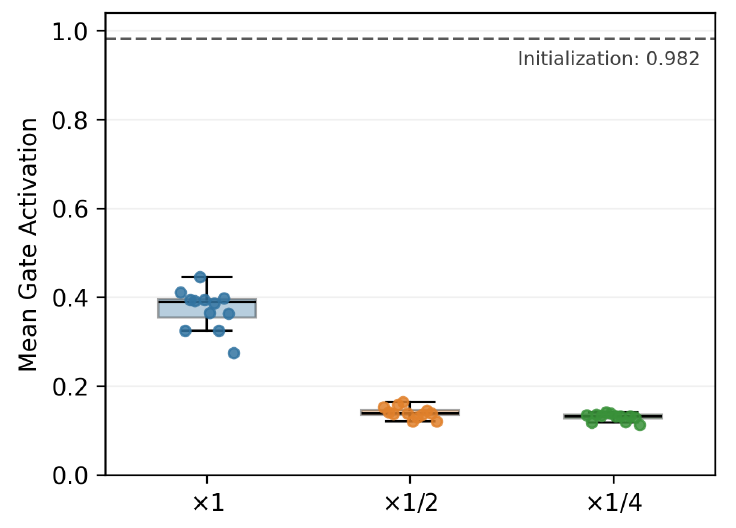}
        {\small (a) Learned Gate Activation}
    \end{minipage}
    \begin{minipage}[t]{0.49\linewidth}
        \vspace{0pt}
        \centering
        \includegraphics[width=\linewidth]{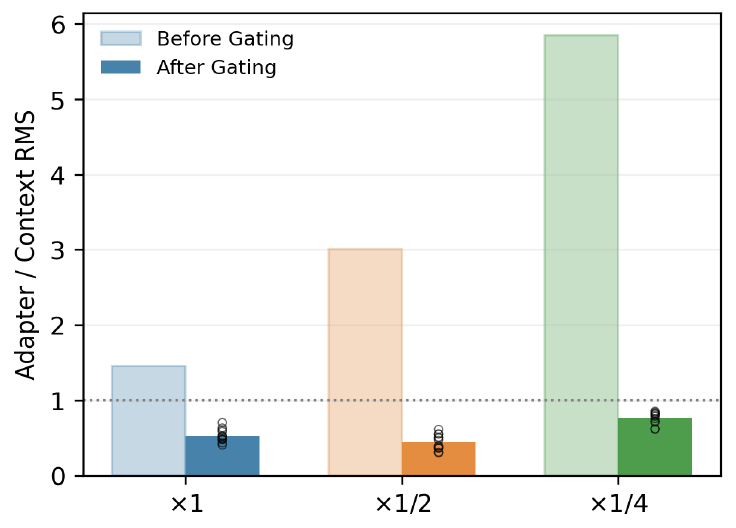}
        {\small (b) Relative Feature Magnitude}
    \end{minipage}
    \vspace{-0.5em}
    \caption{
        \textbf{Event Gating at $Q_1$.}
        (a) Mean gate activations.
        (b) Adapter-to-reference-context RMS ratios before and after gating.
    }
    \label{fig:gate_hierarchy}
    \vspace{-1.0em}
\end{wrapfigure}

The mean gate activations at $\times 1$, $\times 1/2$, and $\times 1/4$ are 0.373, 0.140, and 0.130, respectively (Fig.~\hyperref[fig:gate_hierarchy]{\ref*{fig:gate_hierarchy}a}).
All three means are substantially below the initialization of 0.982, showing that ENVC learns to attenuate the adapter outputs.
Fig.~\hyperref[fig:gate_hierarchy]{\ref*{fig:gate_hierarchy}b} compares adapter-feature root-mean-square (RMS) magnitudes before and after gating, normalized by the RMS of the aligned reference
contexts before injection.
After gating, these ratios are 0.533, 0.444, and 0.765, respectively.

Although the ungated adapter features exceed the reference contexts in RMS magnitude at all three scales, gating brings their magnitudes below those of the reference contexts while retaining substantial event-conditioned signals.
The $\times 1/4$ branch has the largest adapter-to-context RMS ratio both before and after gating, despite its smallest mean gate activation.
Mean gate activation alone therefore does not characterize the magnitude of event injection, which also depends on the adapter features being modulated.
We also replace all injection gates with constants $G=0$ or $G=1$ at inference, without retraining.
Across both interventions, PSNR decreases by 2.30--3.89~dB on four tested sequences, indicating that the trained model relies on the learned modulation for reconstruction.
Additional analyses and detailed intervention results are provided in Apps.~\ref{app:gate} and \ref{app:event_perturbation}.

\vspace{-0.5em}
\subsection{Effects of Motion Magnitude and Intra Period}
\label{sec:exp_effect}

\paragraph{Motion Magnitude.}

We examine how our coding gains vary with the motion magnitude.
Following DCVC-RT~\citep{jia2025towards}, we measure motion using SEA-RAFT~\citep{wang2024sea} optical flow between consecutive original frames.
On 30 MCL-JCV sequences (96 frames each, intra period 32), ENVC achieves an overall BD-rate saving of 52.28\% over DCMVC.

\begin{figure*}[t]
    \centering
    \begin{subfigure}[t]{0.31\textwidth}
        \centering
        \includegraphics[width=\linewidth]{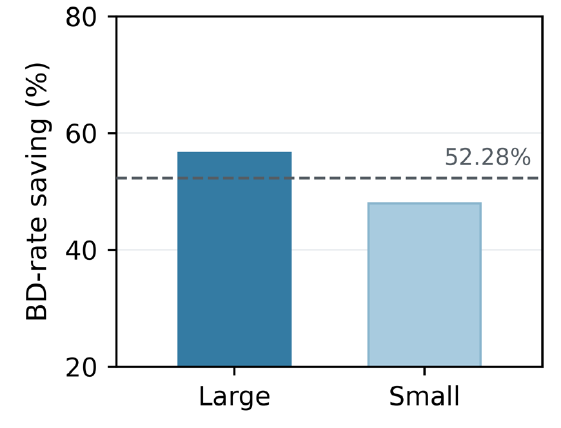}
        \vspace{-1.5em}
        \caption{Sequence-level Motion}
        \label{fig:motion_intra_a}
    \end{subfigure}\hspace{0.2em}
    \begin{subfigure}[t]{0.31\textwidth}
        \centering
        \includegraphics[width=\linewidth]{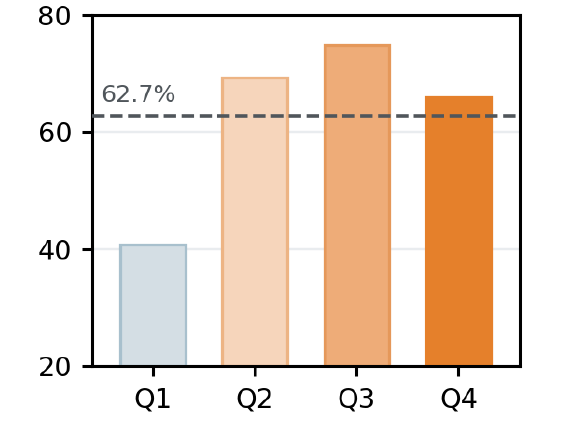}
        \vspace{-1.5em}
        \caption{Frame-level Motion}
        \label{fig:motion_intra_b}
    \end{subfigure}
    \begin{subfigure}[t]{0.31\textwidth}
        \centering
        \includegraphics[width=\linewidth]{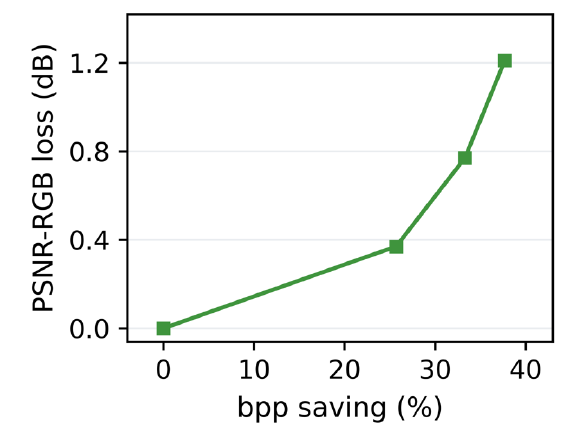}
        \vspace{-1.5em}
        \caption{Intra Period}
        \label{fig:motion_intra_c}
    \end{subfigure}\hspace{0.2em}
    \label{fig:motion_intra}
    \vspace{-0.5em}
    \caption{
        \textbf{Motion-dependent BD-rate Savings and Longer Intra Periods.}
        (a--b) BD-rate savings over DCMVC. Dashed lines indicate the sequence mean and the mean over four quartiles, respectively.
        The frame-level analysis in (b) includes only P-frames.
        (c) Relative bpp saving and PSNR-RGB loss for single-GOP coding, using a 32-frame GOP as the reference. Points from left to right correspond to GOP lengths of 32, 64, 96, and 192 frames, each containing one I-frame.
    }
    \vspace{-1.0em}
\end{figure*}

At the sequence level, the large-motion group achieves a saving of 56.60\%, compared with 47.96\% for the small-motion group (Fig.~\ref{fig:motion_intra_a}).
We further examine this relationship at the frame level by grouping only P-frames into four motion quartiles.
From the lowest to the highest motion quartile, BD-rate savings are 40.8\%, 69.3\%, 74.8\%, and 65.9\%, respectively (Fig.~\ref{fig:motion_intra_b}).
Although the largest saving occurs in the third quartile, the highest-motion quartile still shows substantial gains.
Together, these results indicate that event guidance remains useful across different motion magnitudes, including sequences and frames with large motion.

\vspace{-0.8em}
\paragraph{Longer Intra Periods.}
On seven UVG sequences, we examine how longer intra periods affect RGB coding rate and reconstruction quality. 
Relative to $K=32$, increasing $K$ to 64, 96, and 192 reduces mean bpp by 25.7\%, 33.3\%, and 37.7\%, with PSNR-RGB losses of 0.37, 0.77, and 1.21 dB, respectively (Fig.~\ref{fig:motion_intra_c}).
Increasing $K$ from 32 to 64 provides the largest incremental rate saving. 
Further increases yield smaller additional savings while reconstruction quality continues to decline.

\begin{wrapfigure}{r}{0.45\textwidth}
    \centering
    \vspace{-1.5em}
    \small
    \captionof{table}{
        \textbf{Complexity Analysis.}
        Model size and encoding/decoding speed on UVG Beauty using a single RTX 3090.
    }
    \label{tab:complexity}
    \resizebox{\linewidth}{!}{%
    \begin{tabular}{lrrr}
        \toprule
        Model & Params. & Enc. FPS & Dec. FPS \\
        \midrule
        DCVC            & 7.94M  & 0.153  & 0.023  \\
        DCVC-TCM        & 10.71M & 1.883  & 2.749  \\ 
        DCVC-DC         & 19.78M & 1.691  & 2.082  \\
        DCVC-FM         & 18.34M & 2.719  & 3.298  \\
        DCMVC           & 21.02M & 1.061  & 1.208  \\
        DCVC-RT         & 20.69M & 57.221 & 51.198 \\
        ENVC            & 33.52M & 0.323  & 0.386  \\
        \bottomrule
    \end{tabular}%
    }
\end{wrapfigure}

\vspace{-0.5em}
\subsection{Complexity Analysis}
Tab.~\ref{tab:complexity} reports model size and encoding/decoding throughput measured on the 1920$\times$1080 UVG Beauty sequence using a single RTX 3090.
ENVC processes events in addition to video frames and has 33.52M parameters, with encoding and decoding speeds of 0.323 and 0.386 FPS, respectively.
Both encoder and decoder run EFP and EME modules to construct the shared motion prior, adding computation to both coding stages.
The additional event-processing modules increase model size and computational cost, making efficiency a limitation of the current implementation.

%% file: sec/6_discussion.tex
\section{Discussion}
\label{sec:discussion}

Our results suggest that events are a promising complementary modality to guide RGB video compression.
This motivates further research on efficient event coding and joint compression of RGB frames and event streams.
Jointly modeling the dependencies between these modalities could further exploit cross-modal redundancy and improve overall coding efficiency, extending the present framework toward joint RGB--event compression.

As an initial exploration of RGB coding with shared events, ENVC focuses on integrating event information into motion coding and temporal context construction.
The current implementation requires event-conditioned prediction and motion estimation at both the encoder and decoder, introducing additional computational overhead.
Developing lightweight event-guidance modules and efficient coding pipelines to reduce complexity and end-to-end latency remains an important direction for future work.

%% file: sec/7_conclusion.tex
\section{Conclusion}

In this work, we presented ENVC, which incorporated events as a complementary modality for RGB video compression through event-guided residual motion coding and gated temporal context refinement.
Across six benchmarks with synthetic events, ENVC achieved average BD-rate savings of 39.13\% using PSNR-RGB and 67.63\% using LPIPS relative to DCMVC.
Further analysis showed scale-dependent event injection, while longer intra periods revealed a growing rate–quality trade-off.
These results demonstrate the potential of shared event observations as a complementary modality for reducing the RGB coding rate.

%% file: sec/a_scenario.tex
\section{Practical Scenarios}
\label{app:scenario}

In this paper, we mainly consider systems that already transmit or retain events for downstream processing and additionally require RGB video.
Event cameras offer high temporal resolution, wide dynamic range, and reduced motion blur~\citep{gallego2022event}, with commercial implementations including DAVIS346\footnote{\url{https://docs.inivation.com/hardware/current-products/davis346.html}.} and Sony--Prophesee-based industrial cameras.\footnote{\url{https://www.prophesee.ai/event-based-cameras-prophesee-partners/}.}

Recent studies demonstrate event-based object detection~\citep{ahmed2025efficient}, continuous-time 3D detection~\citep{kang2025unleashing}, and monocular depth estimation~\citep{bartolomei2025depth}, while TESPEC studies representations across detection, segmentation, and depth estimation~\citep{mohammadi2025tespec}.
These capabilities give applications a reason to preserve event streams independently of RGB compression.

For example, a robotic or industrial system may send events to a server for perception while providing RGB video for human inspection.
An event-based detector can identify a moving object, but an operator may still need its color, appearance, and surrounding context to interpret the prediction.
Displaying detections or trajectories over RGB frames connects machine outputs to an interpretable view of the scene.
Both streams may also support machine processing: EventDrive combines events, RGB, and language for driving-related perception, prediction, and planning~\citep{lu2026eventdrive}.
When these computations share a receiving endpoint, the RGB decoder can reuse the events already delivered for perception.

Related cross-modal compression methods use shared LiDAR depth maps~\citep{gnutti2024lidar} or sparse point clouds~\citep{jiang2024sparse} to assist RGB image coding, while RangeCM uses RGB camera context for LiDAR compression~\citep{song2026lowlatency}.
ENVC applies this principle to shared event information.
With the event stream held fixed across RGB codecs, its transmission or storage cost is common to the compared systems.
Our reported rate therefore measures the additional RGB coding cost.
Both coding ends use the same available event representation, and App.~\ref{app:event_coding} examines sensitivity to compression-induced event degradation.

\vspace{-0.5em}
\subsection{Sensitivity to Event Compression}
\label{app:event_coding}

\paragraph{Motivation.} 
In the transmission and archival scenarios discussed above, the events available at the receiver may themselves have undergone compression.
Existing approaches include temporal aggregation followed by video coding~\citep{khan2021time}, point-cloud-based coding~\citep{huang2023event}, and learned lossless and lossy codecs~\citep{sezavar2024learning,seleem2026event}, and the JPEG XE initiative further reflects the practical need for efficient event representation and interoperability.\footnote{\url{https://jpeg.org/jpegxe/index.html}.}
Lossless coding preserves the coded representation, whereas lossy coding trades event fidelity for rate reduction.
Prior work therefore evaluates compressed events through their effects on downstream tasks, including recognition, video reconstruction, optical flow, and depth estimation~\citep{huang2023evaluation,seleem2025double}.
In our setting, the decoded event stream can likewise be reused as shared side information at both RGB coding ends.
This motivates evaluating how event coding quality affects its usefulness for RGB reconstruction in ENVC.

\vspace{-0.8em}
\paragraph{Settings.}
To examine this setting, we evaluate a fixed ENVC checkpoint on the first 96 frames of seven UVG sequences, using synthetic events, an intra period of 32, and quality $Q_0$.
For each inter-frame interval, positive and negative events are represented as separate $(x,y,t)$ point clouds~\citep{huang2023event} and compressed using G-PCC (TMC13)\footnote{\url{https://github.com/MPEGGroup/mpeg-pcc-tmc13}.}.
Timestamps are mapped to an integer grid before coding, and the decoded events undergo the original voxelization and normalization and are supplied to both coding ends without fine-tuning.
We vary the event quality factor $q_E$, corresponding to the quantization scale, where smaller values produce coarser coordinate quantization, and $q_E=1$ is lossless for the coded integer-coordinate representation.

\begin{wrapfigure}{r}{0.6\textwidth}
    \centering
    \begin{minipage}[t]{0.49\linewidth}
        \vspace{0pt}
        \centering
        \includegraphics[width=\linewidth]{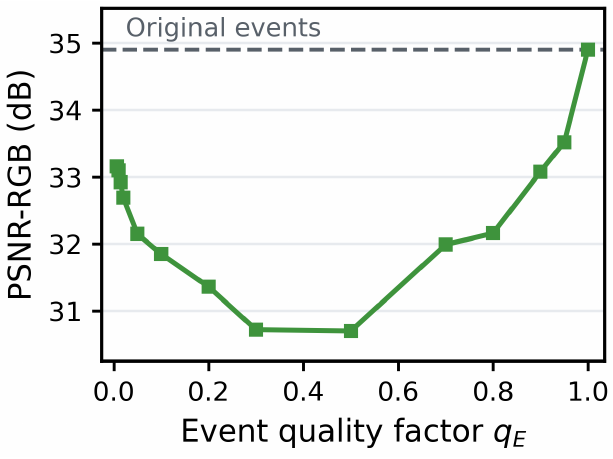}
        \par{\small (a) RGB Quality}
    \end{minipage}
    \begin{minipage}[t]{0.49\linewidth}
        \vspace{0pt}
        \centering
        \includegraphics[width=\linewidth]{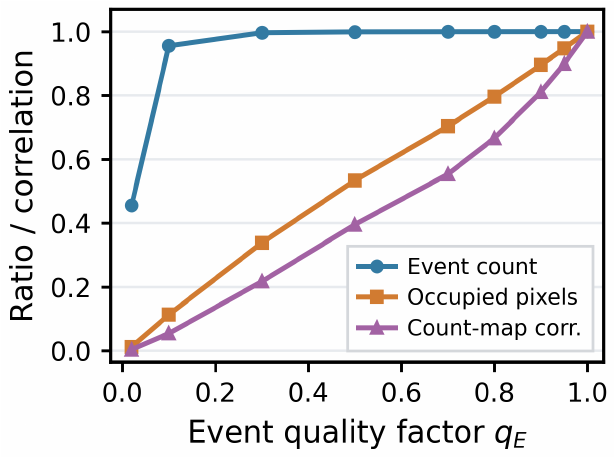}
        \par{\small (b) Event Fidelity}
    \end{minipage}
    \vspace{-0.5em}
    \caption{
        \textbf{Sensitivity to Event Compression.}
        (a) PSNR-RGB on UVG, with original-event performance shown by the dashed line.
        (b) Normalized event and occupied-pixel counts, and event-count map correlation.
        Smaller $q_E$ indicates coarser quantization.
    }
    \label{fig:event_compression}
\end{wrapfigure}

\vspace{-0.8em}
\paragraph{Results.}
Fig.~\ref{fig:event_compression} shows that $q_E=1$ reproduces the original-event PSNR-RGB of 34.90~dB.
At $q_E=0.95$, PSNR-RGB decreases to 33.52~dB, and at $q_E=0.5$ it reaches 30.70~dB, a 4.20~dB reduction.
Event statistics over 21 sampled intervals help characterize this sensitivity.
At $q_E=0.5$, 99.9\% of events remain, but the number of occupied pixels falls to 53.4\% of its original value and the count-map correlation is only 0.396.
Thus, preserving event count does not ensure preservation of the spatial structure used for guidance.
The response is also non-monotonic: more aggressive quantization at $q_E=0.02$ yields 32.69~dB despite poorer event fidelity.
One possible explanation is that strongly suppressed guidance can be less disruptive than structured but inaccurate guidance, although this experiment does not isolate that mechanism.

These results reveal sensitivity of the current checkpoint to coordinate-quantized events.
They motivate further research into efficient event codecs that preserve task-relevant spatiotemporal structure, together with compression-aware training of event-guided video codecs.
Such developments will be important for realizing the value of events in practical multimodal systems.

%% file: sec/b_training_details.tex
\section{Training Details}
\label{app:training}

\paragraph{Progressive Training.}
We adopt five training stages to first establish event-based prediction and then optimize its integration into the codec.
First, we train EFP and EME using frame prediction and motion-compensated reconstruction losses, without rate losses ($R$).
Second, we train RMC to balance motion-compensated reconstruction quality against the rate of the coded motion residual.
Third, we train the context refinement modules and conditional frame codec while keeping the event and motion modules fixed.
This stage combines frame reconstruction distortion, frame coding rate, and the decoupling loss of DCMVC~\citep{tang2025dcmvc}.
Fourth, we jointly optimize all inter-frame modules, including CME, under the complete RD objective with both motion and frame coding rates.
The last stage extends this joint optimization to longer sequences (beyond the seven-frame septuplets) through cascaded training.

Both EME and CME are initialized with the SPyNet weights from DCMVC~\citep{tang2025dcmvc}.
We use the pretrained intra-frame codec of DCVC-DC~\citep{li2023dcvcdc} for I-frame coding and keep it fixed throughout training.

\vspace{-0.8em}
\paragraph{Cascaded Training.}
We progressively increase the training sequence length from 8 to 16, 24, and 32 frames and minimize the RD loss averaged over all P-frames in each sequence.
To reduce memory consumption, we use Forward Recomputation Backpropagation (FRB), as in DCMVC~\citep{tang2025dcmvc}.
This approach recomputes intermediate activations during backpropagation instead of retaining them all in memory~\citep{chen2016training}, enabling optimization over longer prediction chains.

\vspace{-0.8em}
\paragraph{Training Data.}
We use Vimeo-90K septuplets~\citep{xue2019video} paired with synthetic events.
For cascaded training, we use 9,220 training clips and 106 validation clips extracted from the original Vimeo videos\footnote{\url{https://github.com/anchen1011/toflow/blob/master/data/original_vimeo_links.txt}.}, each containing 32 frames with paired synthetic events.
Events are represented as signed voxel grids with 10 temporal bins.
Training uses aligned random crops of $256\times256$ pixels, reduced to $128\times128$ for the final 32-frame stage.

\vspace{-0.8em}
\paragraph{Optimization.}
We train a single multi-rate model with $\lambda\in\{85,170,380,840\}$ and periodic frame weights $(0.5,1.2,0.5,0.9)$ following DCVC-DC~\citep{li2023dcvcdc}.
Distortion combines YUV420 and RGB MSE with weights 0.8 and 0.2, respectively.
We use AdamW~\citep{loshchilov2017decoupled} with weight decay 0.01 and train on four NVIDIA A100-SXM4-40GB GPUs.

%% file: sec/c_case_analysis.tex
\section{Sequence-level Gains and Failures}
\label{app:failure}

We examine the 53 sequences in the main benchmarks using the first 96 frames, and an intra period of 32 as in Tab.~\ref{tab:bd_rate}.
ENVC achieves negative PSNR-RGB BD-rates relative to DCMVC on 50 of the 53 sequences, with a median of $-44.9\%$.
Tab.~\ref{tab:per_sequence} reports representative gains and the three sequences with positive overall PSNR-RGB BD-rates, together with P-frame-only results and sequence statistics.

\begin{wrapfigure}{r}{0.55\textwidth}
    \centering
    \vspace{-1.0em}
    \small
    \captionof{table}{
        \textbf{Sequence-level Results against DCMVC.}
        PSNR-RGB BD-rates (\%) use all frames (All) or P-frames only (P).
        Events/px denotes mean event density per P-frame.
        $|\Delta Y|$ denotes mean absolute inter-frame BT.709 luma change on the 0--255 scale.
    }
    \label{tab:per_sequence}
    \setlength{\tabcolsep}{3pt}
    \renewcommand{\arraystretch}{1.05}
    \resizebox{0.9\linewidth}{!}{%
    \begin{tabular}{@{}llrrrr@{}}
        \toprule
        Dataset & Sequence & All & P & Events/px & $|\Delta Y|$ \\
        \midrule
        \multicolumn{6}{l}{\textit{Representative gains}} \\
        UVG     & Beauty       & $-82.4$ & $-89.1$ & 0.146 & 4.93 \\
        MCL-JCV & SRC15        & $-77.0$ & $-94.7$ & 0.340 & 7.06 \\
        MCL-JCV & SRC22        & $-73.5$ & $-79.1$ & 1.693 & 25.44 \\
        HEVC-B  & Kimono1      & $-66.7$ & $-78.2$ & 0.435 & 9.52 \\
        HEVC-C  & RaceHorses   & $-64.7$ & $-74.9$ & 0.820 & 21.15 \\
        \midrule
        \multicolumn{6}{l}{\textit{Limited gains and losses}} \\
        UVG     & HoneyBee     & $-14.7$ & $-39.0$ & 0.008 & 2.21 \\
        HEVC-E  & Johnny       & $-6.7$  & $-30.3$ & 0.022 & 2.32 \\
        MCL-JCV & SRC13        & $+1.2$  & $-4.8$  & 0.034 & 3.36 \\
        UVG     & Bosphorus    & $+1.8$  & $-7.9$  & 0.017 & 3.21 \\
        UVG     & Jockey       & $+56.6$ & $+42.9$ & 0.282 & 11.94 \\
        \bottomrule
    \end{tabular}%
    }
    \vspace{-0.8em}
\end{wrapfigure}

\vspace{-0.8em}
\paragraph{Success Cases.}
Larger savings are generally associated with denser events and greater temporal intensity changes.
Across the 53 sequences, the Spearman correlation between event density and PSNR-RGB BD-rate is $-0.47$.
The 13 sequences with the largest savings average $-71.1\%$ BD-rate, compared with $-7.1\%$ for the 13 with the smallest savings.
Their mean event densities are 0.582 and 0.160 events per pixel, and their mean absolute luma changes are 11.34 and 5.39, respectively.
MCL-JCV SRC22 and HEVC-C RaceHorses exemplify substantial savings on sequences with pronounced temporal changes.
Event density alone, however, does not determine the gain.
UVG Beauty achieves the largest saving ($-82.4\%$) with only 0.146 events per pixel, showing that large gains also occur at relatively low event densities.

\vspace{-0.8em}
\paragraph{Limited Gains and Failure Cases.}
Low-activity sequences such as HoneyBee, Bosphorus, and SRC13 generate few events and show limited gains or slight losses.
Bosphorus and SRC13 have slightly positive overall BD-rates, but retain negative P-frame-only BD-rates of $-7.9\%$ and $-4.8\%$.
The largest loss occurs on Jockey, with $+56.6\%$ overall and $+42.9\%$ P-frame-only BD-rate despite an event density of 0.282 events per pixel.
Its complex motion, occlusions, and repetitive textures are plausible sources of difficulty, although this analysis does not isolate their contributions.
Jockey nevertheless achieves an LPIPS-based BD-rate of $-52.4\%$, demonstrating that its relative performance depends on the quality metric.

\vspace{-0.8em}
\paragraph{Limited Gains on HEVC-E.}
HEVC-E consists of video-conferencing scenes with low temporal activity, producing only 0.027 events per pixel on average.
ENVC achieves a P-frame-only PSNR-RGB BD-rate of $-32.0\%$, but the overall saving is limited to $-7.8\%$.
This difference is partly explained by the substantial contribution of intra coding.
At $Q_0$, I-frames account for $56.2\%$ of the total RGB rate of ENVC, compared with $30.1\%$ for DCMVC, diluting the overall benefit of P-frame rate savings.

The allocation of bits to P-frames also differs between the two codecs.
At $Q_0$, ENVC uses $66\%$ fewer P-frame bits than DCMVC, with a mean P-frame PSNR of 37.71~dB versus 38.80~dB.
Both codecs achieve the same mean I-frame PSNR of 38.02~dB.
Thus, the mean P-frame quality of DCMVC exceeds its intra-frame quality, whereas that of ENVC remains lower.
This pattern is consistent with further refinement of the intra reconstruction by DCMVC in these low-activity scenes.
The results suggest that improving the reconstruction of largely static content, together with adapting event guidance to sparse-event inputs, is a promising direction for increasing coding efficiency on HEVC-E.

%% file: sec/d_real_event.tex
\section{Generalization to Real Events}
\label{app:real_event}

We evaluate whether ENVC, trained with synthetic events, can be applied to real sensor events without additional fine-tuning.
By comparing real and synthetic events for the same RGB sequences, we examine how the event source affects RGB coding performance.

\vspace{-0.8em}
\paragraph{Evaluation Protocol.}
We use 15 test sequences from HS-ERGB~\citep{tulyakov2021time} and 17 validation sequences from BS-ERGB~\citep{tulyakov2022time}.
Each sequence is evaluated over 32 frames with an intra period of 32.
The real-event condition uses sensor recordings, while the synthetic-event condition uses events generated from the same RGB frames following Sec.~\ref{sec:dataset}.

\begin{wrapfigure}{r}{0.50\textwidth}
    \centering
    \begin{minipage}[t]{0.49\linewidth}
        \vspace{0pt}
        \centering
        \includegraphics[width=\linewidth]{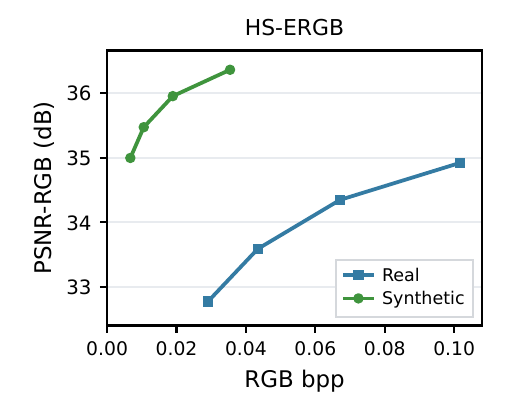}
        \vspace{-1.5em}
        \par{\small (a) PSNR-RGB}
    \end{minipage}
    \begin{minipage}[t]{0.49\linewidth}
        \vspace{0pt}
        \centering
        \includegraphics[width=\linewidth]{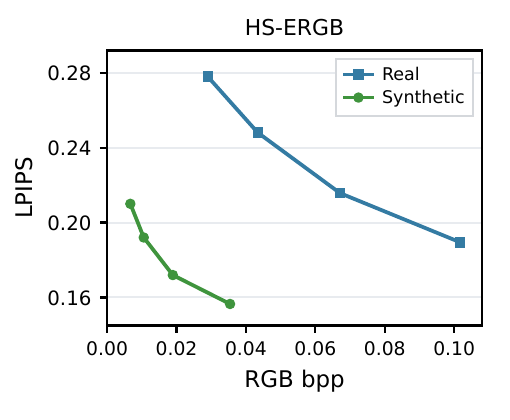}
        \vspace{-1.5em}
        \par{\small (b) LPIPS}
    \end{minipage}
    \par\vspace{0.4em}
    \begin{minipage}[t]{0.49\linewidth}
        \vspace{0pt}
        \centering
        \includegraphics[width=\linewidth]{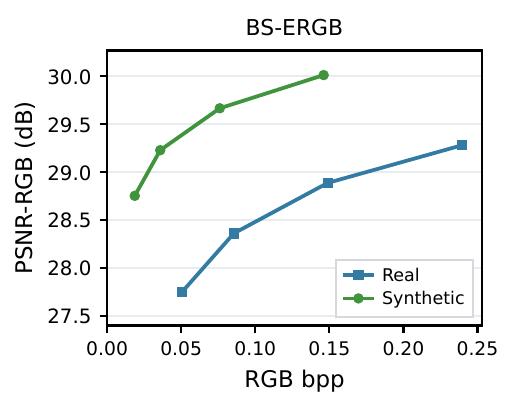}
        \vspace{-1.5em}
        \par{\small (c) PSNR-RGB}
    \end{minipage}
    \begin{minipage}[t]{0.49\linewidth}
        \vspace{0pt}
        \centering
        \includegraphics[width=\linewidth]{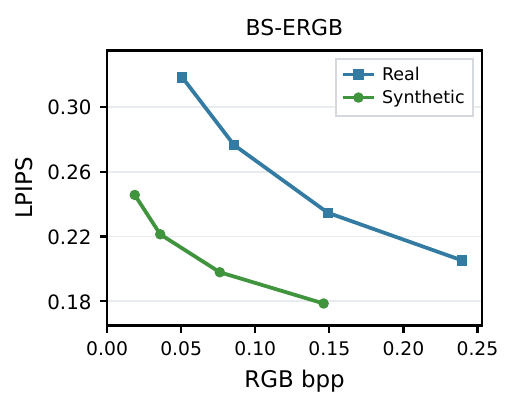}
        \vspace{-1.5em}
        \par{\small (d) LPIPS}
    \end{minipage}
    \caption{
        \textbf{Generalization to Real Events.}
        RD curves with real and synthetic events on HS-ERGB (top) and BS-ERGB (bottom).
    }
    \label{fig:real_events}
    \vspace{-0.8em}
\end{wrapfigure}

\vspace{-0.8em}
\paragraph{RD Performance.}
At corresponding quality settings, real events yield higher RGB rates, lower PSNR-RGB, and higher LPIPS than synthetic events on both datasets (Fig.~\ref{fig:real_events}).
The degradation in both rate and distortion suggests that the synthetic-trained codec uses real observations less effectively for compression.
Thus, ENVC can operate on real sensor events without fine-tuning, but its performance under synthetic guidance does not fully transfer to real recordings.
This comparison demonstrates applicability to real inputs while highlighting a remaining gap in generalization across event domains.

\vspace{-0.8em}
\paragraph{Understanding the Domain Gap.}
Synthetic events are denser on average ($\times$1.54 and $\times$1.59 in Tab.~\ref{tab:event_quality}).
However, App.~\ref{app:failure} shows that event density alone does not determine RGB coding gains.
Differences in spatiotemporal event statistics and compatibility with the training distribution may also contribute to the real--synthetic performance gap.

\vspace{-0.8em}
\paragraph{Implications.}
ENVC can be applied to real sensor events without fine-tuning.
The observed performance difference suggests room to improve transfer from synthetic training data to real sensor recordings.
Training with more diverse event statistics and adaptation to real data are promising directions for improving this transfer.

%% file: sec/e_gate_analysis.tex
\section{Analysis of Learned Event Gating}
\label{app:gate}

\begin{wrapfigure}{r}{0.48\textwidth}
    \centering
    \vspace{-1.0em}

    \small
    \captionof{table}{\textbf{Selected Sequences for the Gate Activation Analyses.} Each sequence contributes its first 15 P-frames.}
    \label{tab:gate_sequences}
    \vspace{-0.5em}
    \begin{tabular}{ll}
        \toprule
        Dataset & Sequences \\
        \midrule
        UVG & Beauty, Jockey \\
        MCL-JCV & SRC02, SRC16 \\
        HEVC-B & BasketballDrive, ParkScene \\
        HEVC-C & BasketballDrill, PartyScene \\
        HEVC-D & BQSquare, BasketballPass \\
        HEVC-E & FourPeople, Johnny \\
        \bottomrule
        \vspace{0.2em}
    \end{tabular}

    \includegraphics[width=0.7\linewidth]{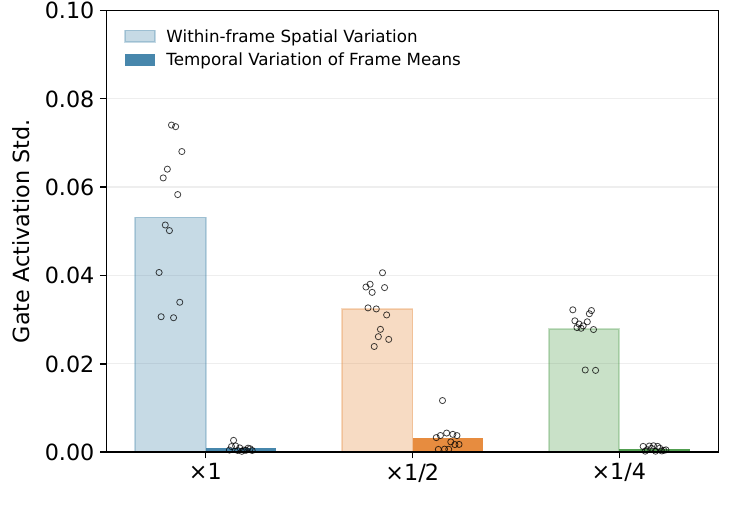}
    \vspace{-1.0em}

    \captionof{figure}{
        \textbf{Spatial and Temporal Variation of Event Gates.}
        Light and dark bars denote spatial and temporal variation, respectively.
    }
    \label{fig:gate_variation}
    \vspace{-2.0em}
\end{wrapfigure}

\paragraph{Analysis Setup.}
We extend the analysis in Sec.~\ref{sec:exp_gate} using the ENVC model at quality factor $Q_1$.
The gate and feature analyses use 12 sequences, with two from each of six dataset groups, as listed in Tab.~\ref{tab:gate_sequences}.
The constant-gate interventions use the four sequences reported in Tab.~\ref{tab:gate_intervention}.
With an intra period of 32, we analyze the first 15 P-frames following the initial I-frame of each sequence.
All gates are initialized to $\sigma(4)\approx0.982$, while the adapters initially output zero.
RMS ratios normalize adapter-feature magnitudes by those of the temporal context before injection.

\vspace{-0.8em}
\paragraph{Spatial and Temporal Variation.}
For each frame, we compute the spatial standard deviation and spatial mean of the gate activations. For each sequence, we average the spatial standard deviations over the first 15 P-frames and compute the temporal standard deviation of the frame-wise means.
Fig.~\ref{fig:gate_variation} shows the averages across 12 sequences as bars and individual sequences as circles. Spatial variation is larger at all three resolutions, indicating spatially nonuniform gating with comparatively stable frame-averaged activation.

\begin{figure}[t]
    \centering
    \vspace{-1.0em}
    \includegraphics[width=\linewidth]{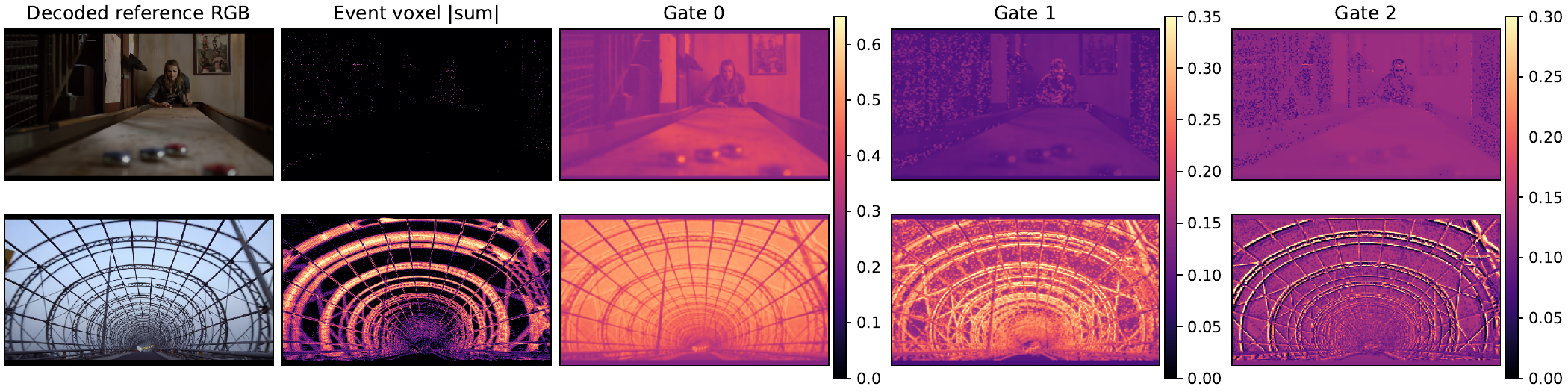}
    \vspace{-1.5em}
    \caption{
        \textbf{Spatial Patterns of Learned Event Gates.}
        MCL-JCV SRC01 (top) and SRC10 (bottom).
        Columns show the decoded reference, event voxel visualization, and gate maps at three scales.
    }
    \label{fig:gate_spatial}
    \vspace{-1.0em}
\end{figure}

\vspace{-0.8em}
\paragraph{Spatial Gate Patterns.}
Fig.~\ref{fig:gate_spatial} visualizes the learned gates on two additional MCL-JCV sequences with contrasting event activity.
The maps exhibit different spatial patterns across the three scales.

In SRC10, Gates 1 ($\times 1/2$) and Gate 2 ($\times 1/4$) show
pronounced responses along curved scene structures,
while Gate 0 ($\times 1$) exhibits a broader response.
These examples complement the variation statistics by illustrating how the learned modulation is spatially organized.

\vspace{-0.8em}
\paragraph{Constant-gate Interventions.}
Tab.~\ref{tab:gate_intervention} reports the results of replacing all learned gates with $G=0$ or $G=1$ at inference, without retraining.
With $G=0$, PSNR drops by 2.42--2.94~dB and estimated bpp increases on all four sequences; on Jockey, bpp rises from 0.005 to 0.021.
With $G=1$, PSNR drops by 2.30--3.89~dB, while bpp increases on three sequences. BQSquare is the exception: its bpp decreases from 0.030 to 0.021, alongside a PSNR decrease from 31.779 to 28.937~dB.
For BasketballDrill, $G=1$ yields nearly the same bpp as the learned gates (0.017), but PSNR falls by 2.30~dB.
These interventions show that fixed gates change the measured operating
points. 
They do not isolate the effect of spatial variation, since replacing the gates also changes injection magnitude, and moreover, $G=0$ disables only the gated injection branch, leaving other event pathways active.

\begin{table}[h]
    \centering
    \caption{
        \textbf{Constant-gate Interventions.}
        PSNR (dB) and estimated codec bpp over the first 15 P-frames.
        Bpp rates exclude the initial I-frame and event transmission.
    }
    \vspace{-0.5em}
    \label{tab:gate_intervention}
    \resizebox{0.8\columnwidth}{!}{%
    \begin{tabular}{lcccccc}
        \toprule
        & \multicolumn{2}{c}{Learned Gates}
        & \multicolumn{2}{c}{$G=0$}
        & \multicolumn{2}{c}{$G=1$} \\
        \cmidrule(lr){2-3}\cmidrule(lr){4-5}\cmidrule(l){6-7}
        Sequence & PSNR & bpp & PSNR & bpp & PSNR & bpp \\
        \midrule
        BQSquare        & 31.779 & 0.030 & 28.841 (-2.94) & 0.073 & 28.937 (-2.84) & 0.021 \\
        Jockey          & 36.351 & 0.005 & 33.801 (-2.55) & 0.021 & 32.461 (-3.89) & 0.014 \\
        BasketballDrill & 33.102 & 0.017 & 30.679 (-2.42) & 0.032 & 30.799 (-2.30) & 0.017 \\
        HoneyBee        & 36.715 & 0.007 & 33.772 (-2.94) & 0.013 & 33.729 (-2.99) & 0.009 \\
        \bottomrule
    \end{tabular}%
    }
    \vspace{-1.0em}
\end{table}

%% file: sec/f_sensitivity_to_event_perturbations.tex
\section{Learned Use of Spatiotemporally Aligned Events}
\label{app:event_perturbation}

We investigate whether ENVC learns to exploit the spatiotemporal correspondence between events and RGB frames.
By removing or disrupting this correspondence at inference, we examine how aligned event observations contribute to the trained codec's reconstruction quality and coding efficiency.

\vspace{-0.8em}
\paragraph{Evaluation Protocol.}
We evaluate a fixed ENVC checkpoint at $Q_1$ without retraining
on 13 selected sequences from HEVC-B/C/D/E, UVG, and MCL-JCV.
Each sequence is paired with synthetic events and contains one I-frame followed by 31 P-frames.
All conditions use aligned events for the first five P-frames, after which we intervene on the remaining 26 P-frames.
Each condition propagates its own reconstructed state.
We average PSNR-RGB and entropy-estimated P-frame bpp over these 26 frames and then equally across sequences.

\vspace{-0.8em}
\paragraph{Event Interventions.}
The \texttt{aligned} condition supplies events from the interval immediately preceding the target frame.
For \texttt{lag 1} and \texttt{lag 5}, we instead use the event interval from one or five frame periods earlier.
For \texttt{temporal shuffle}, a fixed random permutation reassigns the 26 event intervals to target frames, with none remaining at its original target.
This preserves the set of observations while disrupting their temporal correspondence.
For \texttt{spatial shuffle}, one fixed permutation of the $H \times W$ locations is applied to every temporal bin and intervened frame within a sequence.
This preserves voxel values and temporal-bin membership while disrupting spatial structure and correspondence with RGB.
The \texttt{zero-event} condition sets the input event voxel grid to zero.

\begin{figure}[t]
    \centering
    \begin{minipage}[t]{0.40\textwidth}
        \vspace{0pt}
        \centering
        \captionof{table}{
        \textbf{Contribution of Aligned Event Guidance.}
        Averages over 13 selected sequences.
        $\Delta$PSNR denotes the drop relative to aligned inputs.
        Rate ratios are averaged over sequences.
        }
        \label{tab:event_alignment}
        \setlength{\tabcolsep}{3pt}
        \renewcommand{\arraystretch}{1.15}
        \resizebox{\linewidth}{!}{%
        \begin{tabular}{lrrrr}
            \toprule
            Input &
            \shortstack{PSNR\\(dB)} &
            \shortstack{P-frame\\bpp} &
            \shortstack{$\Delta$PSNR\\(dB)} &
            \shortstack{Rate\\ratio} \\
            \midrule
            Aligned
            & 34.794 & 0.01085 & 0.000 & $1.00\times$ \\
            \midrule
            Zero event
            & 32.844 & 0.07611 & 1.950 & $6.52\times$ \\
            Lag 1
            & 32.446 & 0.07752 & 2.348 & $6.75\times$ \\
            Lag 5
            & 32.122 & 0.09238 & 2.672 & $8.06\times$ \\
            Temporal shuffle
            & 32.027 & 0.10400 & 2.767 & $9.08\times$ \\
            Spatial shuffle
            & 29.634 & 0.14133 & 5.159 & $13.83\times$ \\
            \bottomrule
        \end{tabular}%
        }
    \end{minipage}
    \hfill
    \begin{minipage}[t]{0.58\textwidth}
        \vspace{0pt}
        \centering
        \captionsetup{type=figure}
        \captionsetup[subfigure]{font=small, skip=3pt}

        \begin{subfigure}[t]{0.48\linewidth}
            \centering
            \includegraphics[width=\linewidth]{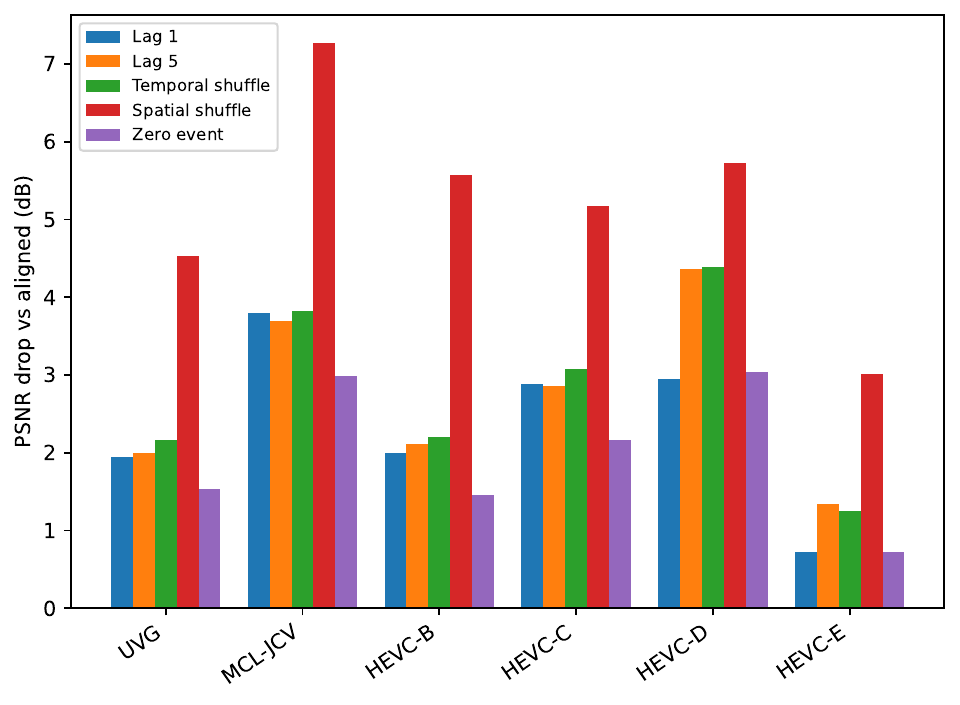}
            \caption{PSNR-RGB Drop}
            \label{fig:event_alignment_psnr}
        \end{subfigure}
        \hfill
        \begin{subfigure}[t]{0.48\linewidth}
            \centering
            \includegraphics[width=\linewidth]{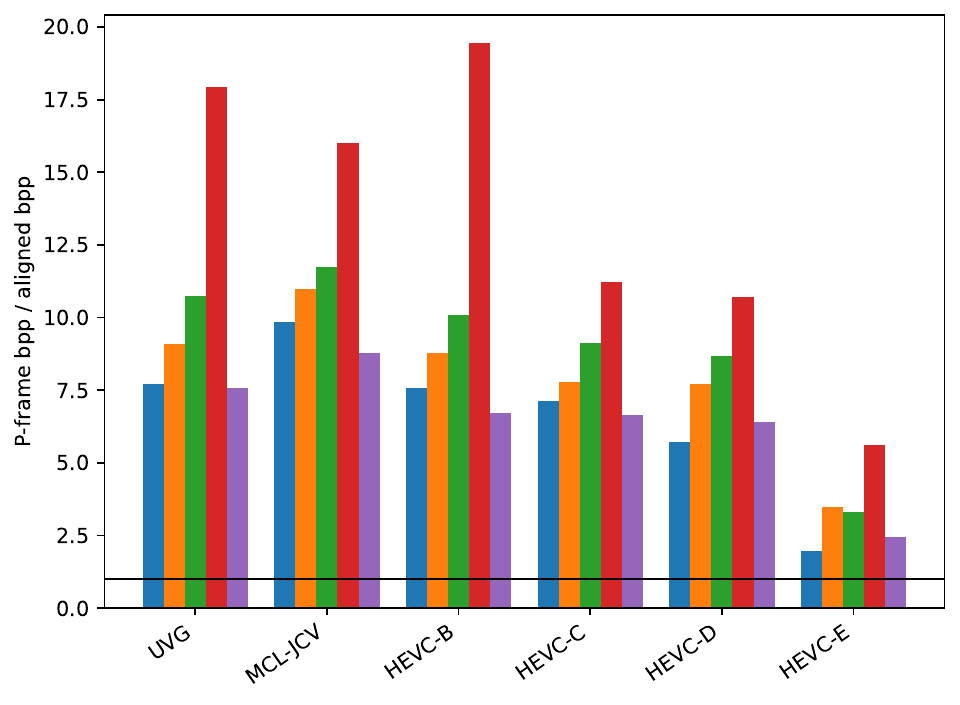}
            \caption{Relative Bpp}
            \label{fig:event_alignment_bpp}
        \end{subfigure}

        \vspace{-0.5em}
        \caption{
        \textbf{Effects of Disrupting Event Correspondence.}
        Selected-sequence averages within each dataset.
        (a) PSNR-RGB drop relative to aligned events.
        (b) Mean per-sequence P-frame bpp ratio relative to
        aligned events.
        }
        \label{fig:event_alignment_datasets}
    \end{minipage}
    \vspace{-1.0em}
\end{figure}

\vspace{-0.8em}
\paragraph{Learned Use of Event Correspondence.}
Tab.~\ref{tab:event_alignment} and Fig.~\ref{fig:event_alignment_datasets} show that aligned events enable higher reconstruction quality at lower coding rates.
Shifting the input by one frame reduces PSNR-RGB by 0.72--3.79~dB across the selected dataset groups, with mean per-sequence bpp ratios of $1.99$--$9.86$.
Both shuffle conditions perform worse than zero-event inputs in PSNR and bpp on every evaluated sequence.
These results support that ENVC has learned to exploit event observations that correspond to the target frame, with benefits extending to both reconstruction and coding efficiency.

\vspace{-0.8em}
\paragraph{Content-dependent Contribution.}
The contribution of aligned events varies with video content.
Removing or temporally reassigning events has smaller effects on the selected HEVC-E sequences than on the MCL-JCV examples.
Increasing the delay also produces sequence-dependent changes.
These observations are consistent with different degrees of reliance on event guidance across video content.

%% file: sec/g_bit_allocation.tex
\vspace{-0.5em}
\section{Bit Allocation Analysis}
\label{app:bit_allocation}

\paragraph{Evaluation Protocol.}
We evaluate ENVC and DCMVC on seven UVG sequences using the first
96 frames and an intra period of 32.

\begin{figure}[t]
    \centering

    \begin{minipage}[t]{0.35\textwidth}
        \vspace{2pt}
        \centering
        \captionsetup{font=small, skip=4pt}

        \captionof{table}{
            \textbf{RGB Coding Rate Contributions.}
            Rates ($10^{-3}$ bpp) on UVG are averaged over the common PSNR-RGB range after interpolation and normalized over all coded frames.
            Motion and frame coding include their respective hyperlatents.
            Changes are relative to DCMVC.
        }
        \label{tab:bit_allocation}

        \footnotesize
        \setlength{\tabcolsep}{3pt}
        \renewcommand{\arraystretch}{1.15}
        \resizebox{\linewidth}{!}{%
            \begin{tabular}{lrrr}
                \toprule
                Component & DCMVC & ENVC & Change \\
                \midrule
                Motion
                & 2.038 & 0.655 & $-67.9\%$ \\
                Frame coding
                & 22.093 & 13.292 & $-39.8\%$ \\
                I-frame
                & 5.390 & 7.878 & $+46.2\%$ \\
                \midrule
                Total
                & 29.521 & 21.825 & $-26.1\%$ \\
                \bottomrule
            \end{tabular}%
        }
    \end{minipage}
    \hfill
    \begin{minipage}[t]{0.63\textwidth}
        \vspace{5pt}
        \centering
        \captionsetup{type=figure, font=small, skip=4pt}
        \captionsetup[subfigure]{font=small, skip=2pt}

        \begin{subfigure}[t]{0.49\linewidth}
            \centering
            \includegraphics[width=\linewidth]
                {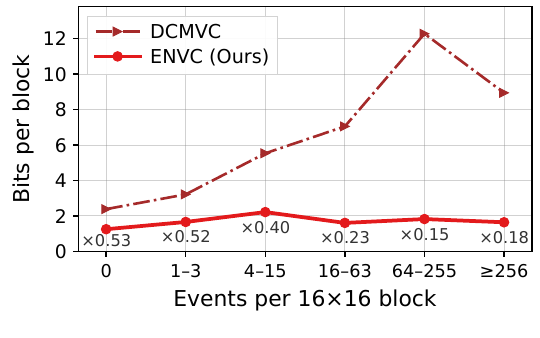}
            \vspace{-1.5em}
            \caption{Frame Latent Bits}
            \label{fig:bit_allocation_frame}
        \end{subfigure}
        \hfill
        \begin{subfigure}[t]{0.49\linewidth}
            \centering
            \includegraphics[width=\linewidth]
                {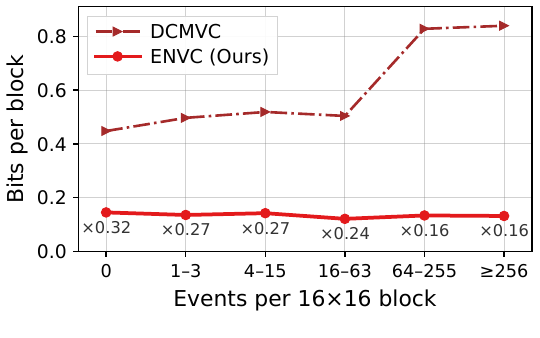}
            \vspace{-1.5em}
            \caption{Motion Bits}
            \label{fig:bit_allocation_motion}
        \end{subfigure}

        \caption{
            \textbf{Spatial Distribution of Coding Rates at $Q_1$.}
            Mean bits per $16\times16$ P-frame block on UVG,
            grouped by event count within the frame interval.
            (a) Frame latent $y$.
            (b) Motion latent and hyperlatent.
            Annotations show ENVC/DCMVC ratios.
            Motion hyperlatent bits are distributed uniformly
            across the corresponding blocks.
        }
        \label{fig:bit_allocation_spatial}
    \end{minipage}
\end{figure}

For Tab.~\ref{tab:bit_allocation}, we linearly interpolate each latent
stream's log-bpp against dataset-average PSNR-RGB using $Q_0$--$Q_3$
and average over 50 equally spaced points within the common range
of 35.445--37.394~dB.
Motion and frame coding rates each include their corresponding hyperlatents.
For the spatial analysis, we group $16\times16$ P-frame blocks by event
count at $Q_1$, where ENVC and DCMVC achieve mean PSNR-RGB values of
35.71 and 36.40~dB, respectively.
This block comparison therefore does not match reconstruction quality.

\vspace{-0.8em}
\paragraph{Component-wise Rate Reduction.}
Over the common quality range, ENVC reduces motion and conditional frame
coding rates by 67.9\% and 39.8\%, respectively
(Tab.~\ref{tab:bit_allocation}).
Frame coding accounts for 86.4\% of the P-frame bit savings,
showing that the larger absolute reduction occurs beyond the motion stream.
These results are consistent with event guidance reducing coding costs
through both motion estimation and temporal context construction.
Matching average quality selects higher quality settings for ENVC,
increasing the shared intra codec's rate and partially offsetting
the P-frame savings.

\vspace{-0.8em}
\paragraph{Spatial Distribution of Savings.}
In Fig.~\ref{fig:bit_allocation_spatial}, the frame-latent rate of ENVC
is 53\% of DCMVC's in blocks without events and 15--18\% in blocks
with at least 64 events.
Blocks with at least 16 events constitute 35.4\% of all evaluated blocks
but account for 69.6\% of the frame-latent bit savings.
ENVC's motion rate remains within 0.12--0.15 bits per block,
while DCMVC uses 0.83--0.84 bits in the two highest event-count groups.
The larger relative reductions in regions with more event activity
are consistent with ENVC exploiting the supplied temporal observations.

%% file: sec/h_additional_datasets_quality_metrics.tex
\section{Additional Datasets and Quality Metrics}
\label{app:results}

\begin{figure*}[t]
    \centering
    \begin{subfigure}[t]{0.24\textwidth}
        \centering
        \includegraphics[width=\linewidth]{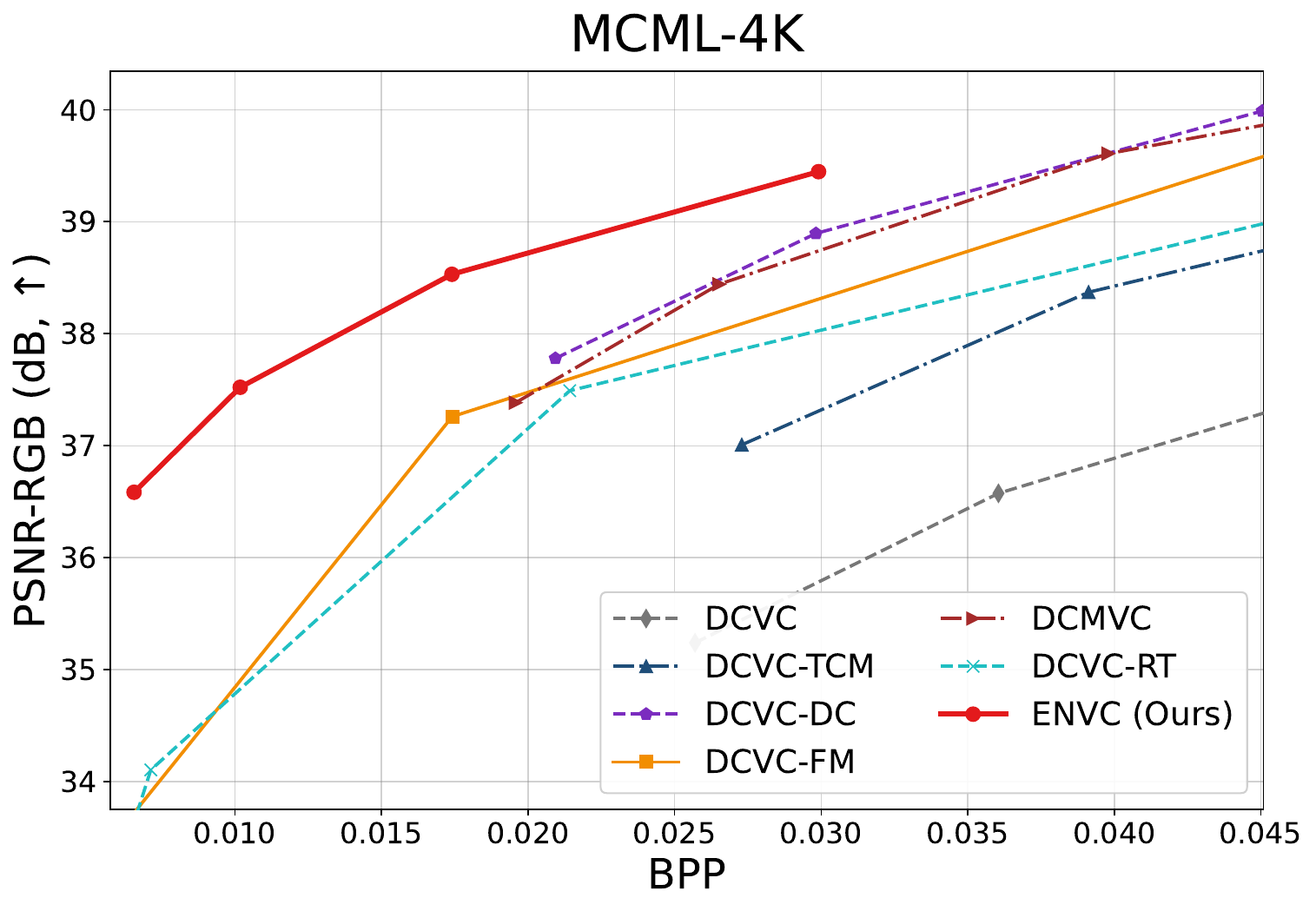}
        \vspace{-1.2em}
        \caption{PSNR-RGB}
        \label{fig:mcml4k_psnr}
    \end{subfigure}
    \begin{subfigure}[t]{0.24\textwidth}
        \centering
        \includegraphics[width=\linewidth]{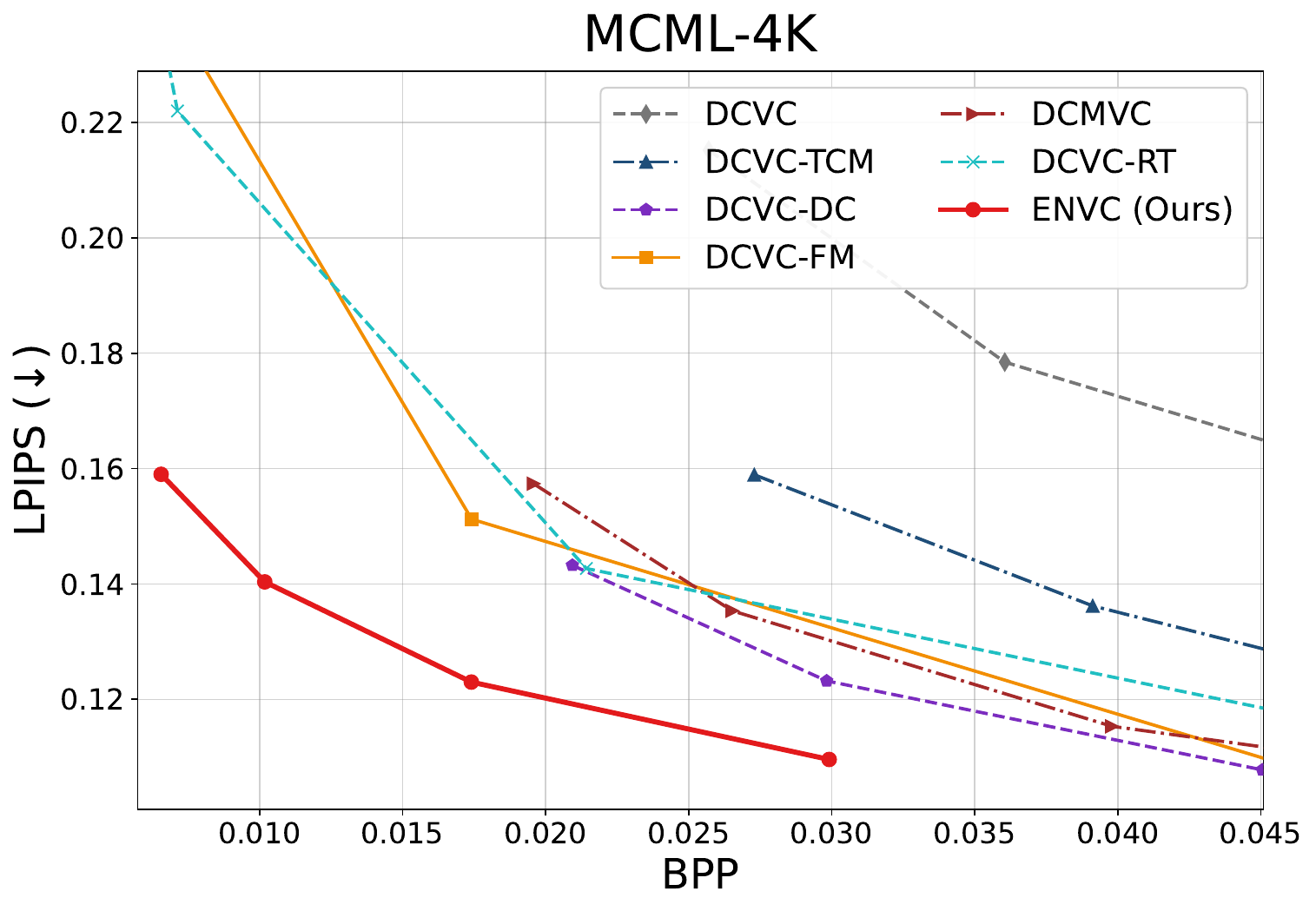}
        \vspace{-1.2em}
        \caption{LPIPS}
        \label{fig:mcml4k_lpips}
    \end{subfigure}
    \begin{subfigure}[t]{0.24\textwidth}
        \centering
        \includegraphics[width=\linewidth]{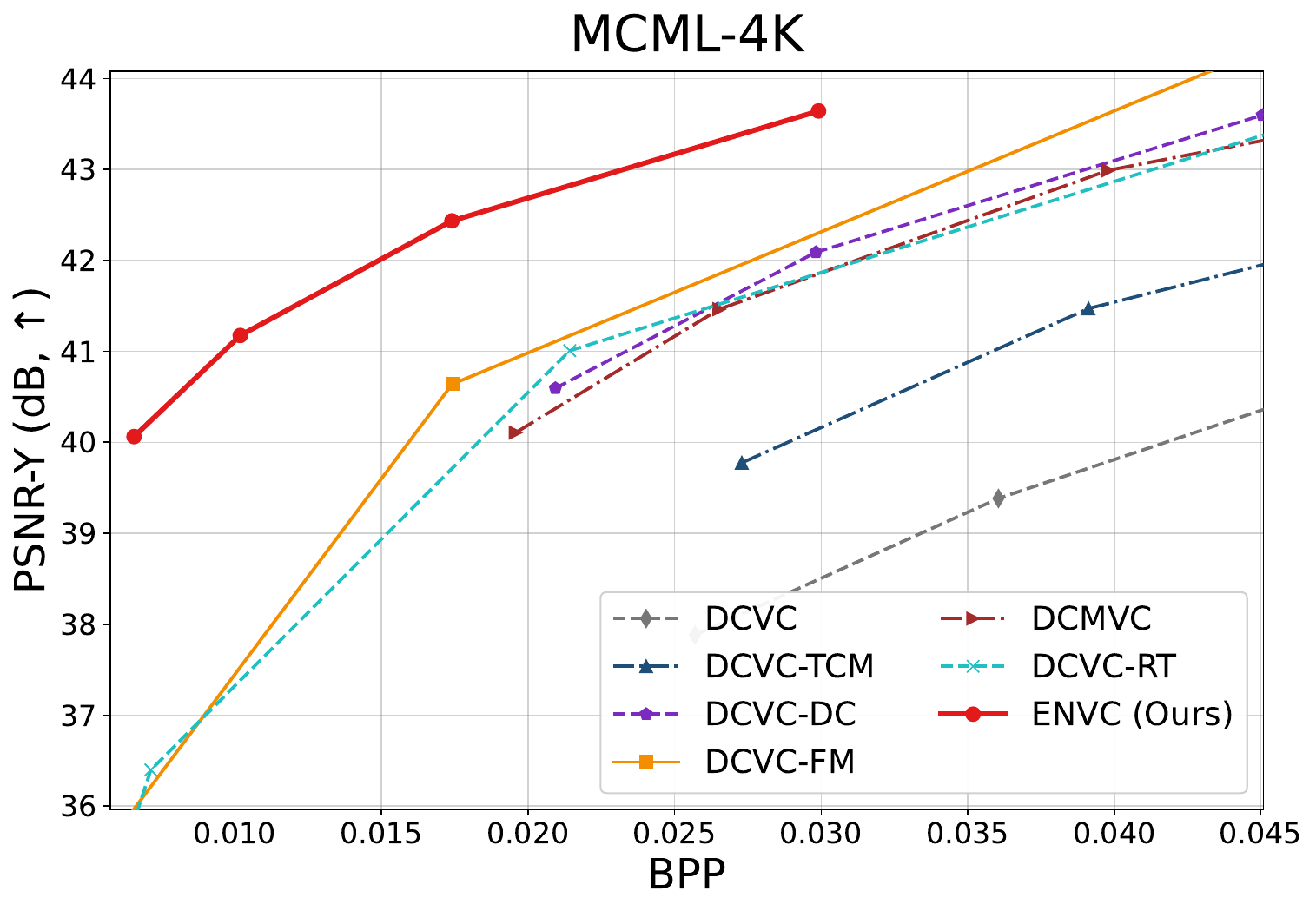}
        \vspace{-1.2em}
        \caption{PSNR-Y}
        \label{fig:mcml4k_psnr_y}
    \end{subfigure}
    \begin{subfigure}[t]{0.24\textwidth}
        \centering
        \includegraphics[width=\linewidth]{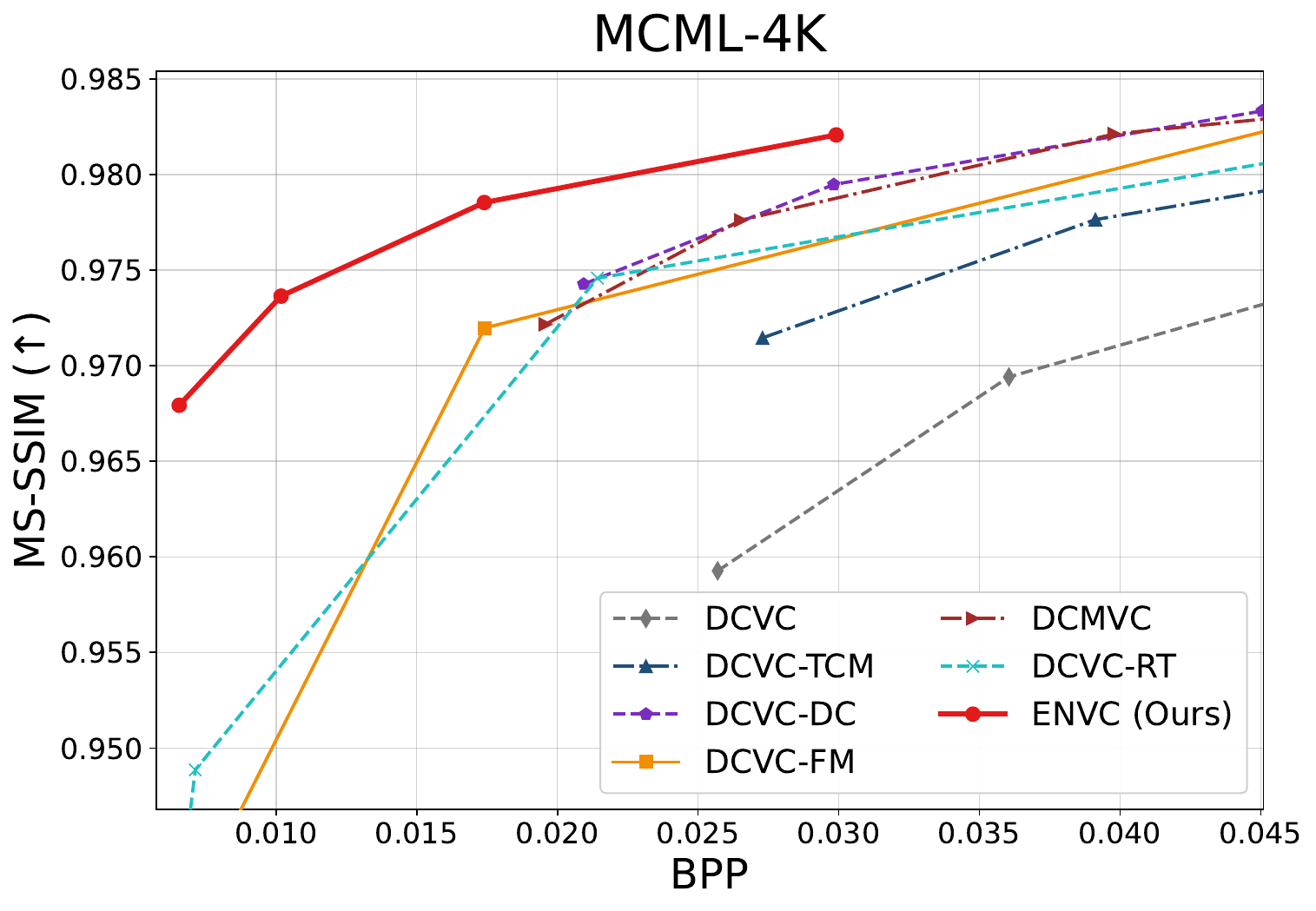}
        \vspace{-1.2em}
        \caption{MS-SSIM}
        \label{fig:mcml4k_ms_ssim}
    \end{subfigure}
    \vspace{-0.5em}
    \caption{
        \textbf{Rate--Distortion Performance on MCML-4K~\citep{cheon2018mcml4k}.}
        Results are reported using PSNR-RGB, LPIPS, PSNR-Y, and MS-SSIM.
    }
    \label{fig:mcml4k_rd}
    \vspace{-1.0em}
\end{figure*}

\paragraph{Evaluation on MCML-4K.}
To assess whether ENVC's compression benefits extend to higher
spatial resolutions, we additionally evaluate it on
MCML-4K~\citep{cheon2018mcml4k}, which contains 10 source
sequences at a resolution of $3840 \times 2160$ pixels.
As shown in Fig.~\ref{fig:mcml4k_rd}, ENVC achieves higher PSNR-RGB, PSNR-Y, and MS-SSIM, together with lower LPIPS, than the compared codecs over the overlapping bitrate range.
These results support the effectiveness of event-guided compression for 4K video in terms of both reconstruction fidelity and perceptual quality.

\vspace{-0.8em}
\paragraph{Additional Quality Metrics.}
We additionally evaluate the compared methods using PSNR-Y and MS-SSIM~\citep{wang2003msssim}.
Figs.~\ref{fig:rd_curves_psnr_y} and~\ref{fig:rd_curves_ms_ssim}
show the corresponding rate--distortion curves, while
Tab.~\ref{tab:bd_rate_additional_metrics} summarizes the BD-rate results relative to DCMVC.
ENVC achieves average BD-rate savings of 53.51\% in PSNR-Y and
40.00\% in MS-SSIM across the six benchmarks.
These results show that the gains of ENVC are not limited to the
PSNR-RGB and LPIPS metrics used in the main evaluation.

%% file: sec/i_additional_examples.tex
\section{Additional Qualitative Results}
\label{app:examples}

\begin{figure*}[t]
    \centering
    \captionsetup[subfigure]{font=scriptsize,skip=2pt}
    {\small\textbf{\texttt{close\_\_spinning\_umbrella}}}
    \par\vspace{0.3em}

    \begin{subfigure}[t]{0.198\textwidth}
        \centering
        \includegraphics[width=\linewidth]{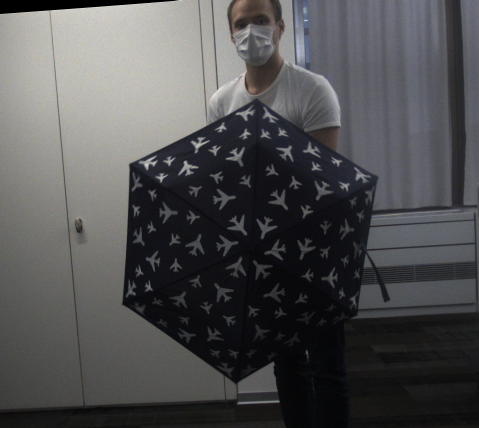}
        \vspace{-1.0em}
        \caption{Target}
    \end{subfigure}%
    \begin{subfigure}[t]{0.198\textwidth}
        \centering
        \includegraphics[width=\linewidth]{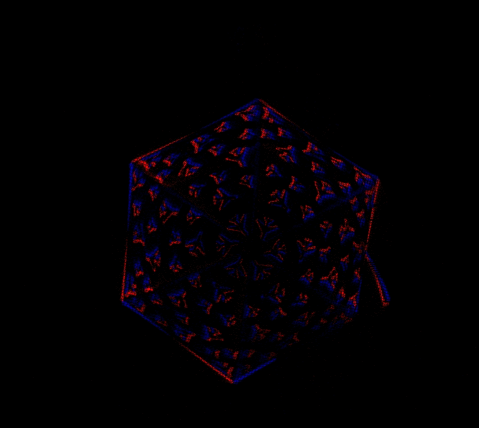}
        \vspace{-1.0em}
        \caption{Real Events}
    \end{subfigure}%
    \begin{subfigure}[t]{0.198\textwidth}
        \centering
        \includegraphics[width=\linewidth]{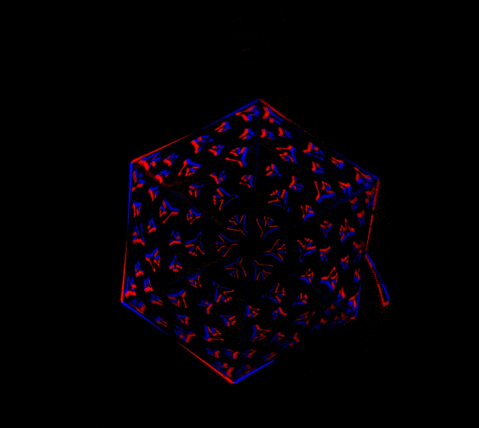}
        \vspace{-1.0em}
        \caption{Synthetic Events}
    \end{subfigure}%
    \begin{subfigure}[t]{0.198\textwidth}
        \centering
        \includegraphics[width=\linewidth]{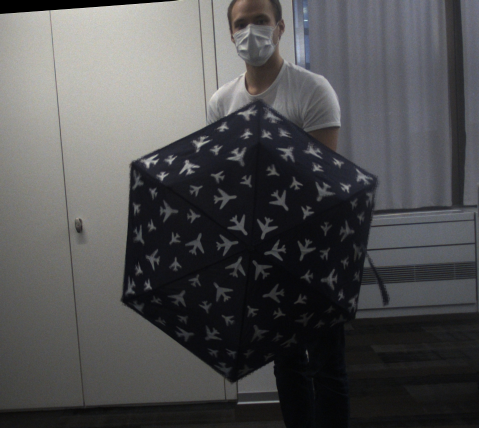}
        \vspace{-1.0em}
        \caption{Pred. (Real)}
    \end{subfigure}%
    \begin{subfigure}[t]{0.198\textwidth}
        \centering
        \includegraphics[width=\linewidth]{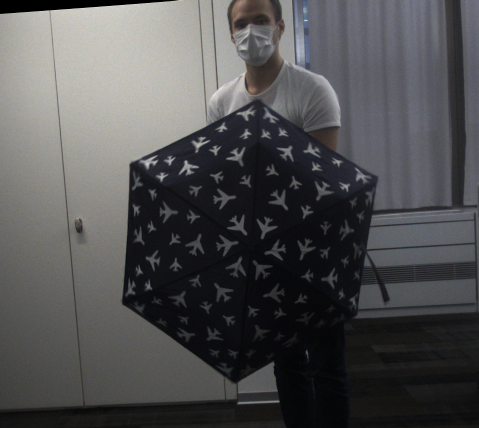}
        \vspace{-1.0em}
        \caption{Pred. (Synthetic)}
    \end{subfigure}

    \par\vspace{0.35em}

    \begin{subfigure}[t]{0.198\textwidth}
        \centering
        \includegraphics[width=\linewidth]{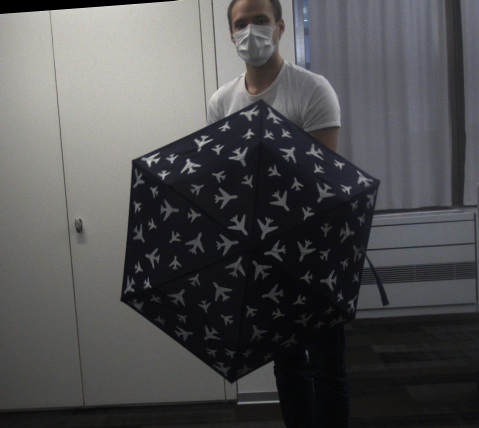}
        \vspace{-1.0em}
        \caption{Previous}
    \end{subfigure}%
    \begin{subfigure}[t]{0.198\textwidth}
        \centering
        \includegraphics[width=\linewidth]{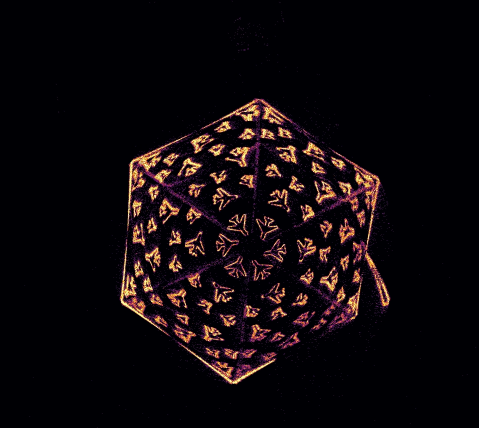}
        \vspace{-1.0em}
        \caption{Event Input Diff.}
    \end{subfigure}%
    \begin{subfigure}[t]{0.198\textwidth}
        \centering
        \includegraphics[width=\linewidth]{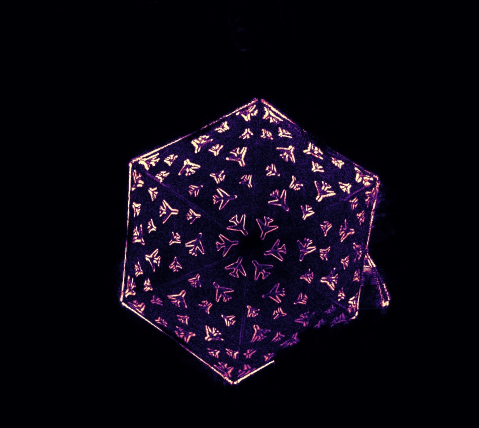}
        \vspace{-1.0em}
        \caption{Prediction Diff.}
    \end{subfigure}%
    \begin{subfigure}[t]{0.198\textwidth}
        \centering
        \includegraphics[width=\linewidth]{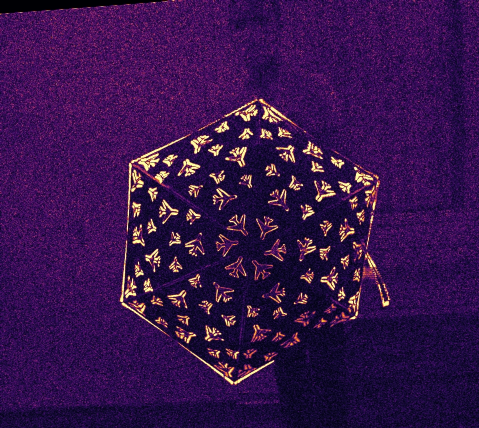}
        \vspace{-1.0em}
        \caption{\mbox{Pred. Error (Real)}}
    \end{subfigure}%
    \begin{subfigure}[t]{0.198\textwidth}
        \centering
        \includegraphics[width=\linewidth]{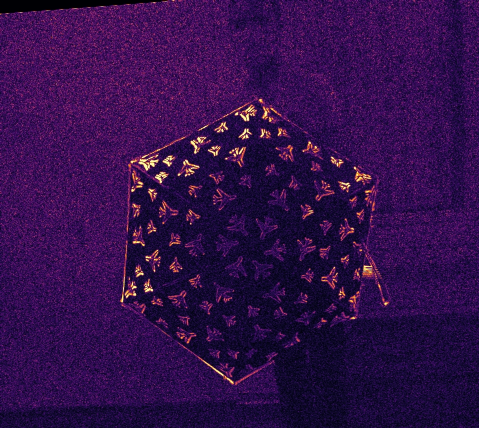}
        \vspace{-1.0em}
        \caption{\mbox{Pred. Error (Synthetic)}}
    \end{subfigure}

    \par\vspace{1.0em}

    {\small\textbf{\texttt{close\_\_fountain\_bellevue2}}}
    \par\vspace{0.3em}

    \setcounter{subfigure}{0}

    \begin{subfigure}[t]{0.198\textwidth}
        \centering
        \includegraphics[width=\linewidth]{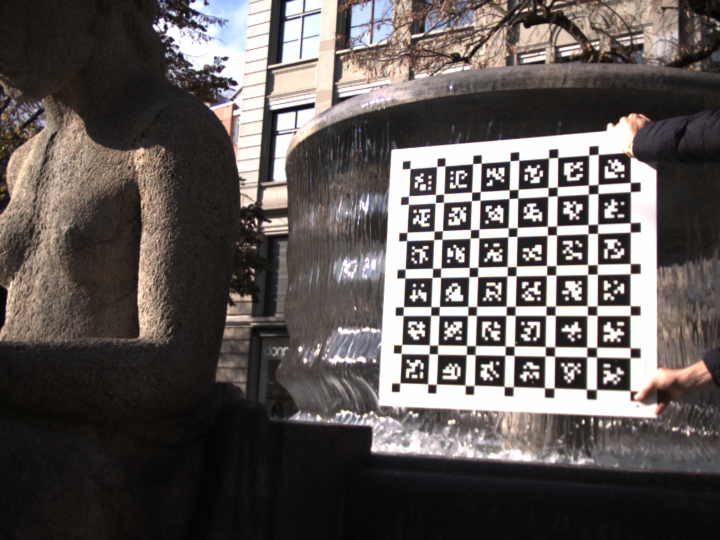}
        \vspace{-1.0em}
        \caption{Target}
    \end{subfigure}%
    \begin{subfigure}[t]{0.198\textwidth}
        \centering
        \includegraphics[width=\linewidth]{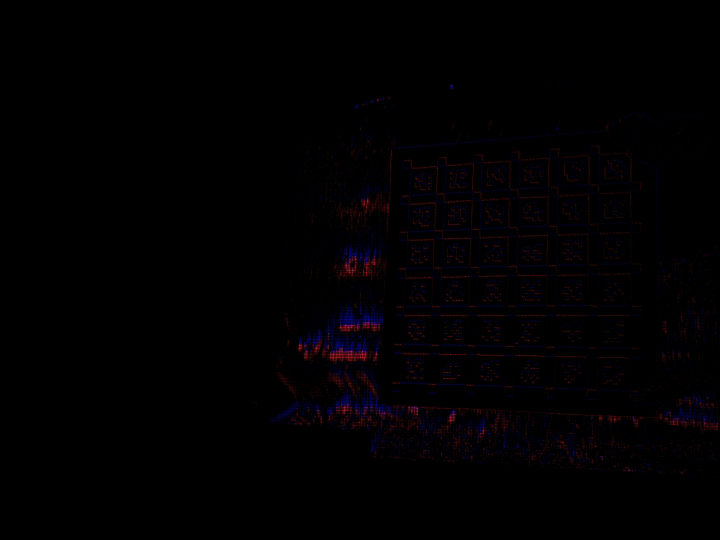}
        \vspace{-1.0em}
        \caption{Real Events}
    \end{subfigure}%
    \begin{subfigure}[t]{0.198\textwidth}
        \centering
        \includegraphics[width=\linewidth]{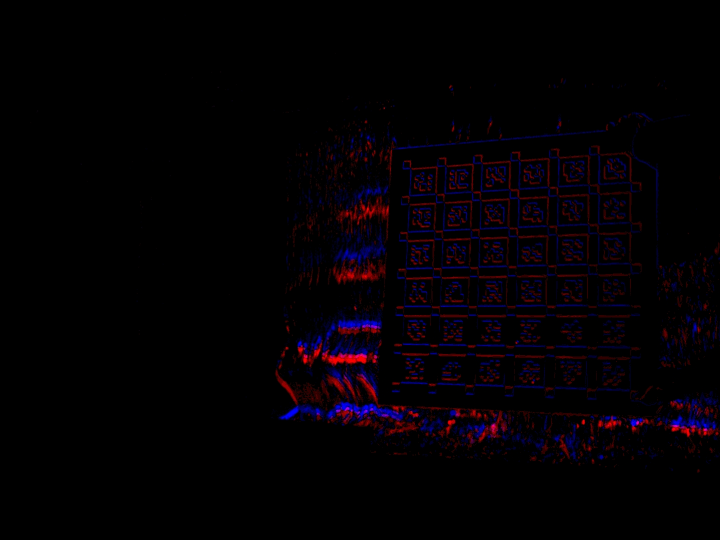}
        \vspace{-1.0em}
        \caption{Synthetic Events}
    \end{subfigure}%
    \begin{subfigure}[t]{0.198\textwidth}
        \centering
        \includegraphics[width=\linewidth]{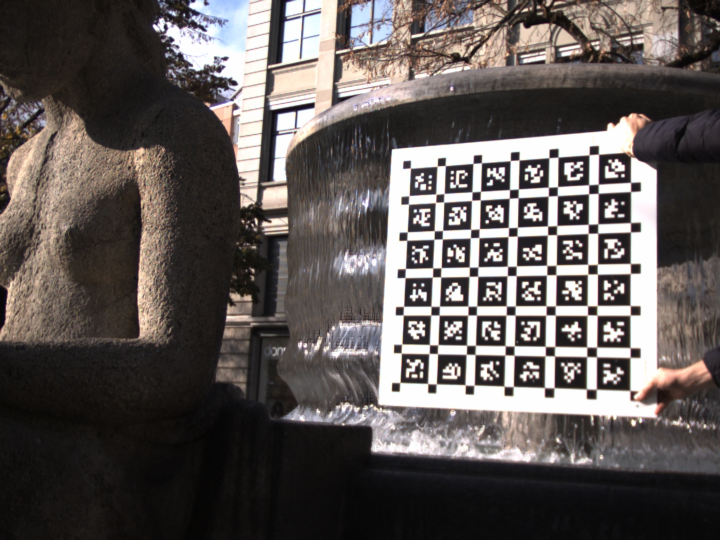}
        \vspace{-1.0em}
        \caption{Pred. (Real)}
    \end{subfigure}%
    \begin{subfigure}[t]{0.198\textwidth}
        \centering
        \includegraphics[width=\linewidth]{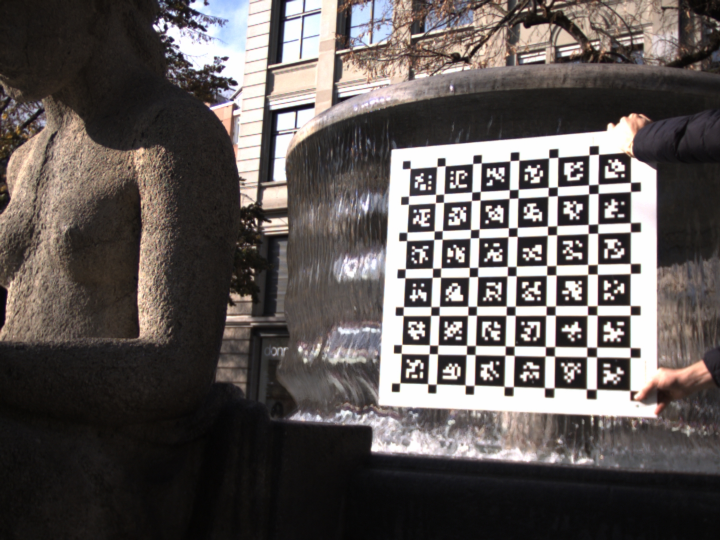}
        \vspace{-1.0em}
        \caption{Pred. (Synthetic)}
    \end{subfigure}

    \par\vspace{2.0em}

    \begin{subfigure}[t]{0.198\textwidth}
        \centering
        \includegraphics[width=\linewidth]{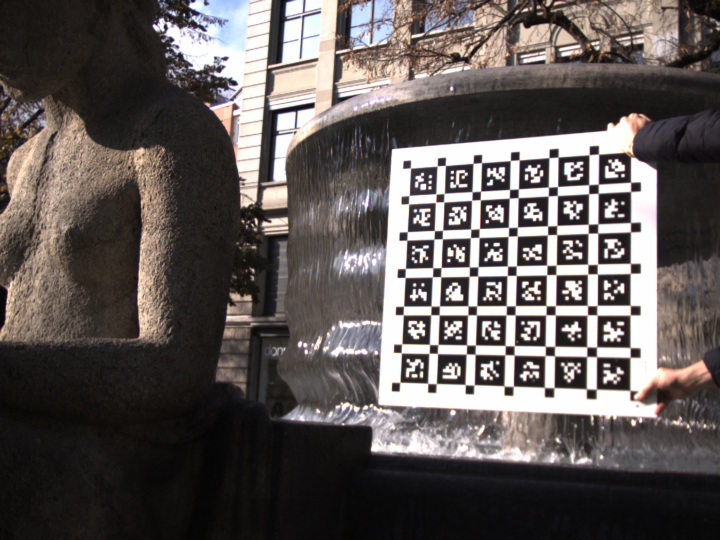}
        \vspace{-1.0em}
        \caption{Previous}
    \end{subfigure}%
    \begin{subfigure}[t]{0.198\textwidth}
        \centering
        \includegraphics[width=\linewidth]{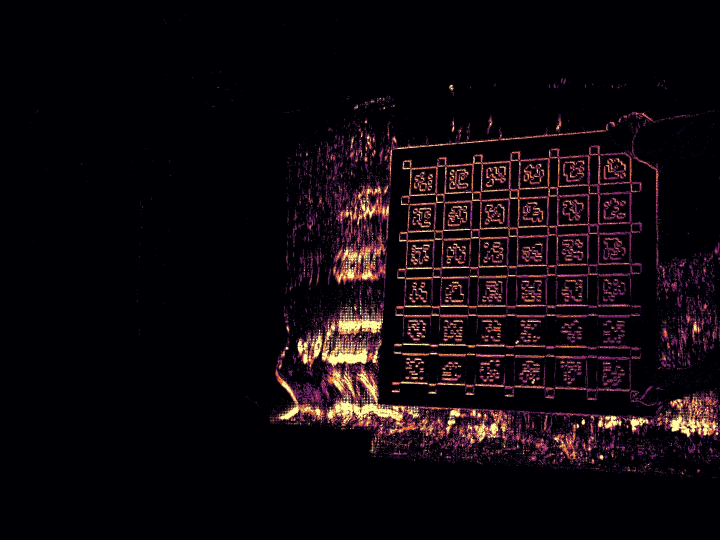}
        \vspace{-1.0em}
        \caption{Event Input Diff.}
    \end{subfigure}%
    \begin{subfigure}[t]{0.198\textwidth}
        \centering
        \includegraphics[width=\linewidth]{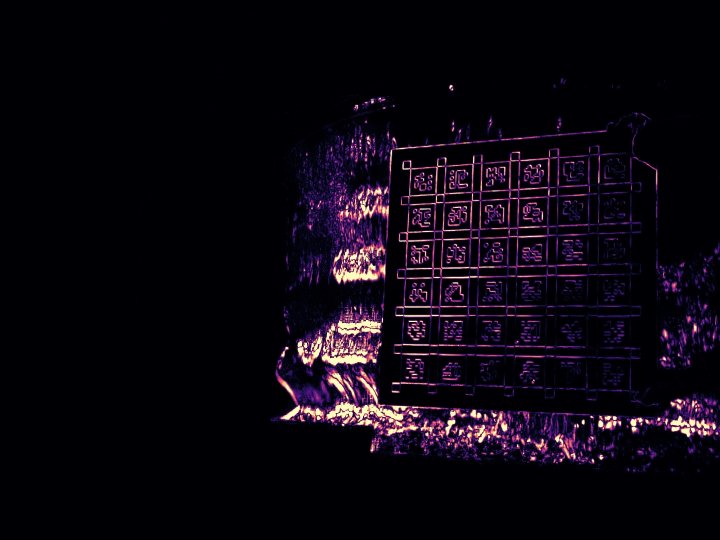}
        \vspace{-1.0em}
        \caption{Prediction Diff.}
    \end{subfigure}%
    \begin{subfigure}[t]{0.198\textwidth}
        \centering
        \includegraphics[width=\linewidth]{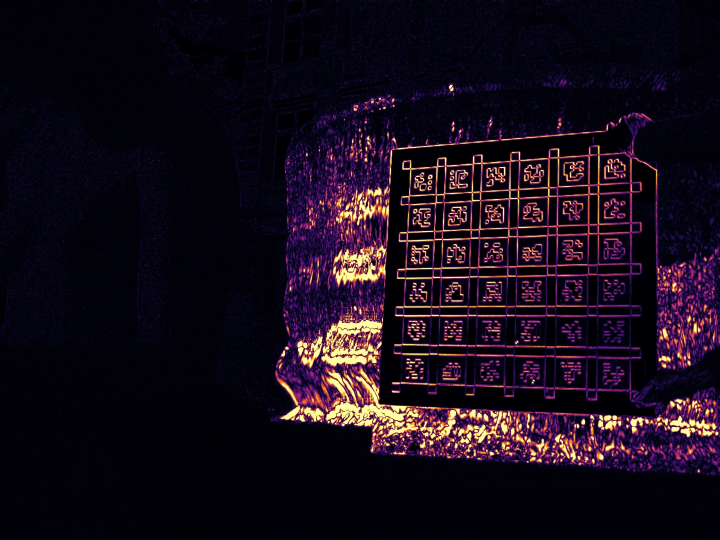}
        \vspace{-1.0em}
        \caption{\mbox{Pred. Error (Real)}}
    \end{subfigure}%
    \begin{subfigure}[t]{0.198\textwidth}
        \centering
        \includegraphics[width=\linewidth]{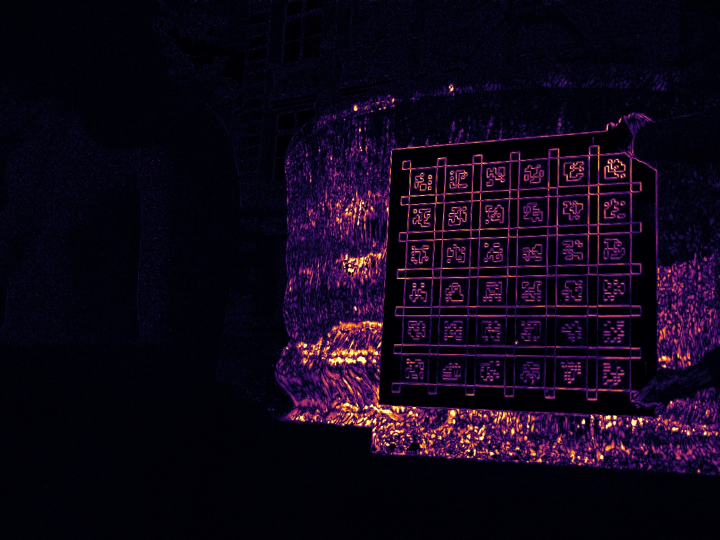}
        \vspace{-1.0em}
        \caption{\mbox{Pred. Error (Synthetic)}}
    \end{subfigure}
    \caption{
        \textbf{Additional Qualitative Comparisons on HS-ERGB~\citep{tulyakov2021time}.}
        For each sequence, we compare the target frame, real and synthetic
        events, predictions obtained from each event source, the previous
        frame, differences between the event inputs and predictions, and
        prediction errors with respect to the target.
    }
    \label{fig:qual_hs_ergb}
\end{figure*}

\begin{figure*}[t]
    \centering
    \captionsetup[subfigure]{font=scriptsize,skip=2pt}

    {\small\textbf{\texttt{horse\_09}}}
    \par\vspace{0.3em}

    \begin{subfigure}[t]{0.198\textwidth}
        \centering
        \includegraphics[width=\linewidth]{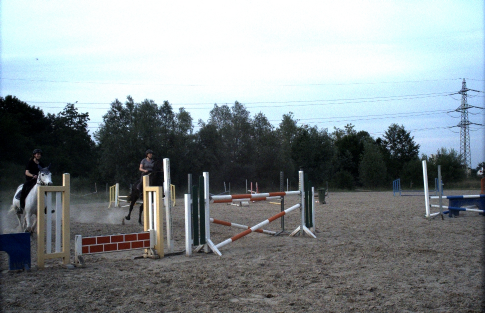}
        \vspace{-1.0em}
        \caption{Target}
    \end{subfigure}%
    \begin{subfigure}[t]{0.198\textwidth}
        \centering
        \includegraphics[width=\linewidth]{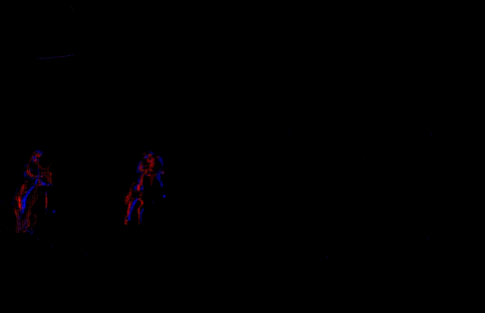}
        \vspace{-1.0em}
        \caption{Real Events}
    \end{subfigure}%
    \begin{subfigure}[t]{0.198\textwidth}
        \centering
        \includegraphics[width=\linewidth]{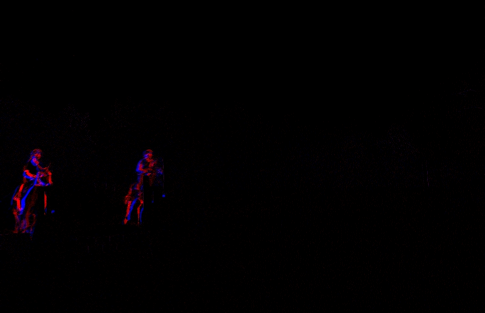}
        \vspace{-1.0em}
        \caption{Synthetic Events}
    \end{subfigure}%
    \begin{subfigure}[t]{0.198\textwidth}
        \centering
        \includegraphics[width=\linewidth]{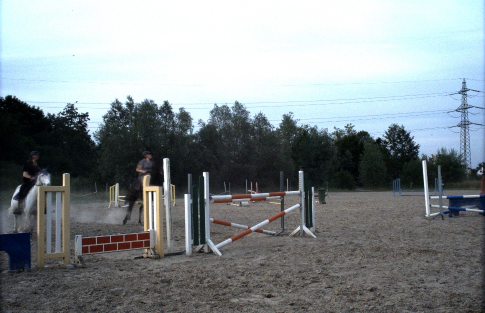}
        \vspace{-1.0em}
        \caption{Pred. (Real)}
    \end{subfigure}%
    \begin{subfigure}[t]{0.198\textwidth}
        \centering
        \includegraphics[width=\linewidth]{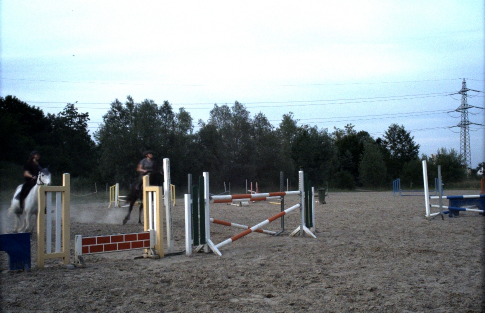}
        \vspace{-1.0em}
        \caption{Pred. (Synthetic)}
    \end{subfigure}

    \par\vspace{0.35em}

    \begin{subfigure}[t]{0.198\textwidth}
        \centering
        \includegraphics[width=\linewidth]{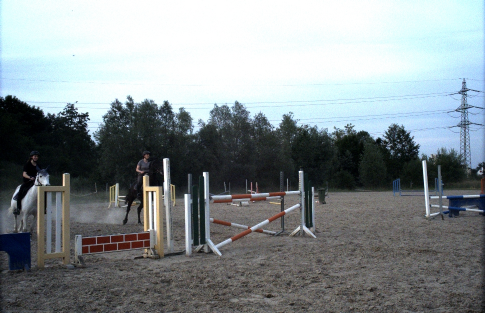}
        \vspace{-1.0em}
        \caption{Previous}
    \end{subfigure}%
    \begin{subfigure}[t]{0.198\textwidth}
        \centering
        \includegraphics[width=\linewidth]{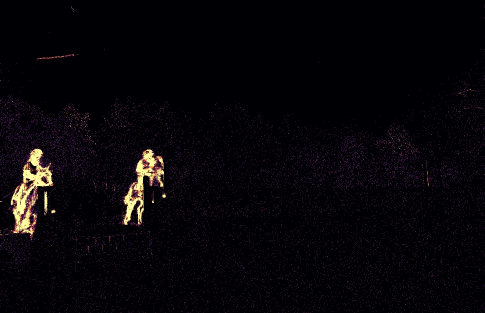}
        \vspace{-1.0em}
        \caption{Event Input Diff.}
    \end{subfigure}%
    \begin{subfigure}[t]{0.198\textwidth}
        \centering
        \includegraphics[width=\linewidth]{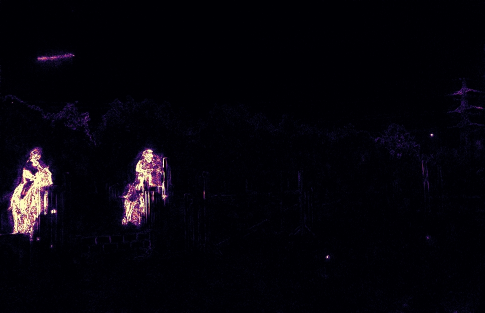}
        \vspace{-1.0em}
        \caption{Prediction Diff.}
    \end{subfigure}%
    \begin{subfigure}[t]{0.198\textwidth}
        \centering
        \includegraphics[width=\linewidth]{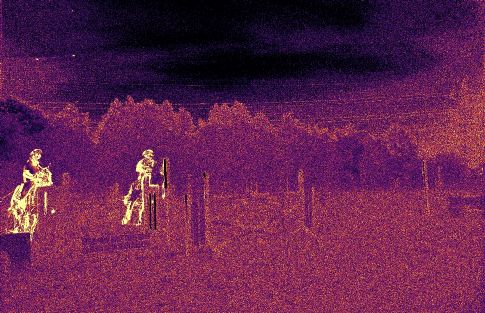}
        \vspace{-1.0em}
        \caption{\mbox{Pred. Error (Real)}}
    \end{subfigure}%
    \begin{subfigure}[t]{0.198\textwidth}
        \centering
        \includegraphics[width=\linewidth]{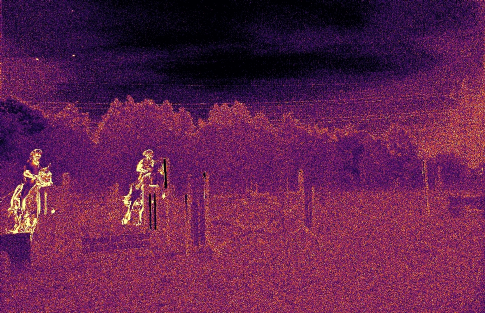}
        \vspace{-1.0em}
        \caption{\mbox{Pred. Error (Synthetic)}}
    \end{subfigure}

    \par\vspace{2.0em}

    {\small\textbf{\texttt{may29\_rooftop\_handheld\_01}}}
    \par\vspace{0.3em}

    \setcounter{subfigure}{0}

    \begin{subfigure}[t]{0.198\textwidth}
        \centering
        \includegraphics[width=\linewidth]{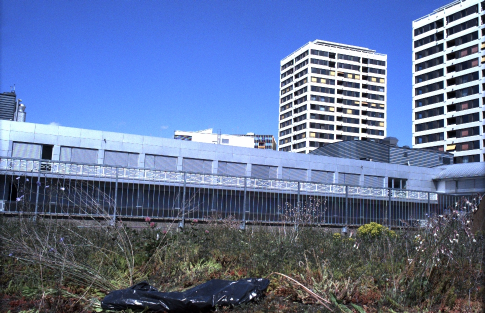}
        \vspace{-1.0em}
        \caption{Target}
    \end{subfigure}%
    \begin{subfigure}[t]{0.198\textwidth}
        \centering
        \includegraphics[width=\linewidth]{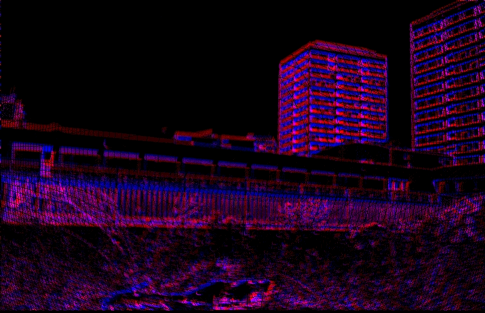}
        \vspace{-1.0em}
        \caption{Real Events}
    \end{subfigure}%
    \begin{subfigure}[t]{0.198\textwidth}
        \centering
        \includegraphics[width=\linewidth]{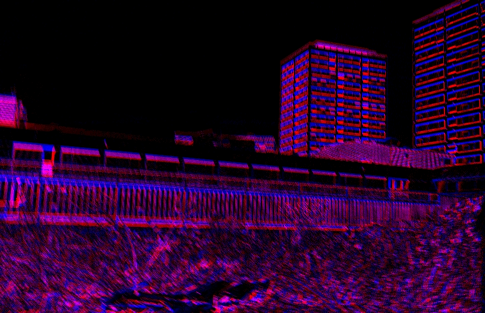}
        \vspace{-1.0em}
        \caption{Synthetic Events}
    \end{subfigure}
    \begin{subfigure}[t]{0.198\textwidth}
        \centering
        \includegraphics[width=\linewidth]{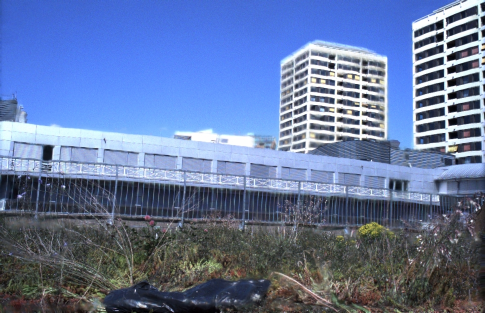}
        \vspace{-1.0em}
        \caption{Pred. (Real)}
    \end{subfigure}%
    \begin{subfigure}[t]{0.198\textwidth}
        \centering
        \includegraphics[width=\linewidth]{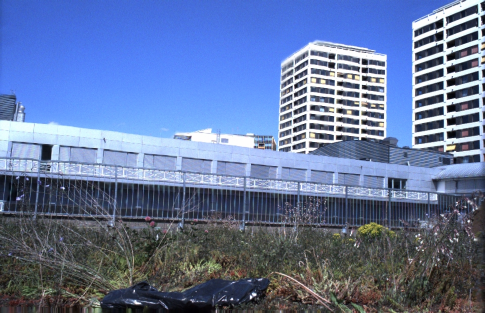}
        \vspace{-1.0em}
        \caption{Pred. (Synthetic)}
    \end{subfigure}
    \par\vspace{0.35em}
    \begin{subfigure}[t]{0.198\textwidth}
        \centering
        \includegraphics[width=\linewidth]{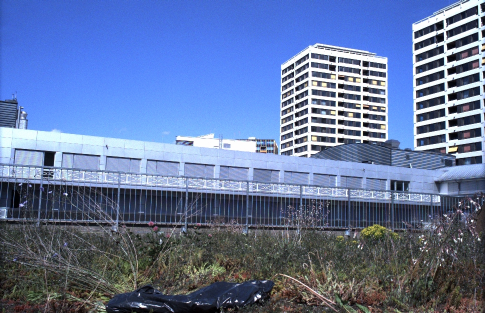}
        \vspace{-1.0em}
        \caption{Previous}
    \end{subfigure}
        \vspace{-1.0em}
    \begin{subfigure}[t]{0.198\textwidth}
        \centering
        \includegraphics[width=\linewidth]{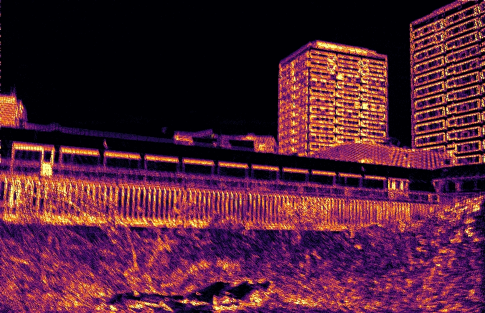}
        \vspace{-1.0em}
        \caption{Event Input Diff.}
    \end{subfigure}%
    \begin{subfigure}[t]{0.198\textwidth}
        \centering
        \includegraphics[width=\linewidth]{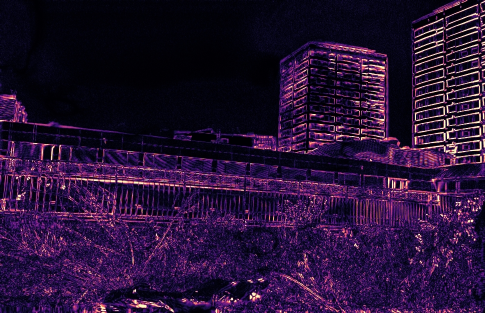}
        \vspace{-1.0em}
        \caption{Prediction Diff.}
    \end{subfigure}%
    \begin{subfigure}[t]{0.198\textwidth}
        \centering
        \includegraphics[width=\linewidth]{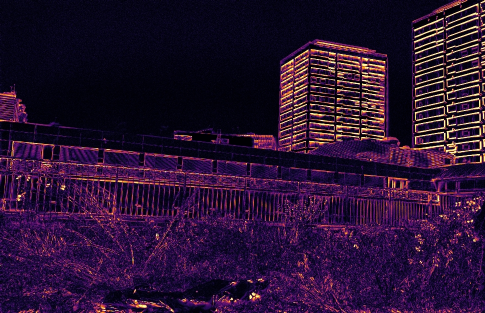}
        \vspace{-1.0em}
        \caption{\mbox{Pred. Error (Real)}}
    \end{subfigure}%
    \begin{subfigure}[t]{0.198\textwidth}
        \centering
        \includegraphics[width=\linewidth]{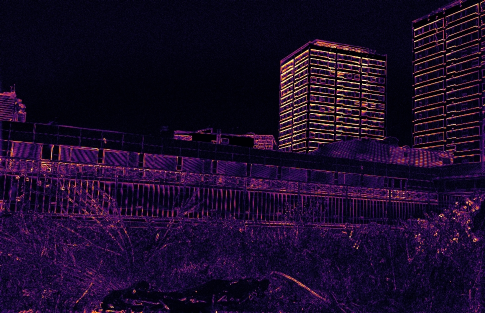}
        \vspace{-1.0em}
        \caption{\mbox{Pred. Error (Synthetic)}}
    \end{subfigure}
    \vspace{0.7em}
    \caption{
        \textbf{Additional Qualitative Comparisons on BS-ERGB~\citep{tulyakov2022time}.}
        For each sequence, we compare the target frame, real and synthetic
        events, predictions obtained from each event source, the previous
        frame, differences between the event inputs and predictions, and
        prediction errors with respect to the target.
    }
    \label{fig:qual_bs_ergb}
\end{figure*}

\paragraph{Real and Synthetic Event Comparison.}
Figs.~\ref{fig:qual_hs_ergb} and~\ref{fig:qual_bs_ergb} provide  qualitative comparisons on HS-ERGB~\citep{tulyakov2021time} and BS-ERGB~\citep{tulyakov2022time}, respectively.
For each sequence, we visualize the real and synthetic events, the corresponding VFPSIE~\citep{zhu2024vfpsie} predictions obtained from the same reference frame, and their difference and prediction-error maps.
These examples complement the qualitative comparison in Fig.~\ref{fig:event_quality} by showing the spatial discrepancies between real and synthetic events and how they propagate to frame prediction.

\begin{figure*}[t]
    \centering

    \newcommand{\centerimg}[1]{%
        \begin{minipage}[c]{0.215\textwidth}
            \centering
            \includegraphics[width=\linewidth]{#1}
        \end{minipage}%
    }

    \centerimg{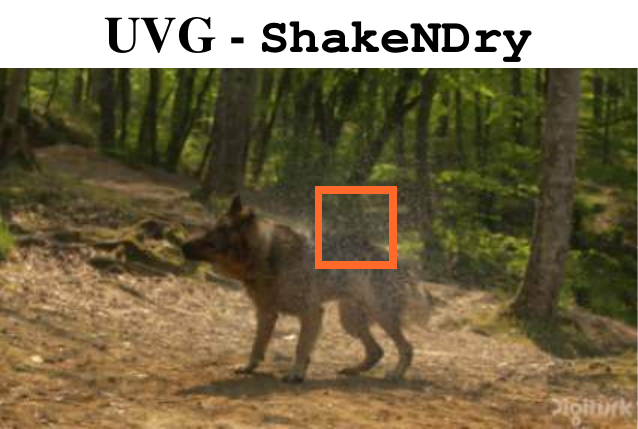}%
    \hspace{0.5em}
    \centerimg{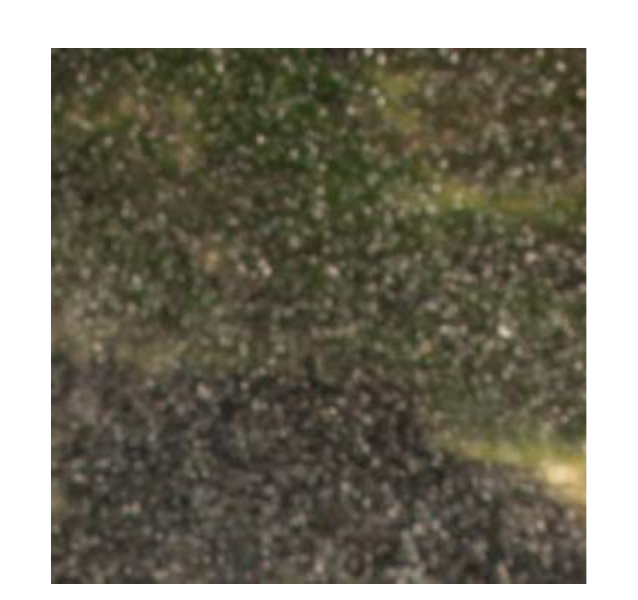}%
    \hspace{0.5em}
    \centerimg{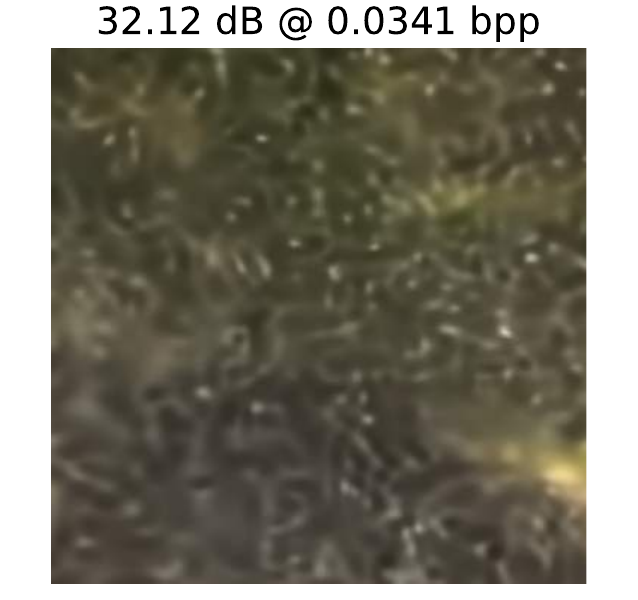}%
    \hspace{0.5em}
    \centerimg{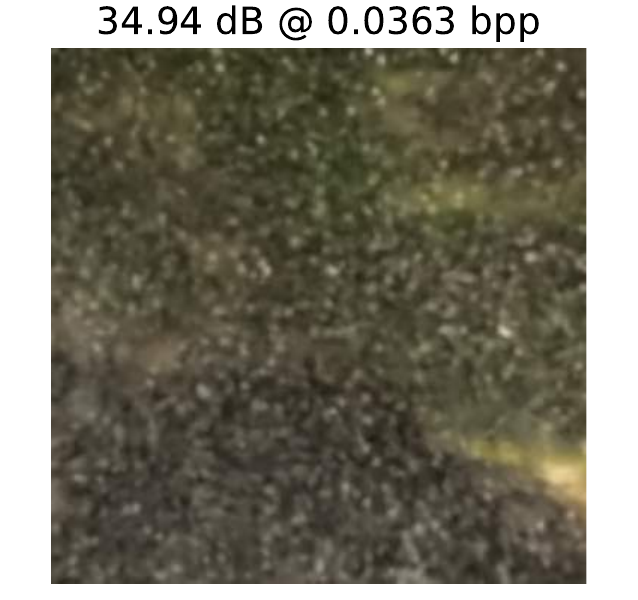}

    \par\vspace{0.04em}

    \centerimg{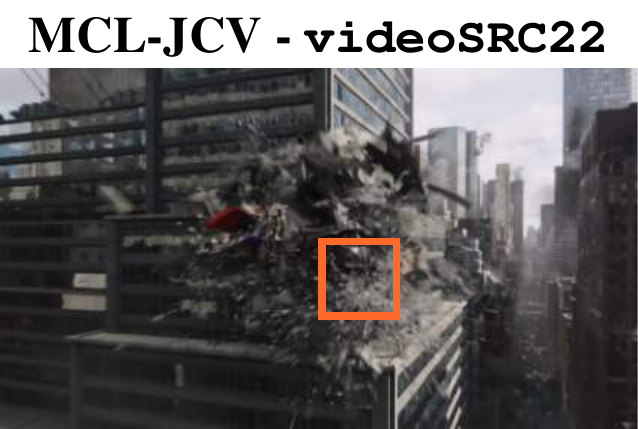}%
    \hspace{0.5em}
    \centerimg{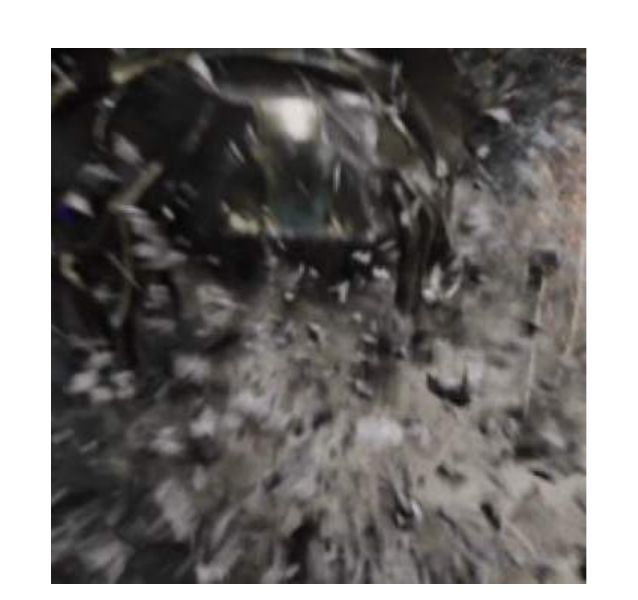}%
    \hspace{0.5em}
    \centerimg{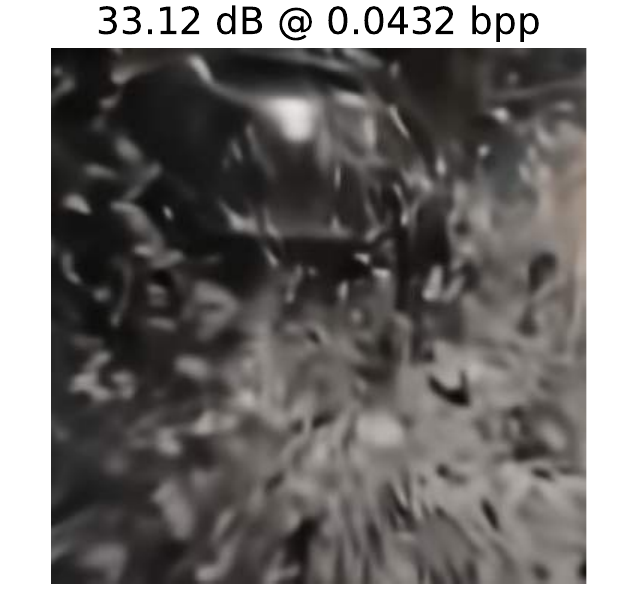}%
    \hspace{0.5em}
    \centerimg{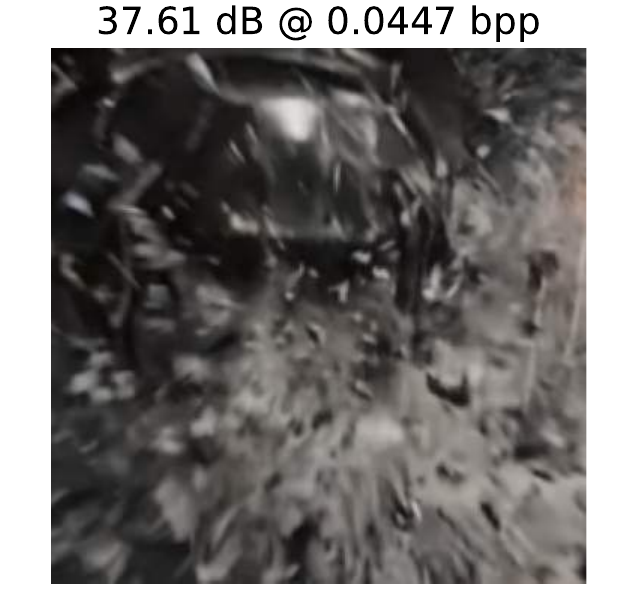}

    \par\vspace{0.04em}

    \centerimg{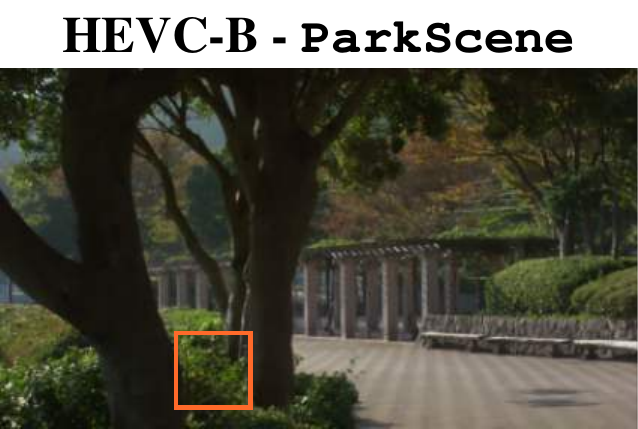}%
    \hspace{0.5em}
    \centerimg{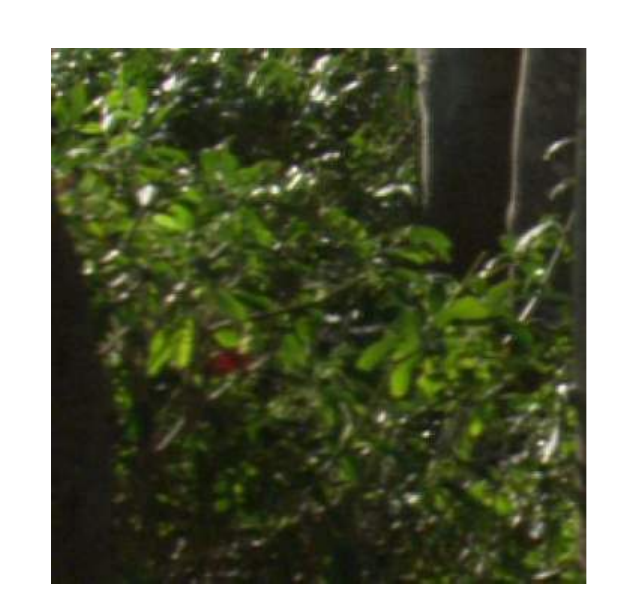}%
    \hspace{0.5em}
    \centerimg{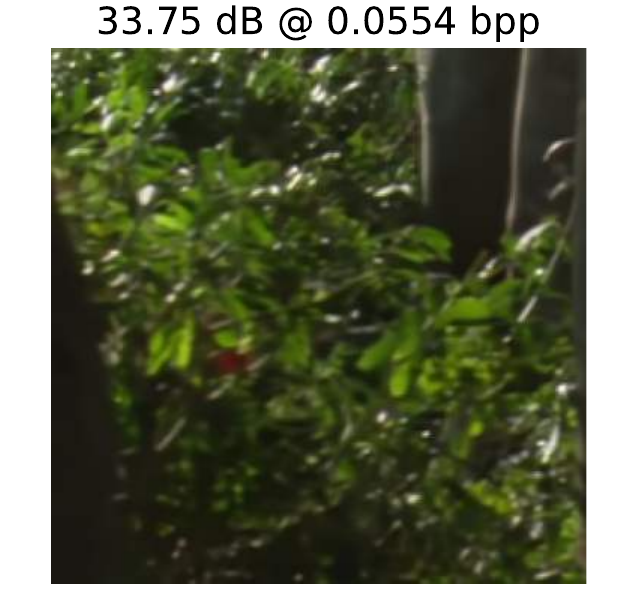}%
    \hspace{0.5em}
    \centerimg{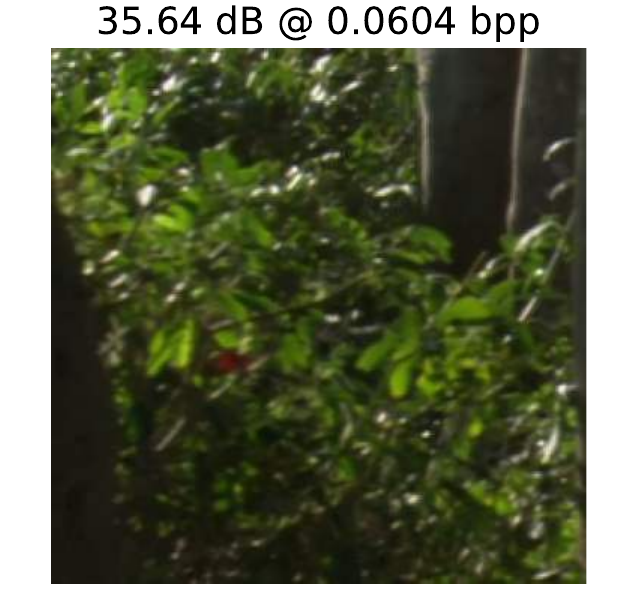}

    \par\vspace{0.04em}

    \centerimg{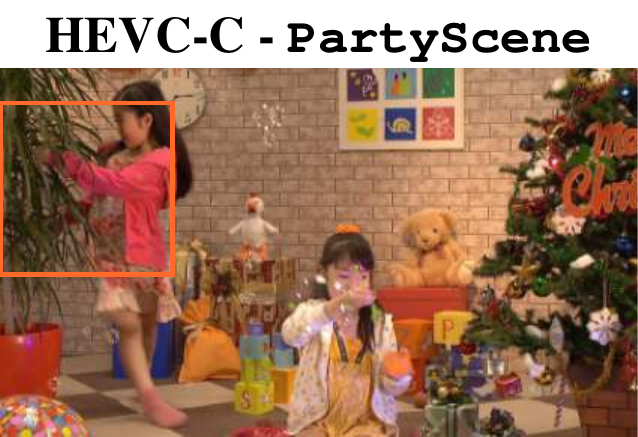}%
    \hspace{0.5em}
    \centerimg{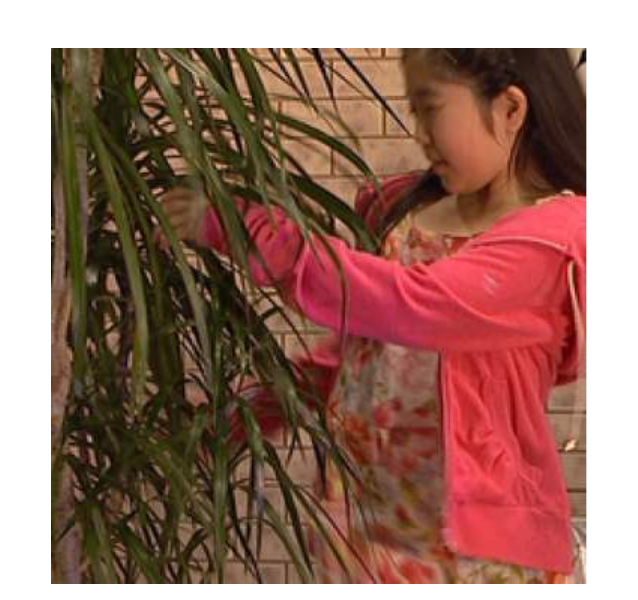}%
    \hspace{0.5em}
    \centerimg{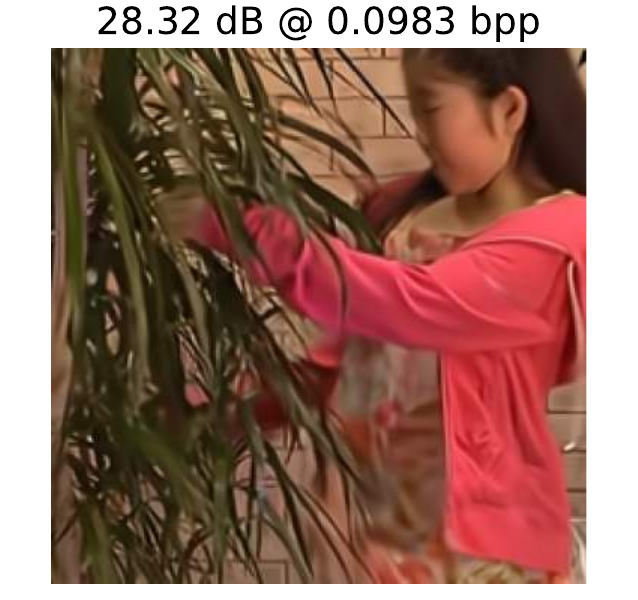}%
    \hspace{0.5em}
    \centerimg{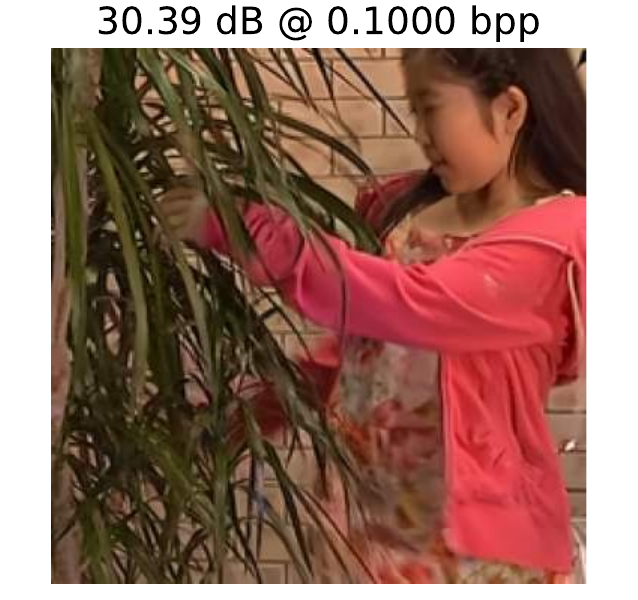}

    \par\vspace{0.04em}

    \centerimg{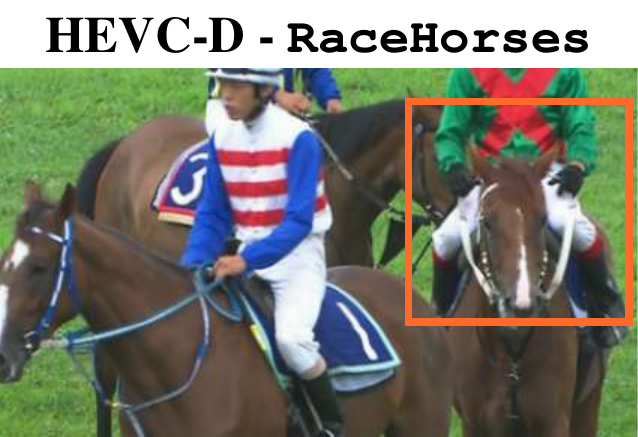}%
    \hspace{0.5em}
    \centerimg{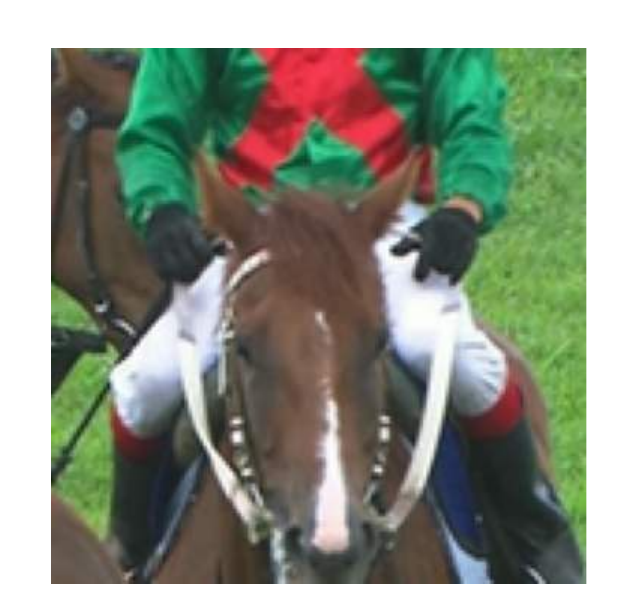}%
    \hspace{0.5em}
    \centerimg{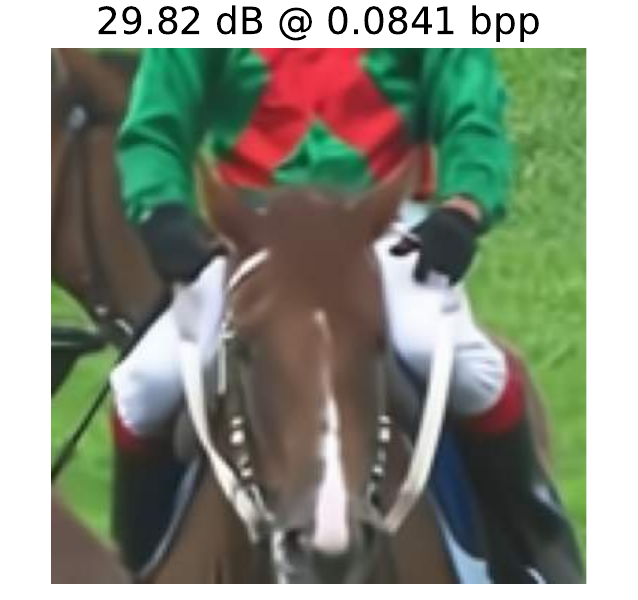}%
    \hspace{0.5em}
    \centerimg{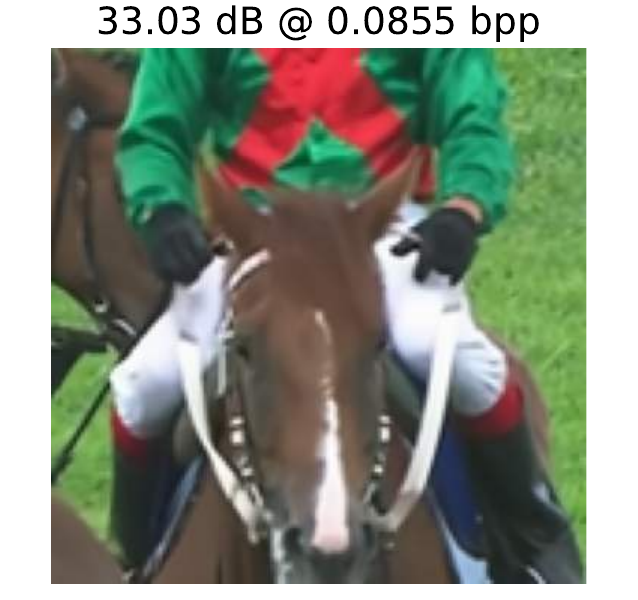}

    \par\vspace{0.04em}

    \centerimg{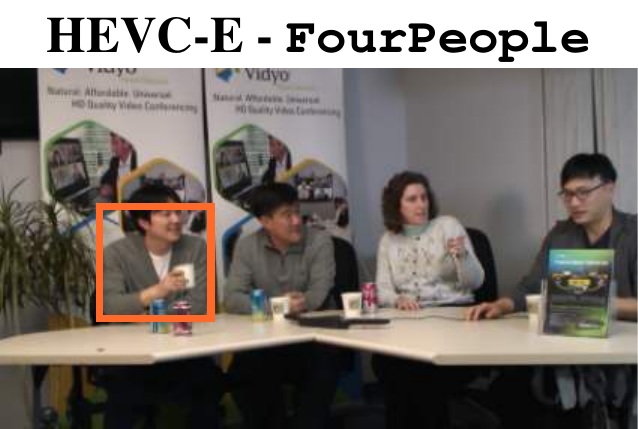}%
    \hspace{0.5em}
    \centerimg{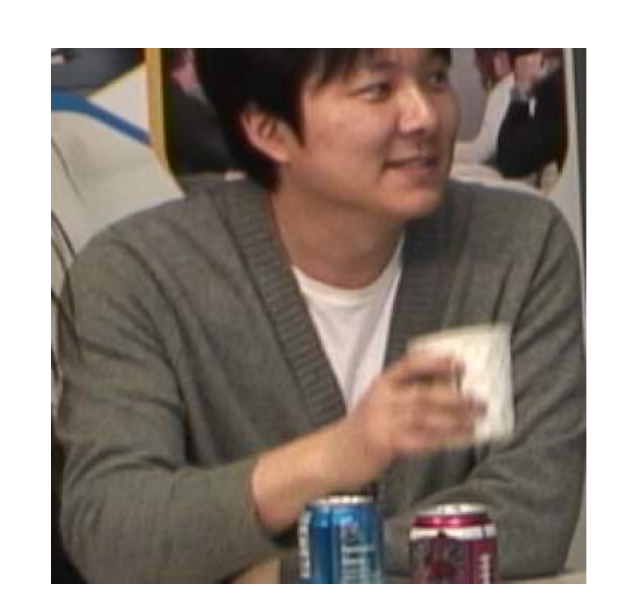}%
    \hspace{0.5em}
    \centerimg{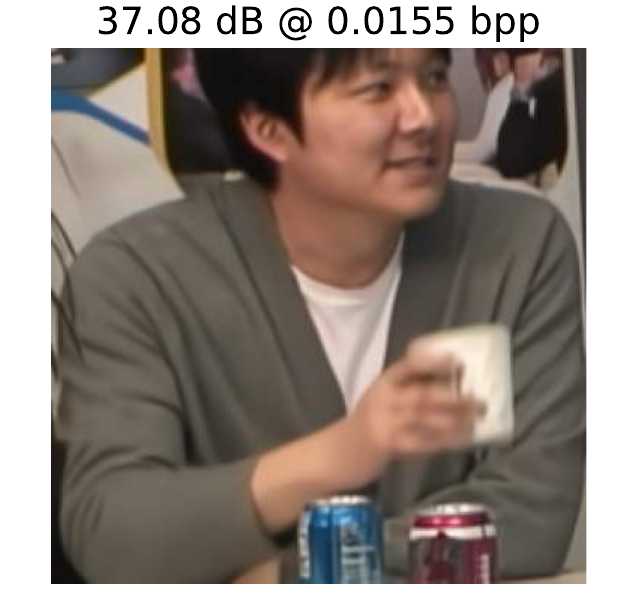}%
    \hspace{0.5em}
    \centerimg{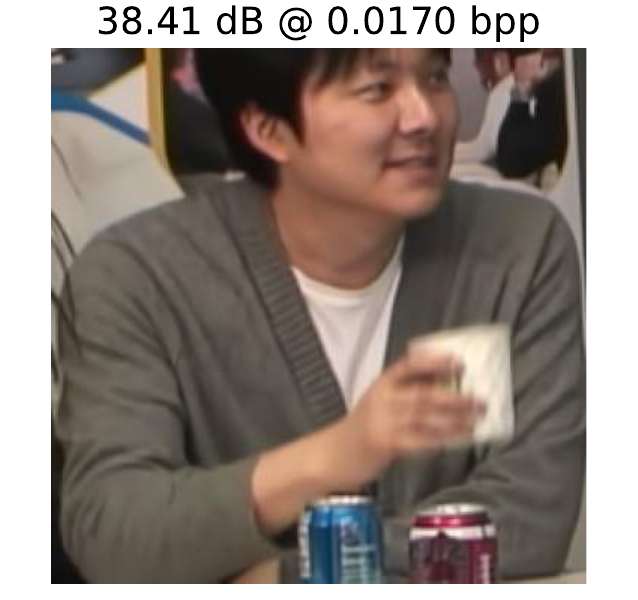}

    \par\vspace{0.04em}

    \centerimg{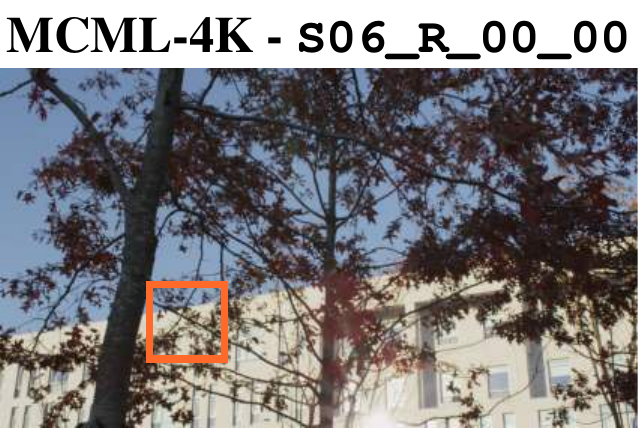}%
    \hspace{0.5em}
    \centerimg{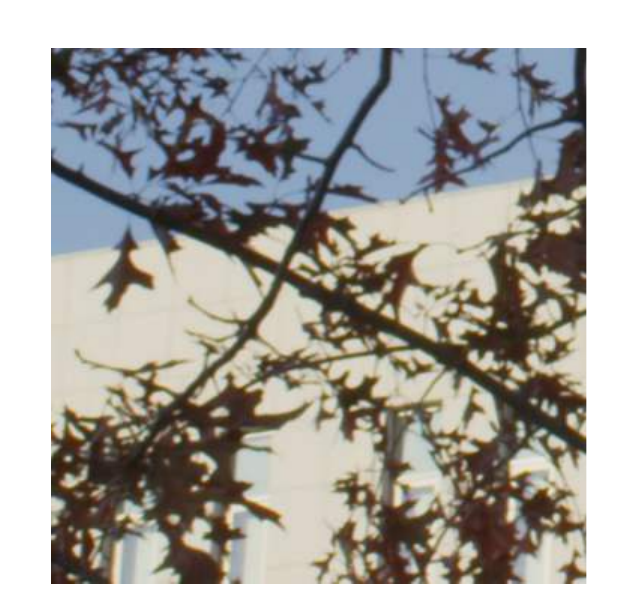}%
    \hspace{0.5em}
    \centerimg{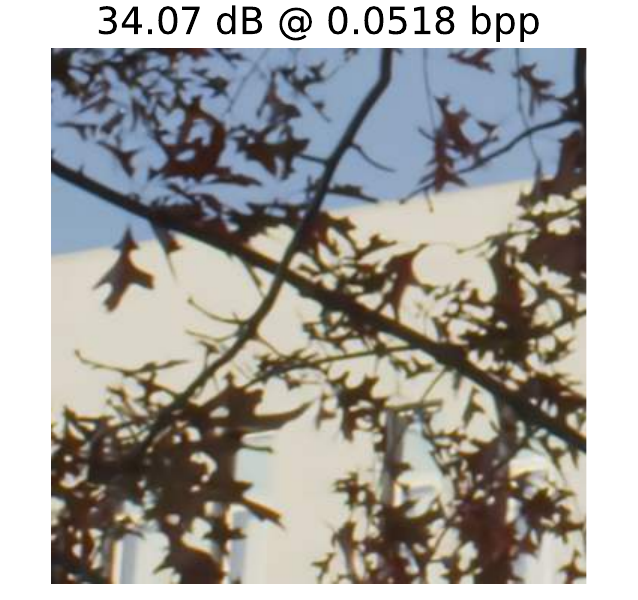}%
    \hspace{0.5em}
    \centerimg{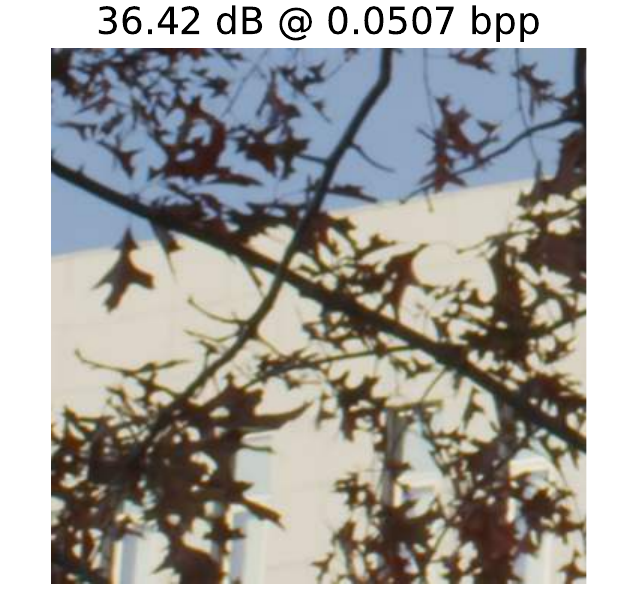}

    \par\vspace{0.2em}

    \noindent
    \makebox[0.215\textwidth][c]{\small (a) Crop location}%
    \hspace{0.5em}
    \makebox[0.215\textwidth][c]{\small (b) Original}%
    \hspace{0.5em}
    \makebox[0.215\textwidth][c]{\small (c) DCMVC}%
    \hspace{0.5em}
    \makebox[0.215\textwidth][c]{\small (d) ENVC (Ours)}

    \vspace{-0.2em}

    \caption{
        \textbf{Closest-rate Qualitative Comparisons.}
        From top to bottom, examples are shown for UVG, MCL-JCV,
        HEVC-B, HEVC-C, HEVC-D, HEVC-E, and MCML-4K, respectively.
        The orange boxes in (a) indicate the regions displayed in
        (b)--(d).
        Note that reported PSNR and bpp are both measured over the full frame.
    }
    \label{fig:crop_comparison}
\end{figure*}

\vspace{-0.8em}
\paragraph{Qualitative Reconstruction Comparison.}
Fig.~\ref{fig:crop_comparison} compares ENVC and DCMVC at similar RGB coding rates.
In these selected examples, ENVC better preserves local textures
and object boundaries, complementing the quantitative RD results.

\begin{figure*}[t]
    \centering
    \newcommand{\centereventimg}[1]{%
        \begin{minipage}[c]{0.215\textwidth}
            \centering
            \includegraphics[width=\linewidth]{#1}
        \end{minipage}%
    }
    {\scriptsize\textbf{UVG - \texttt{ShakeNDry}}\par}
    \vspace{0.3em}
    
    \centereventimg{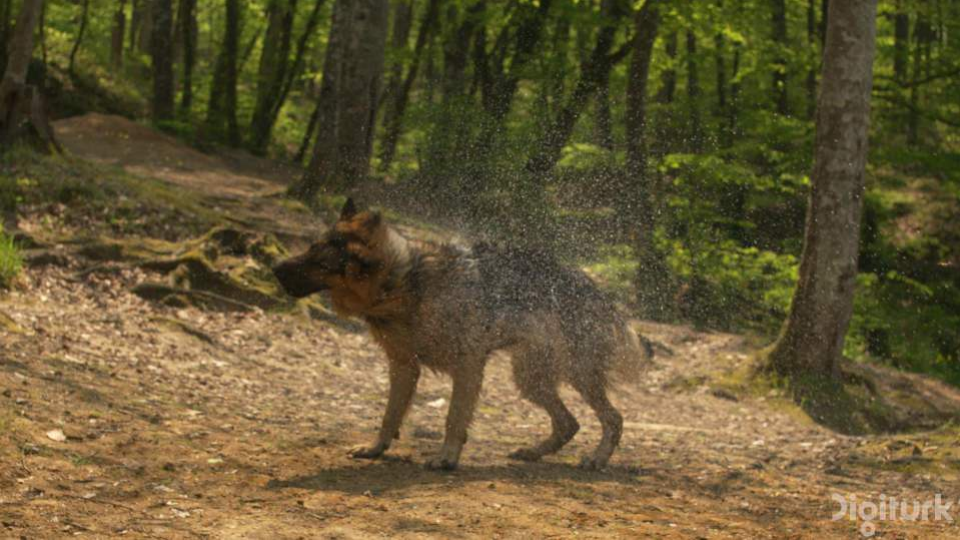}%
    \hspace{0.5em}
    \centereventimg{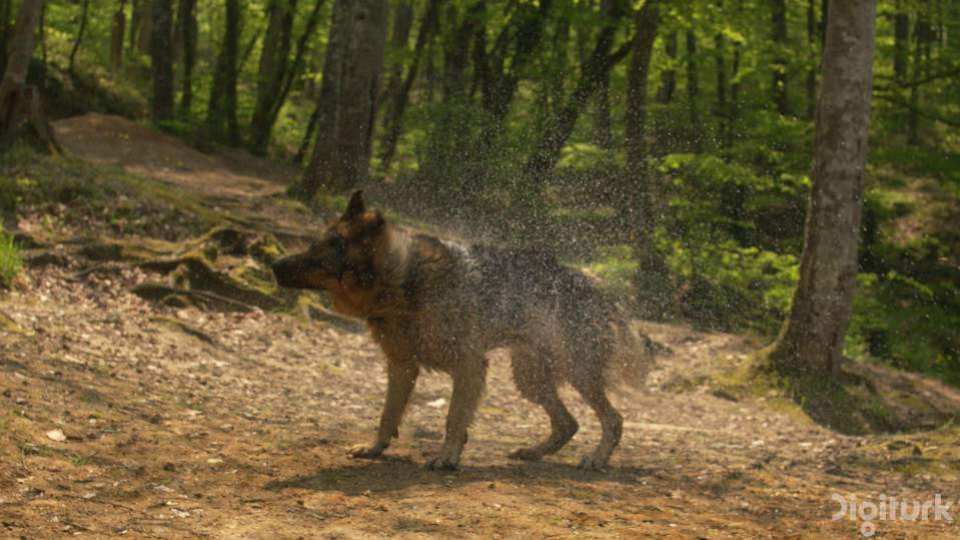}%
    \hspace{0.5em}
    \centereventimg{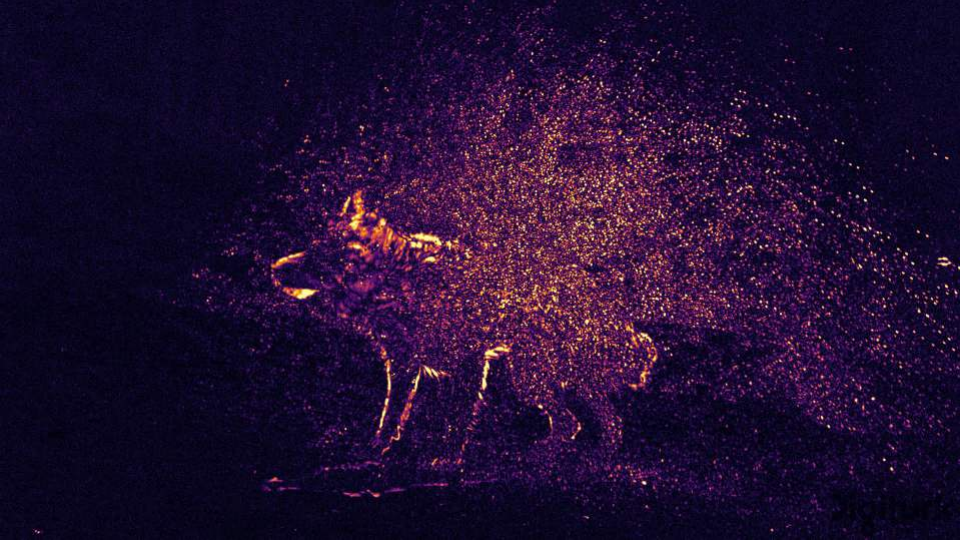}%
    \hspace{0.5em}
    \centereventimg{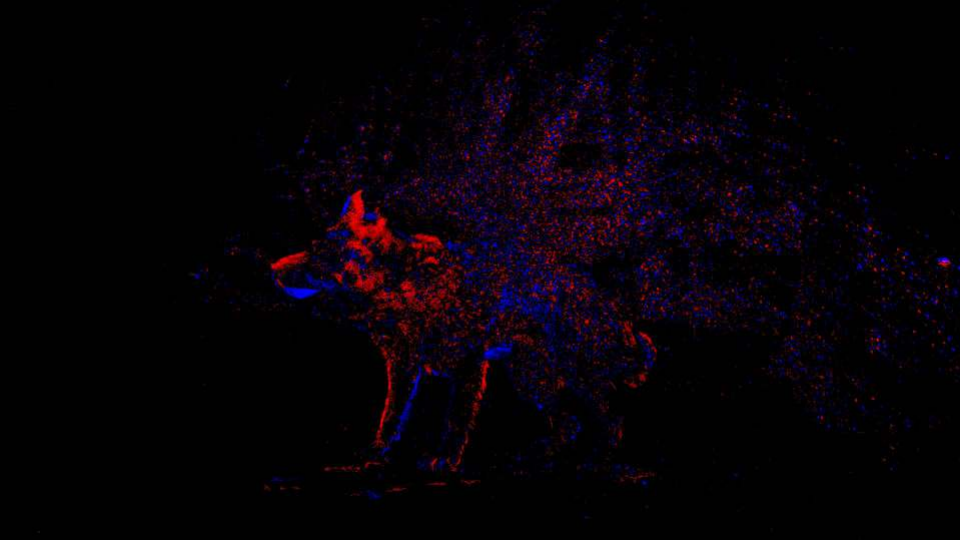}

    \par\vspace{0.3em}

    {\scriptsize\textbf{MCL-JCV - \texttt{videoSRC22}}\par}
    \vspace{0.3em}

    \centereventimg{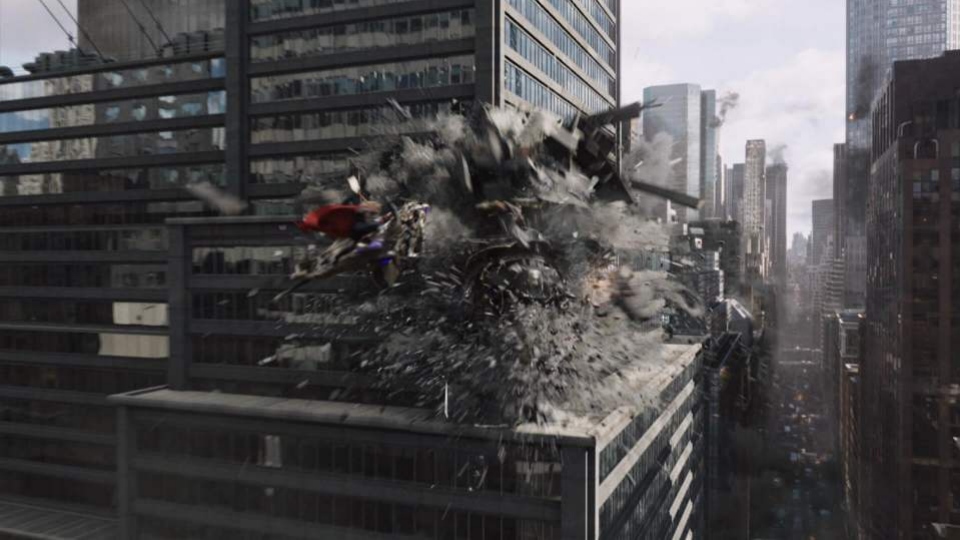}%
    \hspace{0.5em}
    \centereventimg{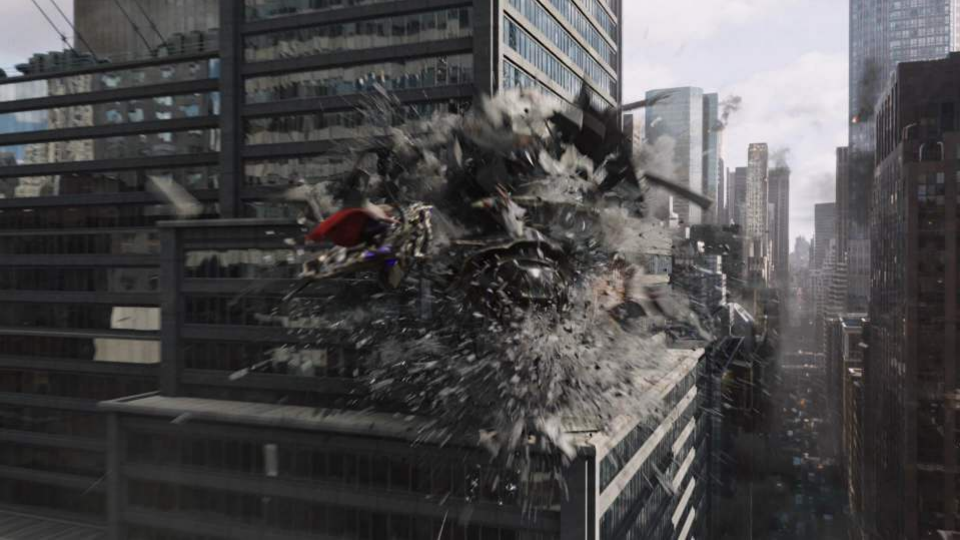}%
    \hspace{0.5em}
    \centereventimg{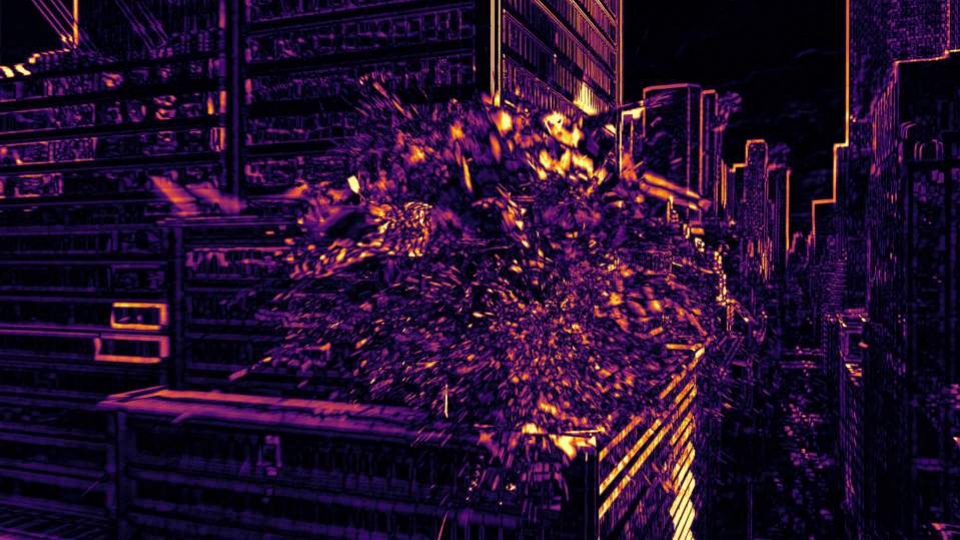}%
    \hspace{0.5em}
    \centereventimg{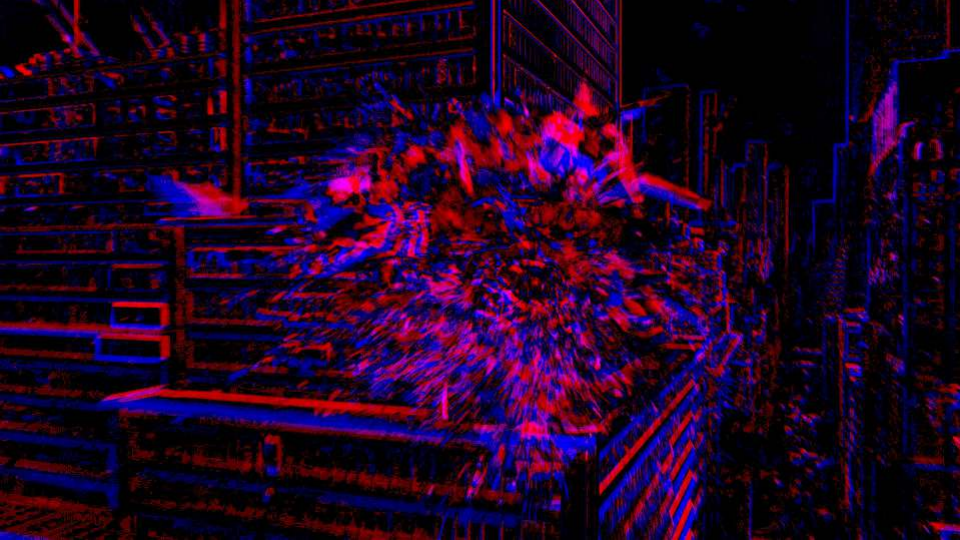}

    \par\vspace{0.3em}

    {\scriptsize\textbf{HEVC-B - \texttt{ParkScene}}\par}
    \vspace{0.3em}

    \centereventimg{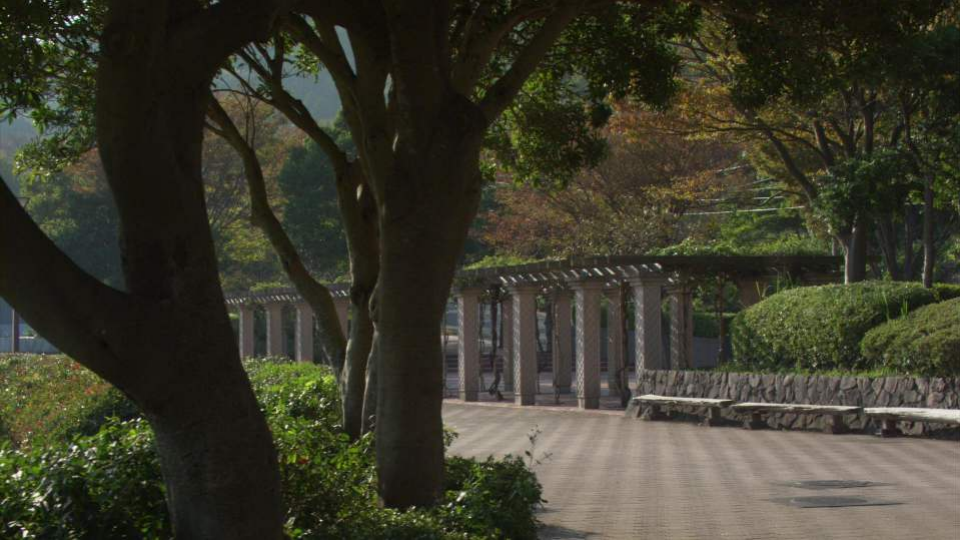}%
    \hspace{0.5em}
    \centereventimg{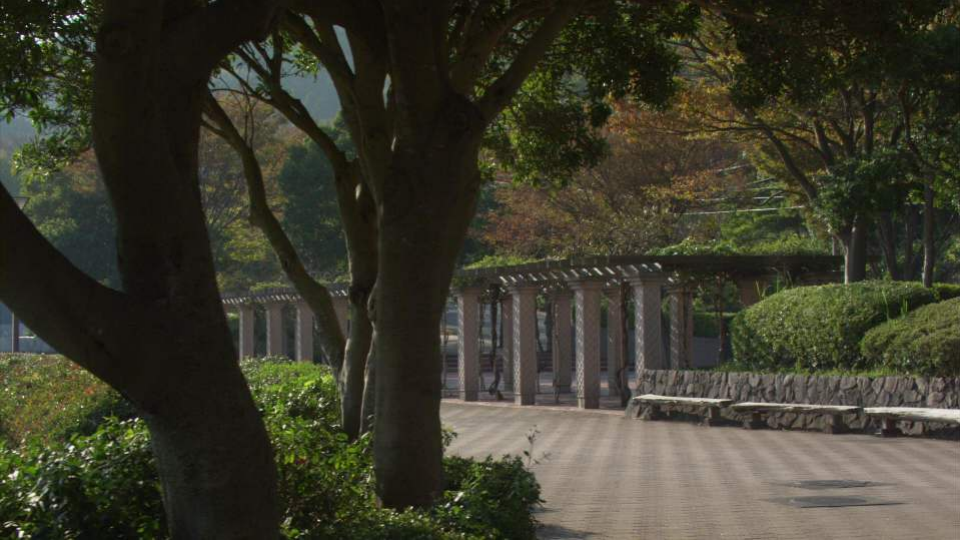}%
    \hspace{0.5em}
    \centereventimg{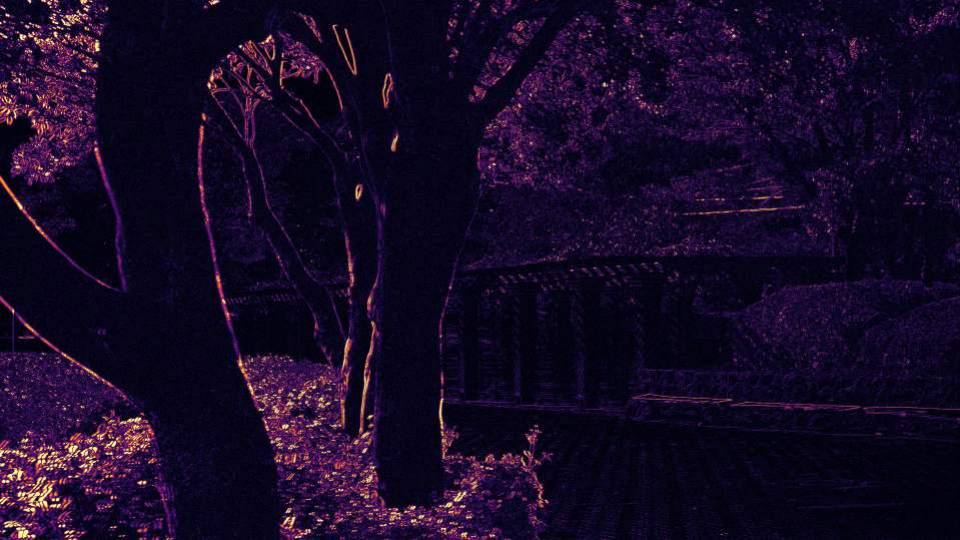}%
    \hspace{0.5em}
    \centereventimg{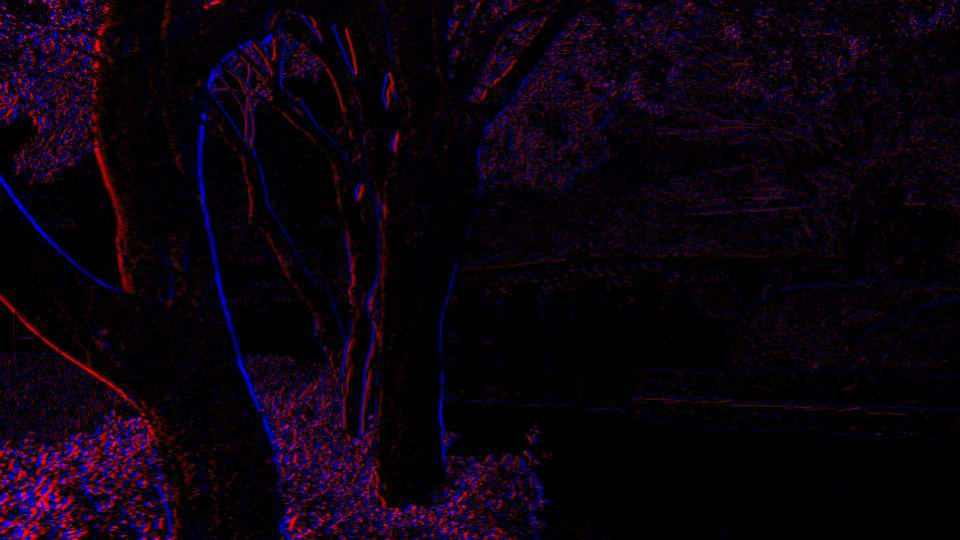}

    \par\vspace{0.3em}

    {\scriptsize\textbf{HEVC-C - \texttt{PartyScene}}\par}
    \vspace{0.3em}

    \centereventimg{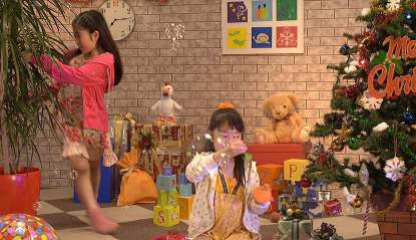}%
    \hspace{0.5em}
    \centereventimg{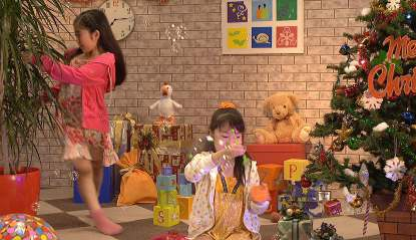}%
    \hspace{0.5em}
    \centereventimg{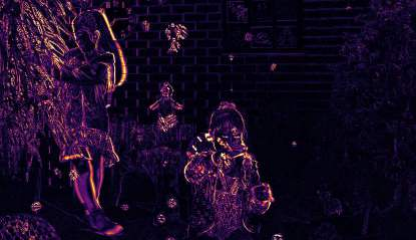}%
    \hspace{0.5em}
    \centereventimg{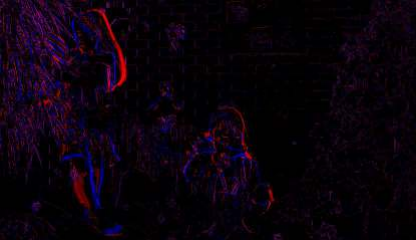}

    \par\vspace{0.3em}

    {\scriptsize\textbf{HEVC-D - \texttt{RaceHorses}}\par}
    \vspace{0.3em}

    \centereventimg{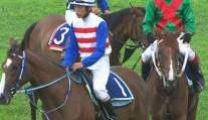}%
    \hspace{0.5em}
    \centereventimg{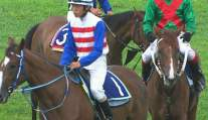}%
    \hspace{0.5em}
    \centereventimg{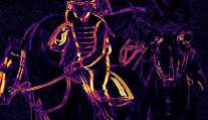}%
    \hspace{0.5em}
    \centereventimg{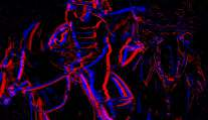}

    \par\vspace{0.3em}

    {\scriptsize\textbf{HEVC-E - \texttt{FourPeople}}\par}
    \vspace{0.3em}

    \centereventimg{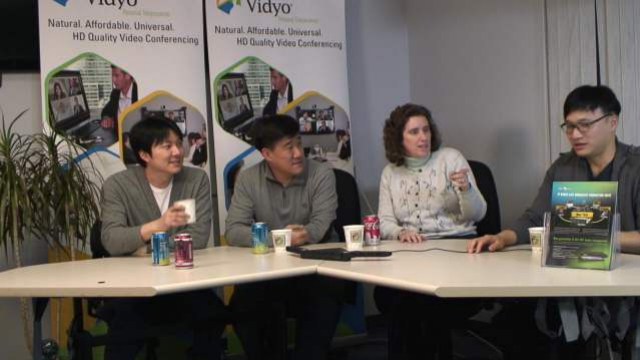}%
    \hspace{0.5em}
    \centereventimg{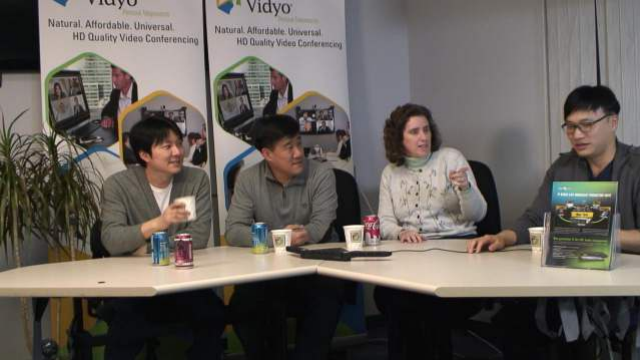}%
    \hspace{0.5em}
    \centereventimg{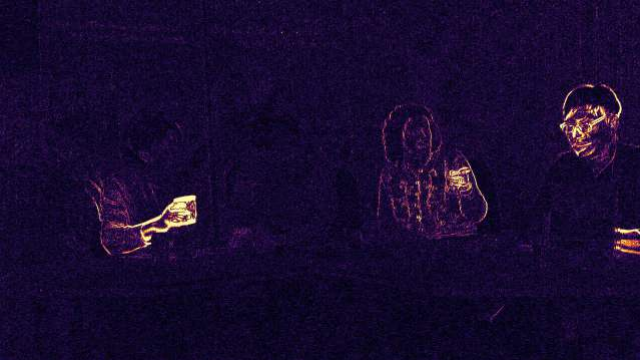}%
    \hspace{0.5em}
    \centereventimg{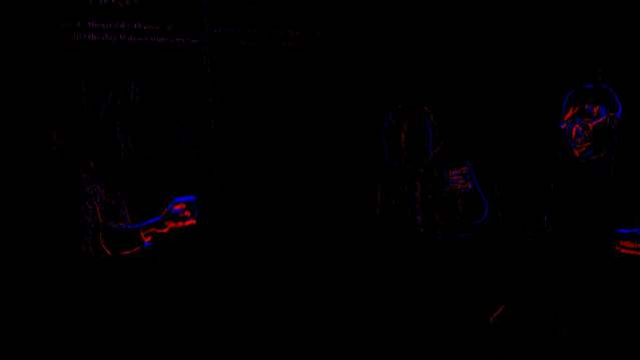}

    \par\vspace{0.3em}
    
    {\scriptsize\textbf{MCML-4K - \texttt{S06\_R\_00\_00}}\par}
    \vspace{0.3em}
    
    \centereventimg{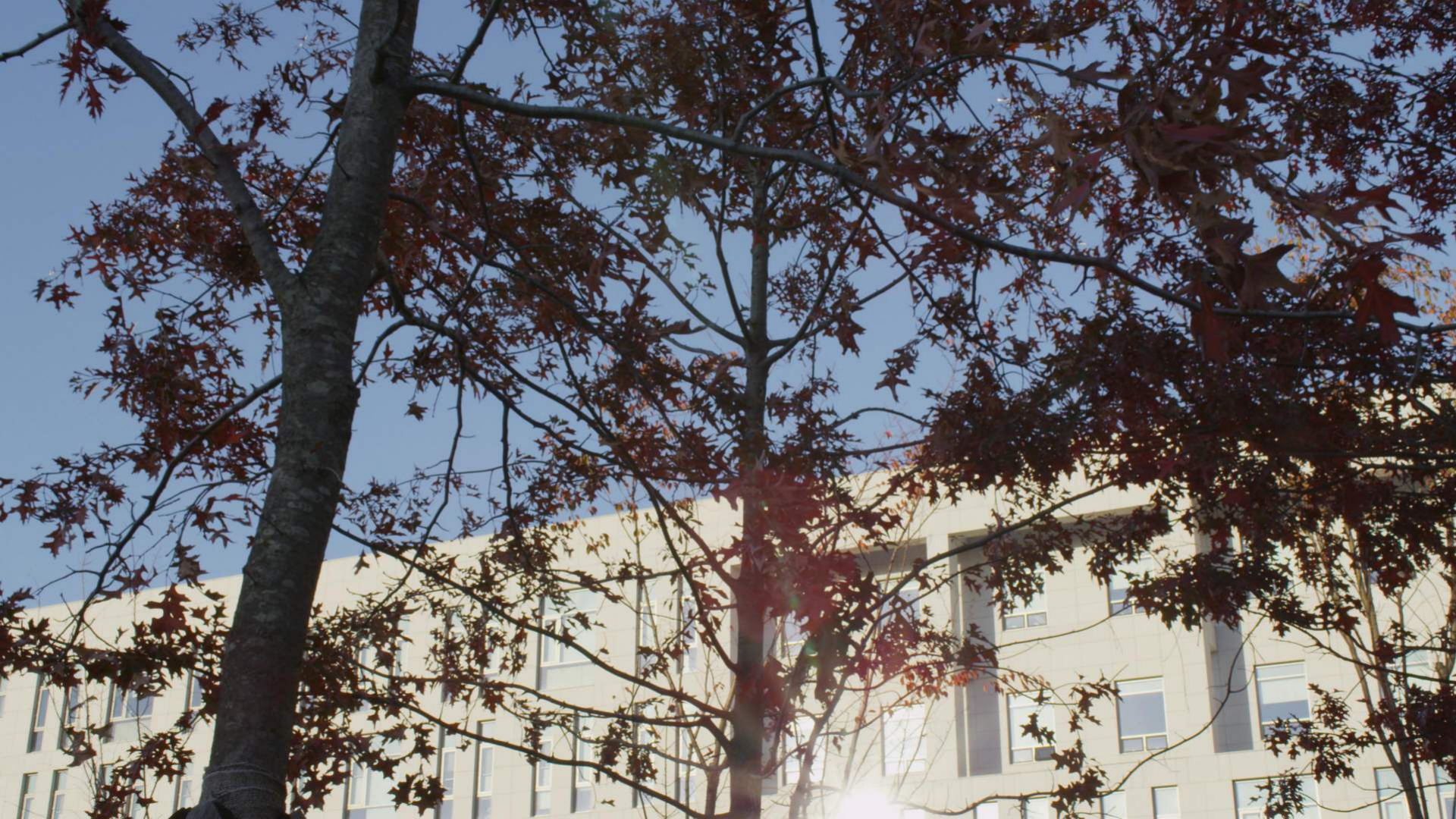}%
    \hspace{0.5em}
    \centereventimg{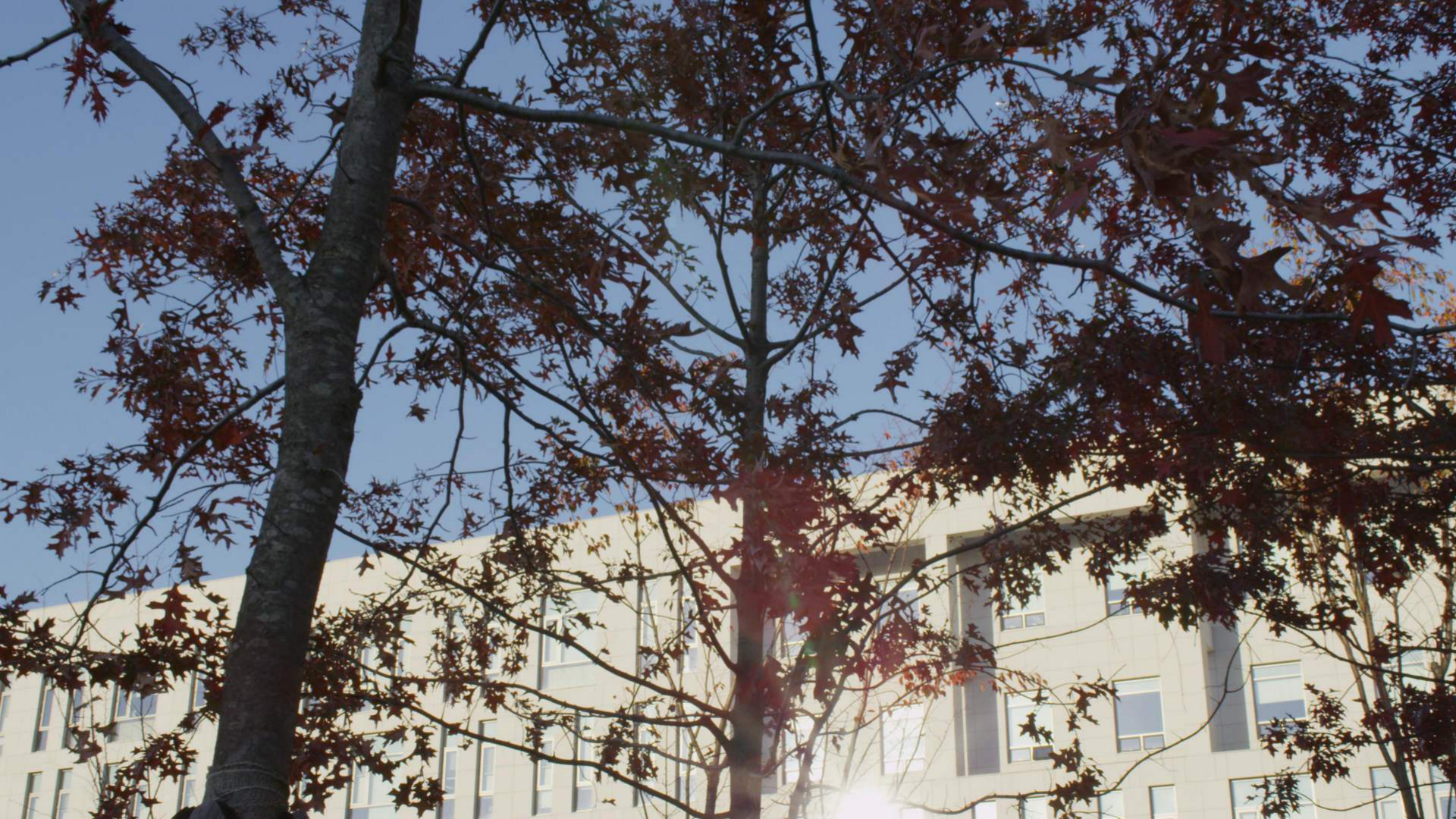}%
    \hspace{0.5em}
    \centereventimg{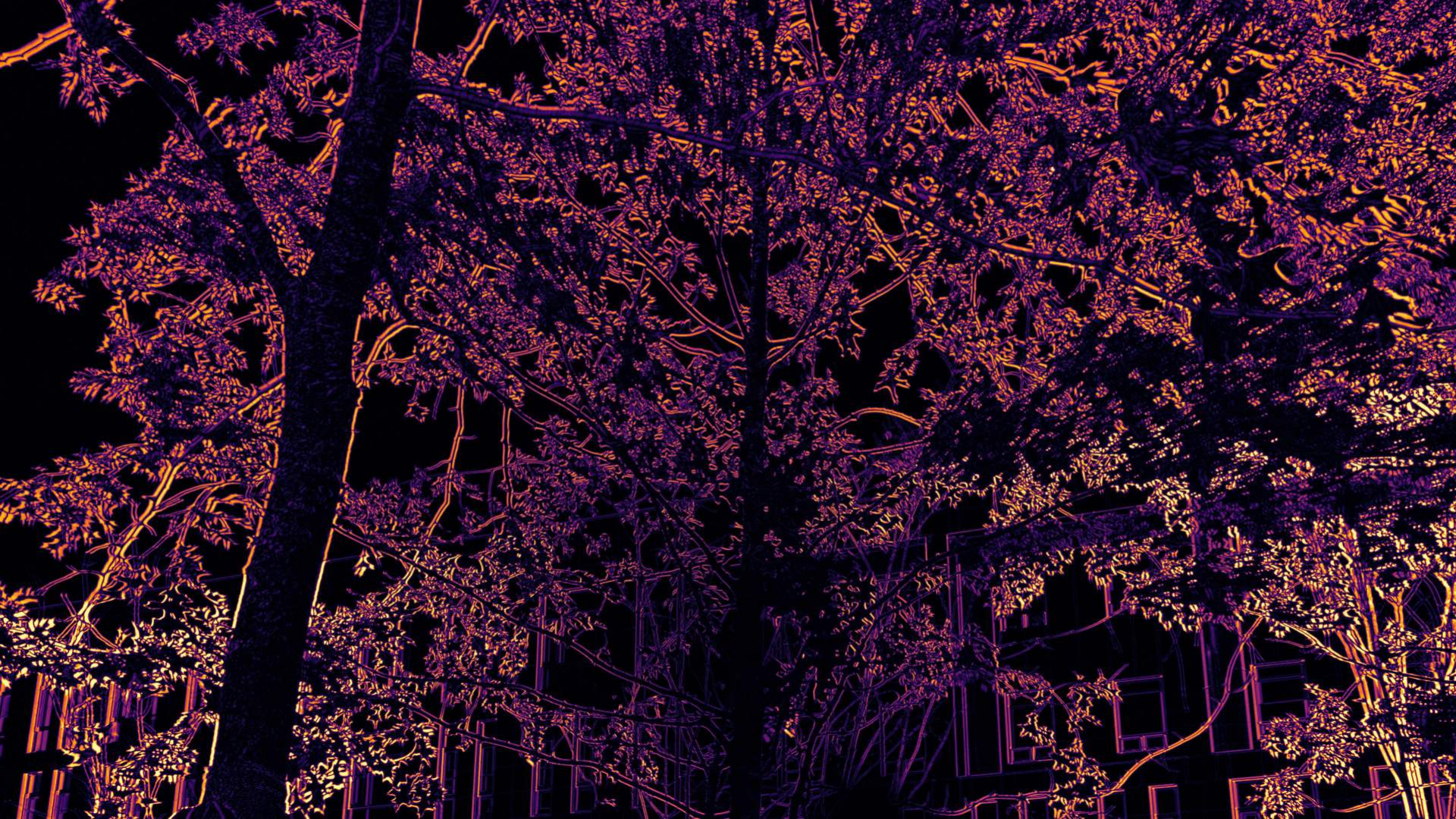}%
    \hspace{0.5em}
    \centereventimg{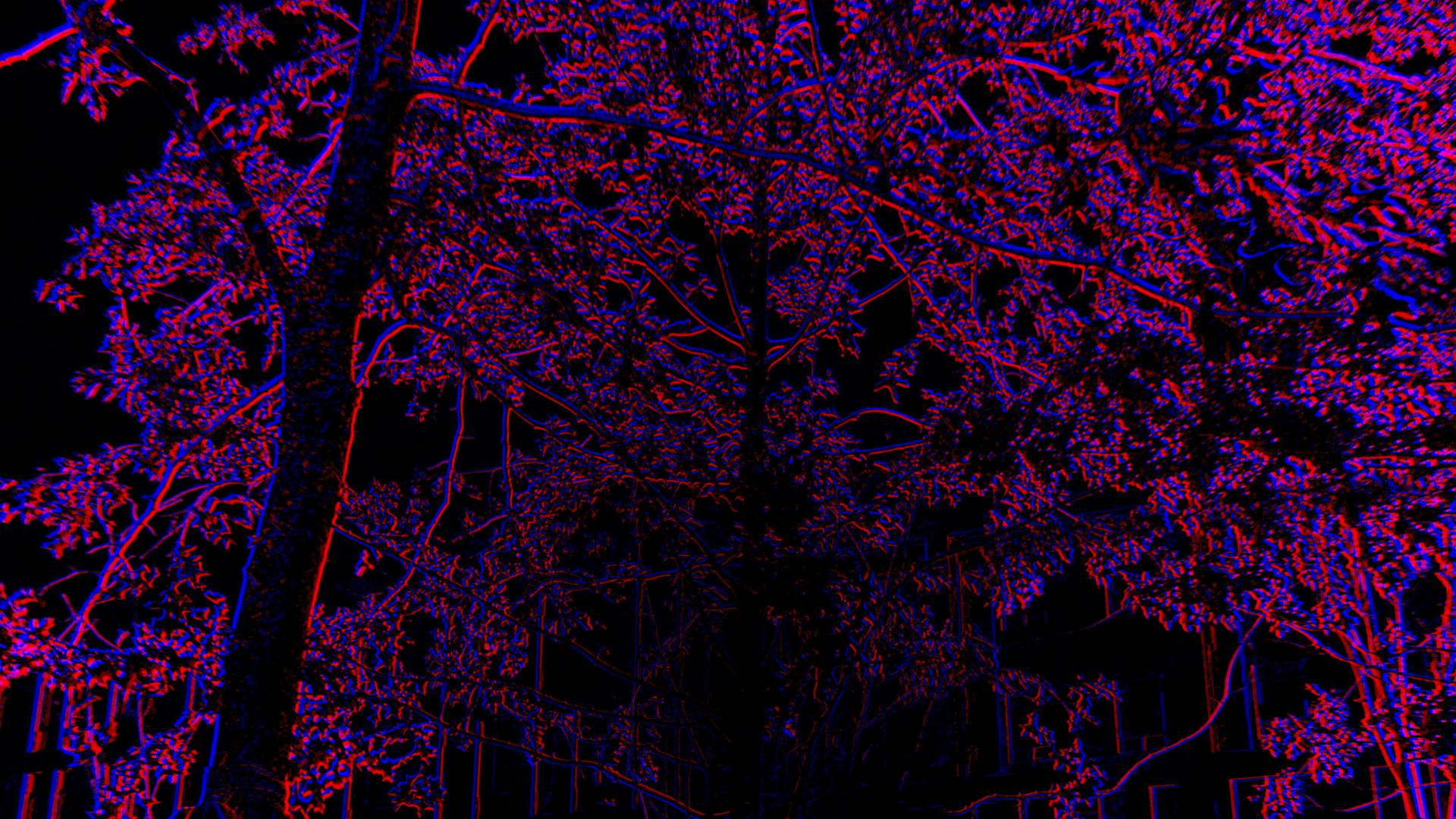}
    
    \par\vspace{0.3em}
    
    \par\vspace{0.3em}
    \noindent
    \makebox[\textwidth][c]{%
        \makebox[0.215\textwidth][c]{\small (a) Previous frame}%
        \hspace{0.5em}%
        \makebox[0.215\textwidth][c]{\small (b) Target frame}%
        \hspace{0.5em}%
        \makebox[0.215\textwidth][c]{\small (c) Frame difference}%
        \hspace{0.5em}%
        \makebox[0.215\textwidth][c]{\small (d) Events}%
    }
    \par
    \vspace{-0.2em}

    \caption{
        \textbf{Frame Changes and Corresponding Events.}
        From top to bottom, examples are shown for UVG, MCL-JCV,
        HEVC-B, HEVC-C, HEVC-D, HEVC-E, and MCML-4K respectively.
        For each sequence, we show the previous frame $\mathbf{X}_{t-1}$, target frame $\mathbf{X}_t$, their frame-difference map, and the synthetic events $\mathbf{E}_{t-1:t}$ generated for the corresponding interval.
    }
    \label{fig:event_frame_comparison}
\end{figure*}

\vspace{-0.8em}
\paragraph{Frame Changes and Event Activity.}
Fig.~\ref{fig:event_frame_comparison} visualizes the RGB frame pair, their frame-difference map, and the intervening events for the same sequences used in Fig.~\ref{fig:crop_comparison}.
The examples illustrate how event activity is concentrated around spatio-temporal changes occurring between consecutive RGB frames.

\begin{table}[h]
    \centering
    \small
    \caption{
        \textbf{BD-Rate (\%) against DCMVC using Additional Quality Metrics.}
        Each comparison uses PSNR-Y (left) and MS-SSIM (right).
        Negative values indicate bitrate savings.
    }
    \vspace{-0.5em}
    \label{tab:bd_rate_additional_metrics}
    \setlength{\tabcolsep}{5pt}
    \renewcommand{\arraystretch}{1.1}
    \resizebox{\linewidth}{!}{
        \begin{tabular}{lcccccc|c}
            \toprule
            Method
            & UVG
            & \shortstack{MCL-JCV}
            & HEVC-B
            & HEVC-C
            & HEVC-D
            & HEVC-E
            & Average \\
            \midrule
            DCMVC
            & 0.00 / 0.00
            & 0.00 / 0.00
            & 0.00 / 0.00
            & 0.00 / 0.00
            & 0.00 / 0.00
            & 0.00 / 0.00
            & 0.00 / 0.00
            \\
            \midrule

            DCVC
            & $+150.24$ / $+128.29$
            & $+117.41$ / $+95.52$
            & $+123.41$ / $+122.59$
            & $+191.51$ / $+125.20$
            & $+122.41$ / $+95.49$
            & $+254.76$ / $+202.22$
            & $+159.96$ / $+128.22$
            \\

            DCVC-TCM
            & $+55.06$ / $+41.91$
            & $+56.62$ / $+44.08$
            & $+45.82$ / $+43.67$
            & $+88.70$ / $+51.40$
            & $+56.26$ / $+37.91$
            & $+84.95$ / $+68.74$
            & $+64.57$ / $+47.95$
            \\

            DCVC-DC
            & $-1.66$ / $-3.19$
            & $-0.59$ / $-0.14$
            & $-4.01$ / $-0.60$
            & $-3.25$ / $-2.29$
            & $-4.30$ / $+1.07$
            & $-11.13$ / $-9.72$
            & $-4.16$ / $-2.48$
            \\

            DCVC-FM
            & $-17.14$ / $+6.74$
            & $-12.00$ / $+8.28$
            & $-21.34$ / $+7.49$
            & $-16.64$ / $+8.31$
            & $-12.98$ / $+4.07$
            & $-24.72$ / $-7.61$
            & $-17.47$ / $+4.55$
            \\

            DCVC-RT
            & $-7.22$ / $+9.95$
            & $+1.75$ / $+12.27$
            & $-8.32$ / $+12.97$
            & $+26.85$ / $+25.06$
            & $+36.53$ / $+24.50$
            & $-10.10$ / $-1.59$
            & $+6.58$ / $+13.86$
            \\
            \midrule
            \textbf{ENVC (Ours)}
            & $\mathbf{-47.47}$ / $\mathbf{-37.10}$
            & $\mathbf{-64.49}$ / $\mathbf{-52.60}$
            & $\mathbf{-69.16}$ / $\mathbf{-46.46}$
            & $\mathbf{-59.86}$ / $\mathbf{-39.46}$
            & $\mathbf{-58.34}$ / $\mathbf{-43.74}$
            & $\mathbf{-21.75}$ / $\mathbf{-20.65}$
            & $\mathbf{-53.51}$ / $\mathbf{-40.00}$
            \\
            \bottomrule
        \end{tabular}
    }
\end{table}

\begin{figure*}[h]
    \centering
    \begin{subfigure}[t]{0.325\textwidth}
        \centering
        \includegraphics[width=\linewidth]{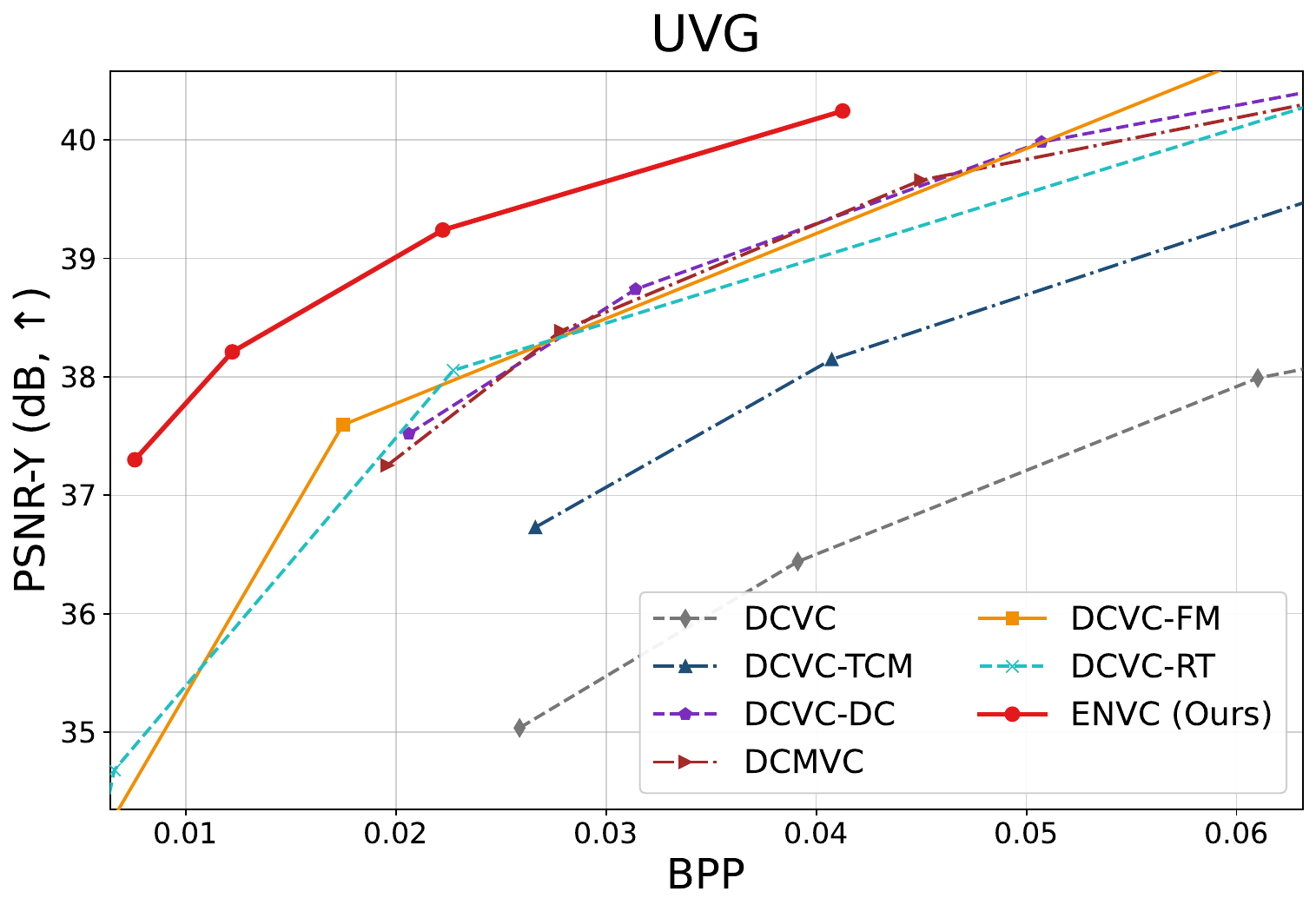}
        \vspace{-1.5em}
        \caption{UVG}
        \label{fig:rd_uvg_psnr_y}
    \end{subfigure}
    \begin{subfigure}[t]{0.325\textwidth}
        \centering
        \includegraphics[width=\linewidth]{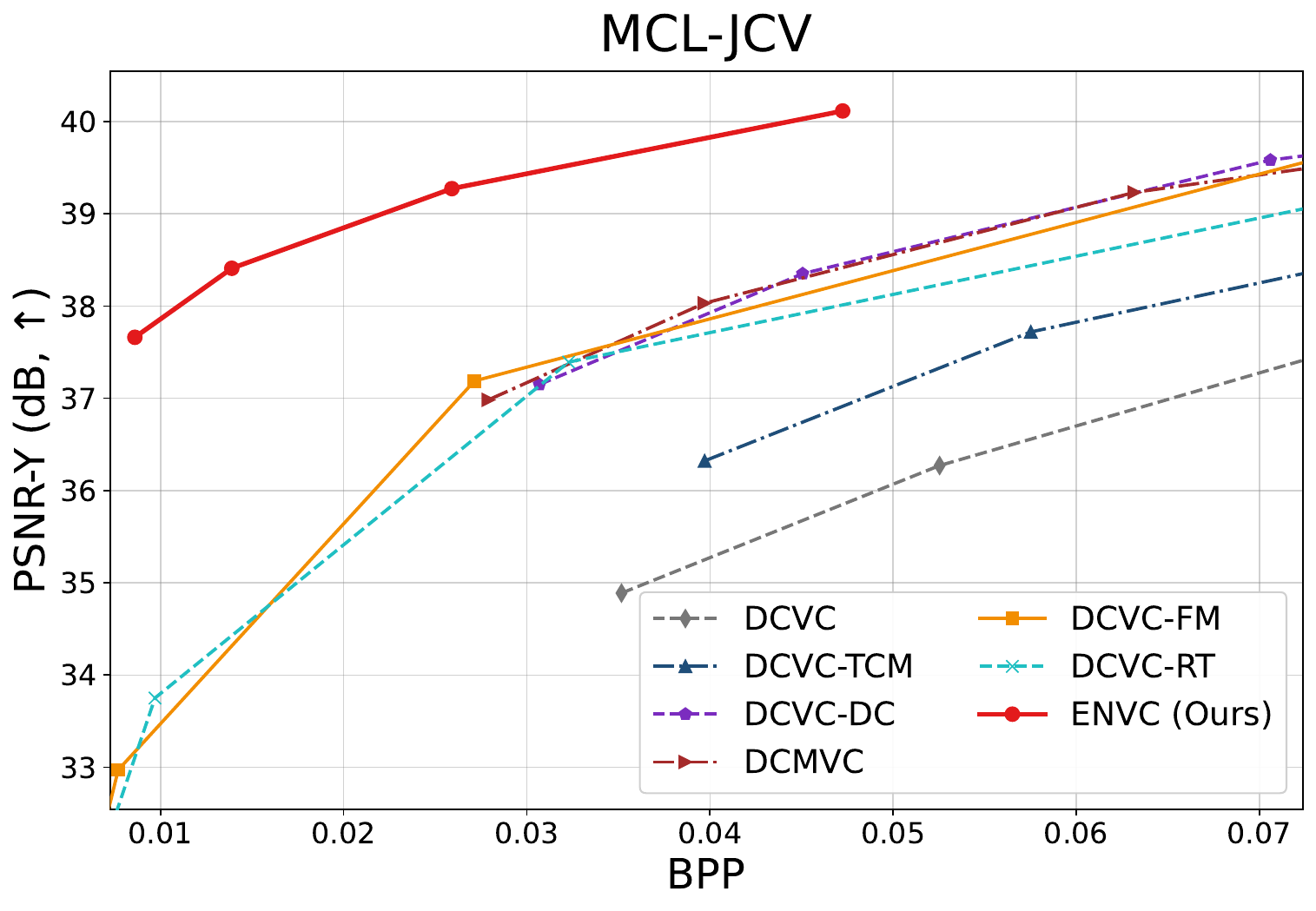}
        \vspace{-1.5em}
        \caption{MCL-JCV}
        \label{fig:rd_mcl_psnr_y}
    \end{subfigure}
    \begin{subfigure}[t]{0.325\textwidth}
        \centering
        \includegraphics[width=\linewidth]{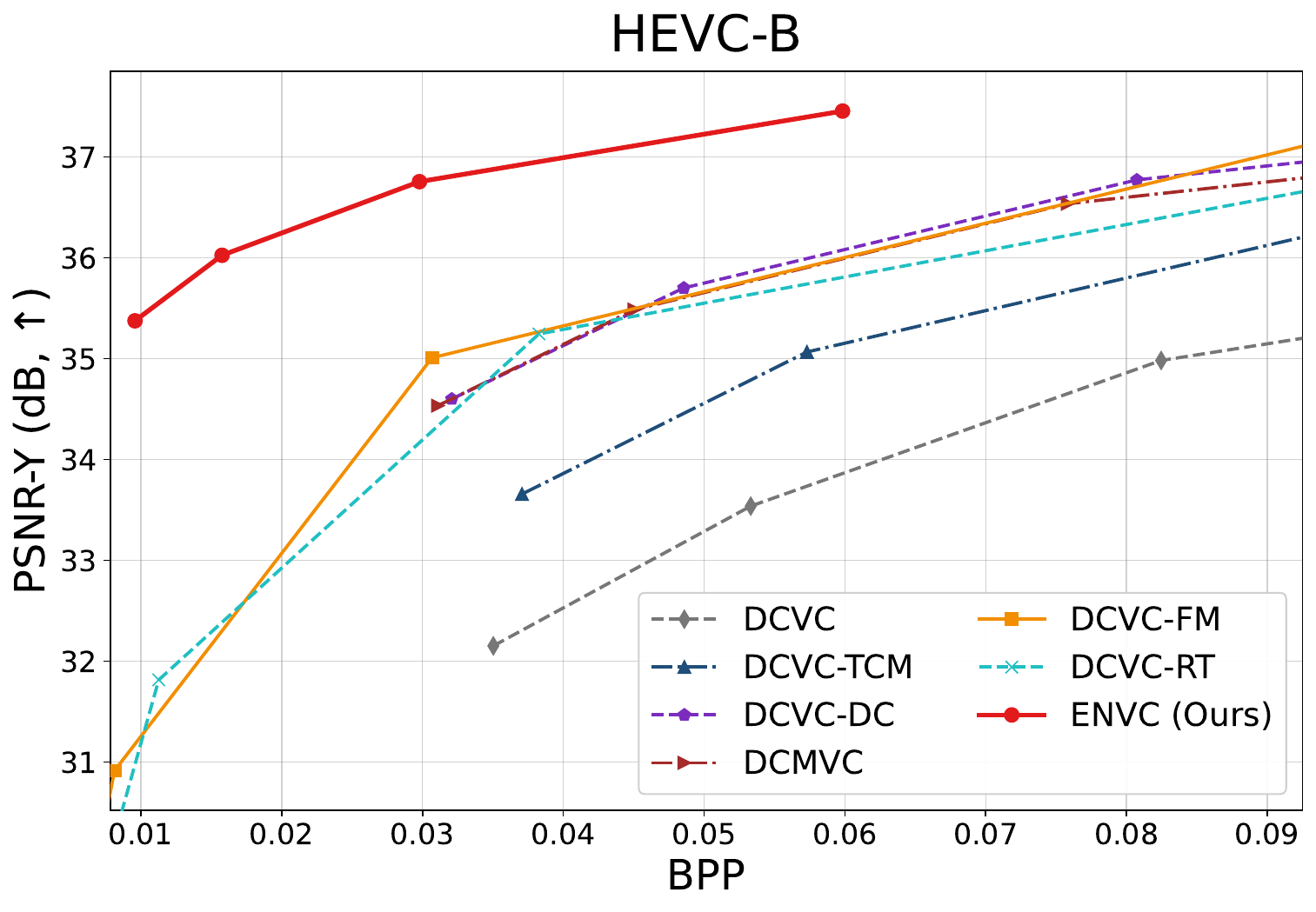}
        \vspace{-1.5em}
        \caption{HEVC-B}
        \label{fig:rd_hevcb_psnr_y}
    \end{subfigure}
    \vspace{0.1em}
    \begin{subfigure}[t]{0.325\textwidth}
        \centering
        \includegraphics[width=\linewidth]{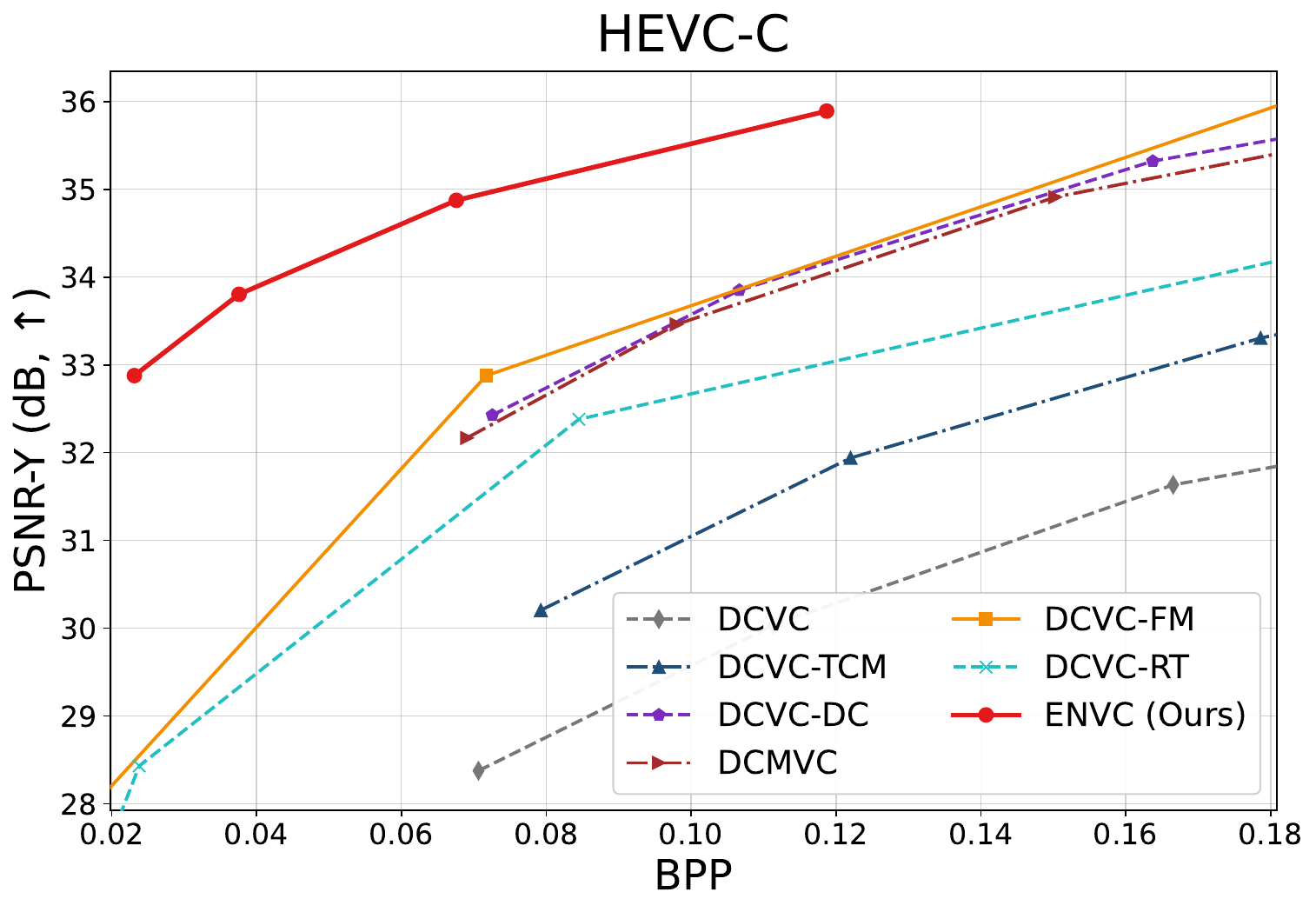}
        \vspace{-1.5em}
        \caption{HEVC-C}
        \label{fig:rd_hevcc_psnr_y}
    \end{subfigure}
    \begin{subfigure}[t]{0.325\textwidth}
        \centering
        \includegraphics[width=\linewidth]{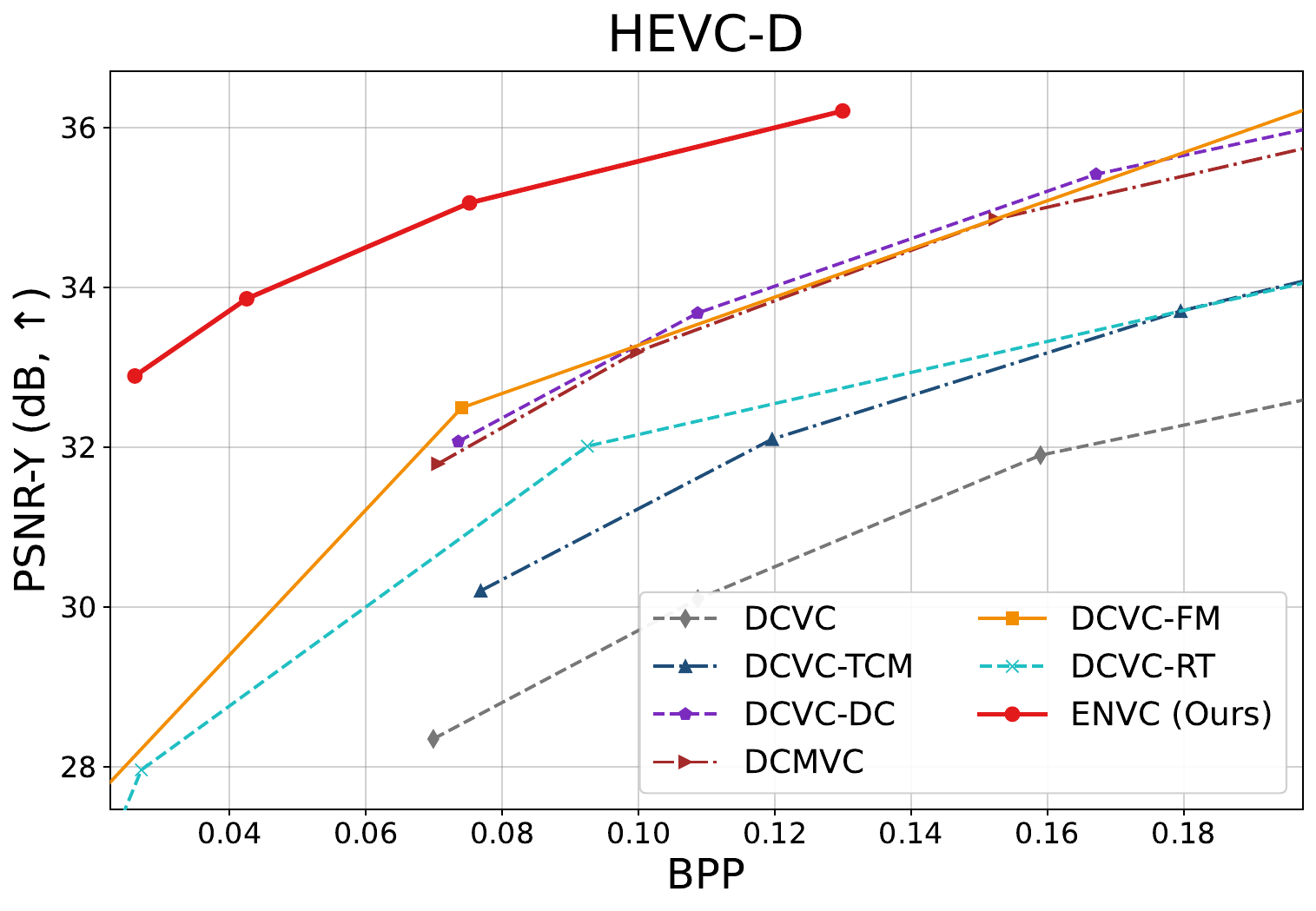}
        \vspace{-1.5em}
        \caption{HEVC-D}
        \label{fig:rd_hevcd_psnr_y}
    \end{subfigure}
    \begin{subfigure}[t]{0.325\textwidth}
        \centering
        \includegraphics[width=\linewidth]{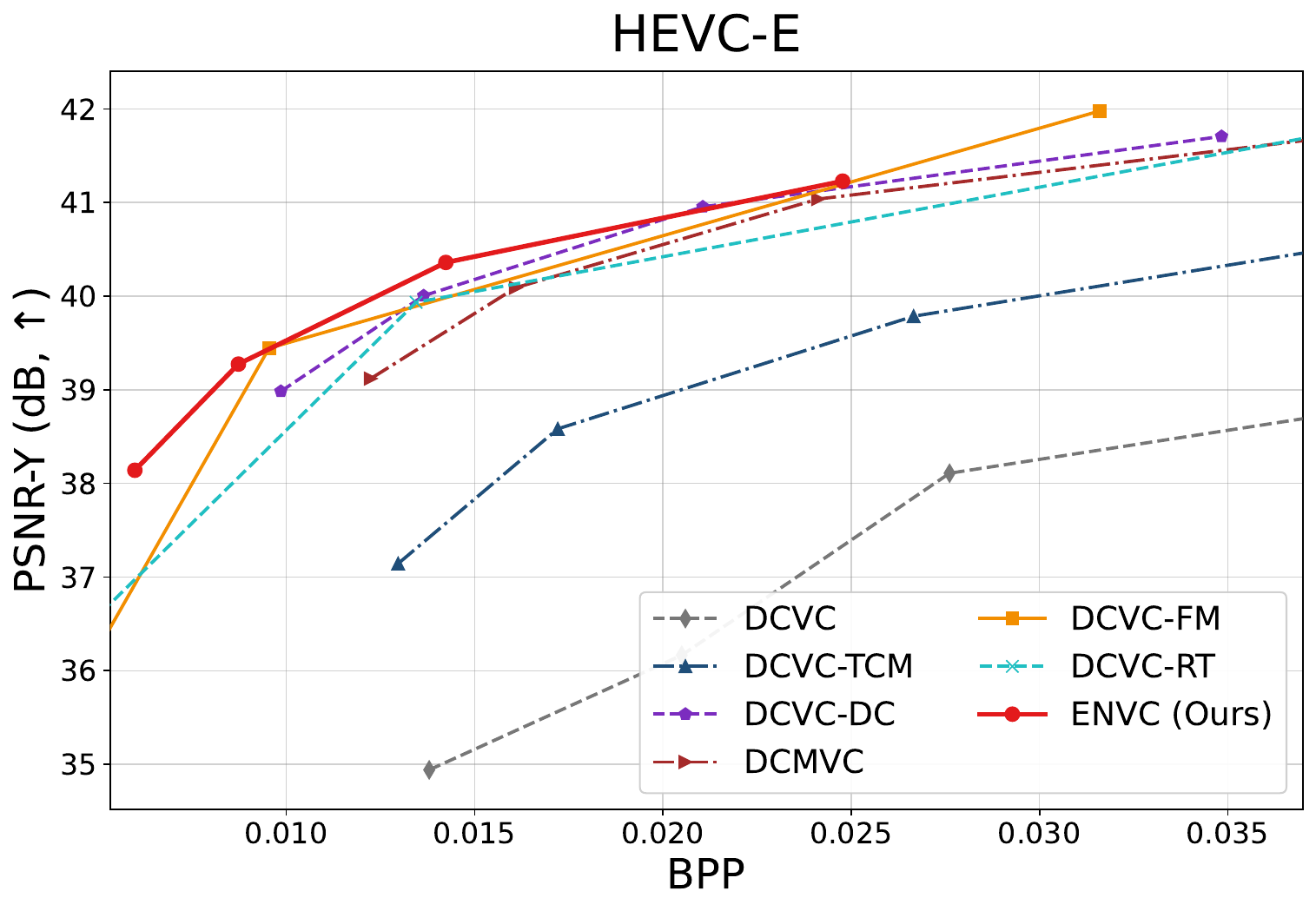}
        \vspace{-1.5em}
        \caption{HEVC-E}
        \label{fig:rd_hevce_psnr_y}
    \end{subfigure}
    \vspace{-0.5em}
    \caption{
        \textbf{Rate--Distortion Performance in PSNR-Y.}
        PSNR-Y ($\uparrow$) versus bitrate on six benchmarks.
    }
    \label{fig:rd_curves_psnr_y}
    \vspace{-1.0em}
\end{figure*}

\begin{figure*}[h]
    \centering
    \begin{subfigure}[t]{0.325\textwidth}
        \centering
        \includegraphics[width=\linewidth]{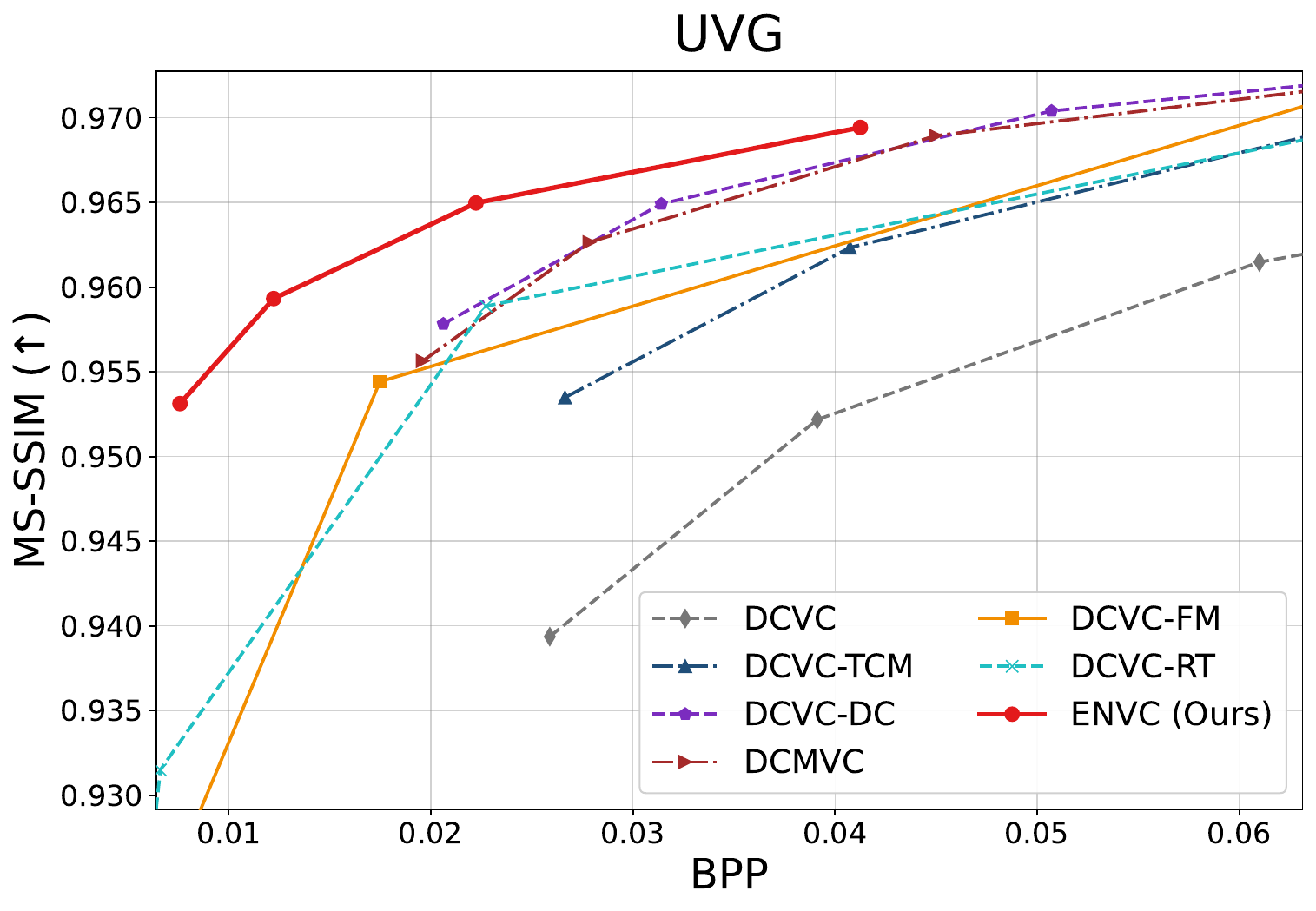}
        \vspace{-1.5em}
        \caption{UVG}
        \label{fig:rd_uvg_ms_ssim}
    \end{subfigure}
    \begin{subfigure}[t]{0.325\textwidth}
        \centering
        \includegraphics[width=\linewidth]{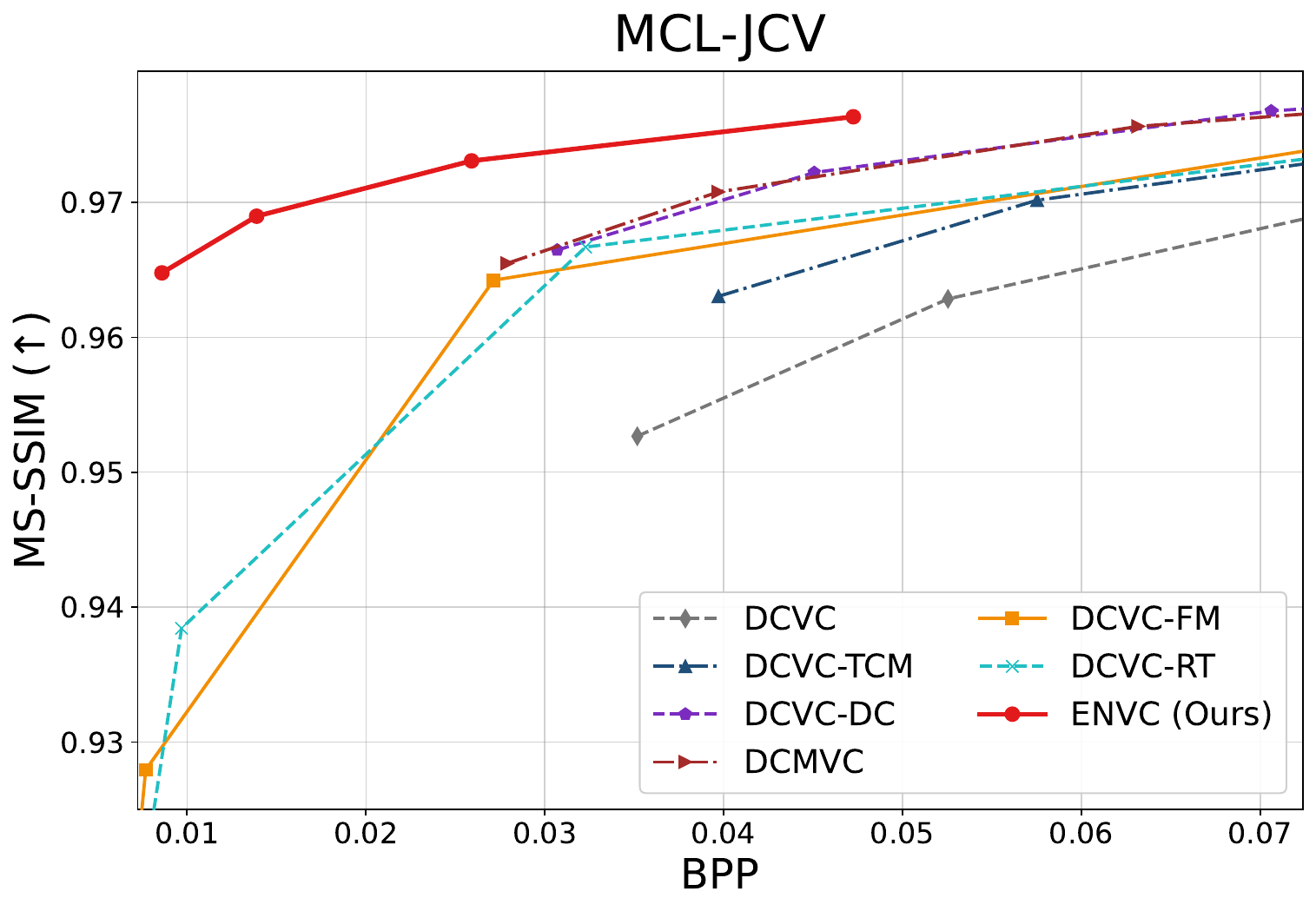}
        \vspace{-1.5em}
        \caption{MCL-JCV}
        \label{fig:rd_mcl_ms_ssim}
    \end{subfigure}
    \begin{subfigure}[t]{0.325\textwidth}
        \centering
        \includegraphics[width=\linewidth]{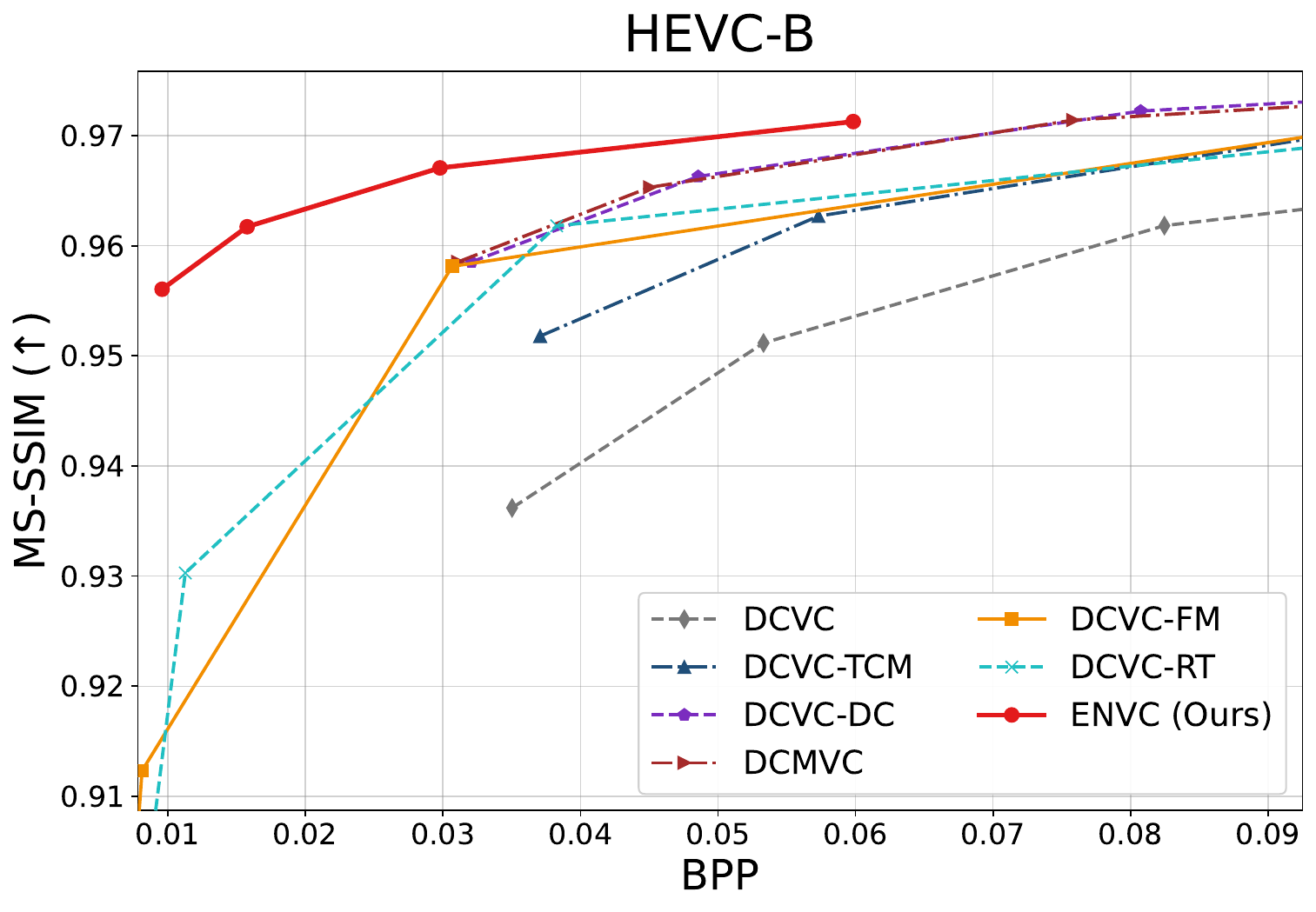}
        \vspace{-1.5em}
        \caption{HEVC-B}
        \label{fig:rd_hevcb_ms_ssim}
    \end{subfigure}
    \vspace{0.1em}
    \begin{subfigure}[t]{0.325\textwidth}
        \centering
        \includegraphics[width=\linewidth]{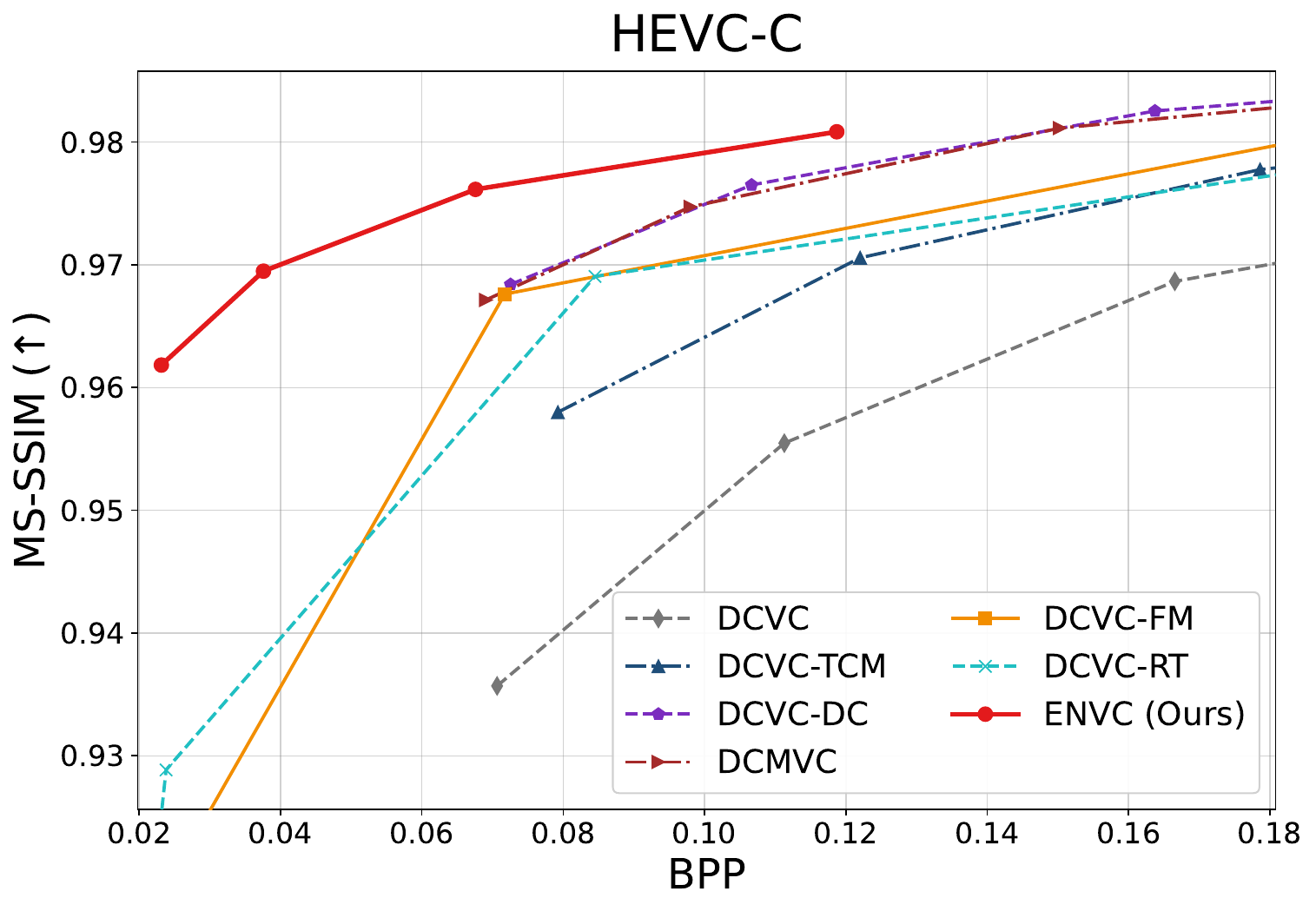}
        \vspace{-1.5em}
        \caption{HEVC-C}
        \label{fig:rd_hevcc_ms_ssim}
    \end{subfigure}
    \begin{subfigure}[t]{0.325\textwidth}
        \centering
        \includegraphics[width=\linewidth]{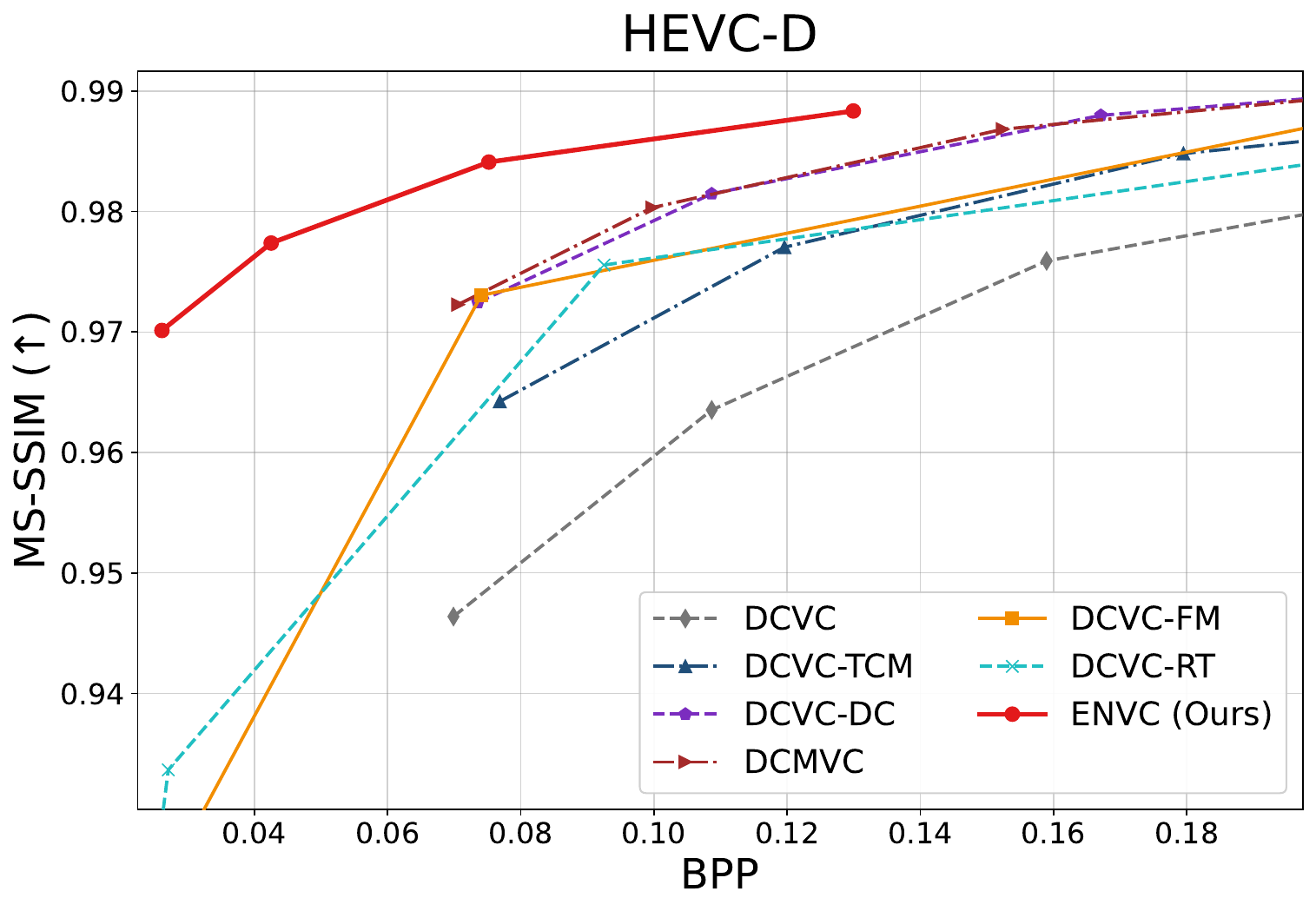}
        \vspace{-1.5em}
        \caption{HEVC-D}
        \label{fig:rd_hevcd_ms_ssim}
    \end{subfigure}
    \begin{subfigure}[t]{0.325\textwidth}
        \centering
        \includegraphics[width=\linewidth]{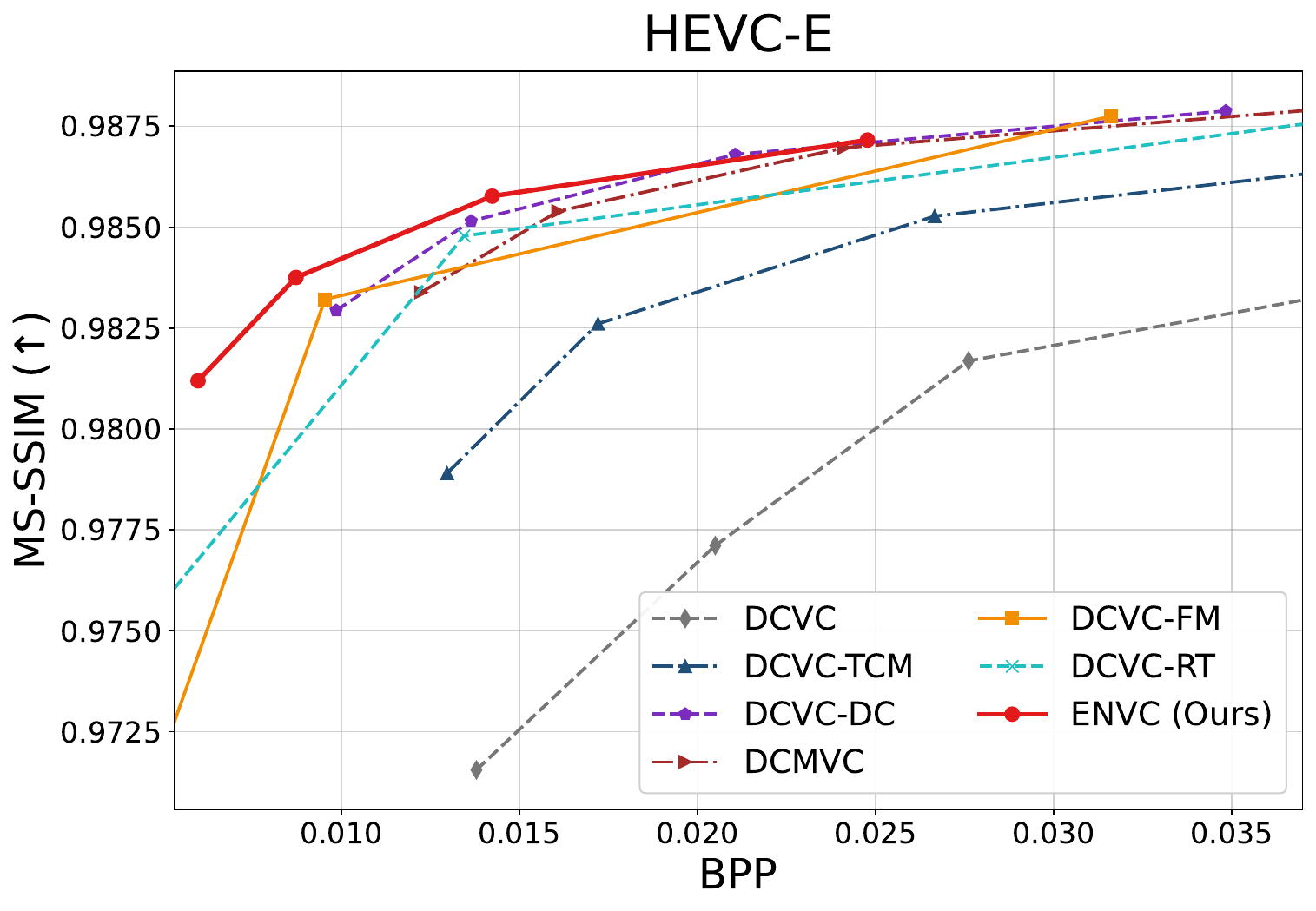}
        \vspace{-1.5em}
        \caption{HEVC-E}
        \label{fig:rd_hevce_ms_ssim}
    \end{subfigure}
    \vspace{-0.5em}
    \caption{
        \textbf{Rate--Distortion Performance in MS-SSIM.}
        MS-SSIM ($\uparrow$) versus bitrate on six benchmarks.
    }
    \label{fig:rd_curves_ms_ssim}
    \vspace{-1.0em}
\end{figure*}

